\documentclass[longauth]{aa}

\usepackage{graphicx}
\usepackage{natbib}
\usepackage{scalerel}
\usepackage{csquotes}
\usepackage[table]{xcolor}
\usepackage[nolist,nohyperlinks]{acronym}
\usepackage{soul}
\usepackage{booktabs}
\usepackage{tabularx}

\usepackage{txfonts}
\usepackage[pdfencoding=auto,psdextra]{hyperref}
\hypersetup{
    colorlinks=true,
    linkcolor=blue,
    filecolor=magenta,      
    urlcolor=blue,
    citecolor=blue
}
\usepackage{orcidlink} 
\newcommand{\orcid}[1]{\orcidlink{#1}}

\makeatletter
\renewcommand*\aa@pageof{, page \thepage{} of \pageref*{LastPage}}
\makeatother
\usepackage[utf8]{inputenc}

\usepackage[switch, modulo]{lineno}
\usepackage[nameinlink,capitalise]{cleveref}
\crefname{section}{Sect.}{Sects.}
\Crefname{section}{Section}{Sections}
\crefname{figure}{Fig.}{Figs.}
\Crefname{figure}{Figure}{Figures}
\crefname{equation}{Eq.}{Eqs.}
\Crefname{equation}{Equation}{Equations}
\crefname{table}{Table}{Tables}
\crefname{appendix}{Appendix}{Appendices}

\usepackage{euclid}
\newcommand*{\gaia}{\textit{Gaia}\xspace}
\newcommand\gdr[1]{\gaia DR#1}

\newcommand\ztemp{\ensuremath{z_\mathrm{temp}}\xspace}
\newcommand\zvi{\ensuremath{z_\mathrm{vi}}\xspace}
\newcommand\zgaia{\ensuremath{z_{\gaia}}\xspace}
\newcommand\zdesi{\ensuremath{z_\mathrm{DESI}}\xspace}
\newcommand\mumaxmag{\ensuremath{\mu_{\mathrm{max}}-\mathrm{mag}}\xspace}

\defcitealias{Q1-SP023}{T25}

\begin{document}

\title{Euclid Quick Data Release (Q1)} \subtitle{\Euclid spectroscopy of quasars. 1. Identification and redshift determination of 3500 bright quasars\thanks{The full catalogue of the identified quasars and the composite spectra are available in electronic form at the CDS via anonymous ftp to \url{cdsarc.u-strasbg.fr} (\url{130.79.128.5}) or via \url{http://cdsweb.u-strasbg.fr/cgi-bin/qcat?J/A+A/}.}}

\author{Euclid Collaboration: Y.~Fu\orcid{0000-0002-0759-0504}\thanks{\email{yfu@strw.leidenuniv.nl}}\inst{\ref{aff1},\ref{aff2}}
\and R.~Bouwens\orcid{0000-0002-4989-2471}\inst{\ref{aff1}}
\and K.~I.~Caputi\orcid{0000-0001-8183-1460}\inst{\ref{aff2},\ref{aff3}}
\and D.~Vergani\orcid{0000-0003-0898-2216}\inst{\ref{aff4}}
\and M.~Scialpi\orcid{0009-0006-5100-4986}\inst{\ref{aff5},\ref{aff6},\ref{aff7}}
\and B.~Margalef-Bentabol\orcid{0000-0001-8702-7019}\inst{\ref{aff8}}
\and L.~Wang\orcid{0000-0002-6736-9158}\inst{\ref{aff8},\ref{aff2}}
\and M.~Bolzonella\orcid{0000-0003-3278-4607}\inst{\ref{aff4}}
\and M.~Banerji\orcid{0000-0002-0639-5141}\inst{\ref{aff9}}
\and E.~Ba\~nados\orcid{0000-0002-2931-7824}\inst{\ref{aff10}}
\and A.~Feltre\orcid{0000-0001-6865-2871}\inst{\ref{aff7}}
\and Y.~Toba\orcid{0000-0002-3531-7863}\inst{\ref{aff11},\ref{aff12}}
\and J.~Calhau\orcid{0000-0003-1803-6899}\inst{\ref{aff13}}
\and F.~Tarsitano\orcid{0000-0002-5919-0238}\inst{\ref{aff14},\ref{aff15}}
\and P.~A.~C.~Cunha\orcid{0000-0002-9454-859X}\inst{\ref{aff16},\ref{aff17},\ref{aff18}}
\and A.~Humphrey\orcid{0000-0002-0510-2351}\inst{\ref{aff18},\ref{aff19}}
\and G.~Vietri\orcid{0000-0001-9155-8875}\inst{\ref{aff20}}
\and F.~Mannucci\orcid{0000-0002-4803-2381}\inst{\ref{aff7}}
\and S.~Bisogni\orcid{0000-0003-3746-4565}\inst{\ref{aff20}}
\and F.~Ricci\orcid{0000-0001-5742-5980}\inst{\ref{aff21},\ref{aff22}}
\and H.~Landt\orcid{0000-0001-8391-6900}\inst{\ref{aff23}}
\and L.~Spinoglio\orcid{0000-0001-8840-1551}\inst{\ref{aff24}}
\and T.~Matamoro~Zatarain\orcid{0009-0007-2976-293X}\inst{\ref{aff25}}
\and D.~Stern\orcid{0000-0003-2686-9241}\inst{\ref{aff26}}
\and M.~J.~Page\orcid{0000-0002-6689-6271}\inst{\ref{aff27}}
\and D.~M.~Alexander\orcid{0000-0002-5896-6313}\inst{\ref{aff23}}
\and G.~Zamorani\orcid{0000-0002-2318-301X}\inst{\ref{aff4}}
\and W.~Roster\orcid{0000-0002-9149-6528}\inst{\ref{aff28}}
\and M.~Salvato\orcid{0000-0001-7116-9303}\inst{\ref{aff28}}
\and Y.~Copin\orcid{0000-0002-5317-7518}\inst{\ref{aff29}}
\and J.~G.~Sorce\orcid{0000-0002-2307-2432}\inst{\ref{aff30},\ref{aff31}}
\and D.~Scott\orcid{0000-0002-6878-9840}\inst{\ref{aff32}}
\and Y.-H.~Zhang\orcid{0000-0003-4916-6346}\inst{\ref{aff1},\ref{aff33}}
\and E.~Lusso\orcid{0000-0003-0083-1157}\inst{\ref{aff5},\ref{aff7}}
\and J.~Wolf\orcid{0000-0003-0643-7935}\inst{\ref{aff10}}
\and D.~Yang\orcid{0000-0002-6769-0910}\inst{\ref{aff1}}
\and H.~J.~A.~Rottgering\orcid{0000-0001-8887-2257}\inst{\ref{aff1}}
\and B.~Laloux\orcid{0000-0001-9996-9732}\inst{\ref{aff13},\ref{aff28}}
\and M.~Siudek\orcid{0000-0002-2949-2155}\inst{\ref{aff34},\ref{aff35}}
\and S.~Belladitta\orcid{0000-0003-4747-4484}\inst{\ref{aff10},\ref{aff4}}
\and Q.~Liu\orcid{0000-0002-7490-5991}\inst{\ref{aff1}}
\and V.~Allevato\orcid{0000-0001-7232-5152}\inst{\ref{aff13}}
\and K.~Kuijken\orcid{0000-0002-3827-0175}\inst{\ref{aff1}}
\and S.~Andreon\orcid{0000-0002-2041-8784}\inst{\ref{aff36}}
\and N.~Auricchio\orcid{0000-0003-4444-8651}\inst{\ref{aff4}}
\and C.~Baccigalupi\orcid{0000-0002-8211-1630}\inst{\ref{aff37},\ref{aff38},\ref{aff39},\ref{aff40}}
\and M.~Baldi\orcid{0000-0003-4145-1943}\inst{\ref{aff16},\ref{aff4},\ref{aff41}}
\and A.~Balestra\orcid{0000-0002-6967-261X}\inst{\ref{aff42}}
\and S.~Bardelli\orcid{0000-0002-8900-0298}\inst{\ref{aff4}}
\and P.~Battaglia\orcid{0000-0002-7337-5909}\inst{\ref{aff4}}
\and A.~Biviano\orcid{0000-0002-0857-0732}\inst{\ref{aff38},\ref{aff37}}
\and E.~Branchini\orcid{0000-0002-0808-6908}\inst{\ref{aff43},\ref{aff44},\ref{aff36}}
\and M.~Brescia\orcid{0000-0001-9506-5680}\inst{\ref{aff45},\ref{aff13}}
\and J.~Brinchmann\orcid{0000-0003-4359-8797}\inst{\ref{aff18},\ref{aff17},\ref{aff46}}
\and S.~Camera\orcid{0000-0003-3399-3574}\inst{\ref{aff47},\ref{aff48},\ref{aff49}}
\and G.~Ca\~nas-Herrera\orcid{0000-0003-2796-2149}\inst{\ref{aff50},\ref{aff1}}
\and V.~Capobianco\orcid{0000-0002-3309-7692}\inst{\ref{aff49}}
\and C.~Carbone\orcid{0000-0003-0125-3563}\inst{\ref{aff20}}
\and J.~Carretero\orcid{0000-0002-3130-0204}\inst{\ref{aff51},\ref{aff52}}
\and S.~Casas\orcid{0000-0002-4751-5138}\inst{\ref{aff53},\ref{aff54}}
\and M.~Castellano\orcid{0000-0001-9875-8263}\inst{\ref{aff22}}
\and G.~Castignani\orcid{0000-0001-6831-0687}\inst{\ref{aff4}}
\and S.~Cavuoti\orcid{0000-0002-3787-4196}\inst{\ref{aff13},\ref{aff55}}
\and K.~C.~Chambers\orcid{0000-0001-6965-7789}\inst{\ref{aff56}}
\and A.~Cimatti\inst{\ref{aff57}}
\and C.~Colodro-Conde\inst{\ref{aff58}}
\and G.~Congedo\orcid{0000-0003-2508-0046}\inst{\ref{aff33}}
\and C.~J.~Conselice\orcid{0000-0003-1949-7638}\inst{\ref{aff59}}
\and L.~Conversi\orcid{0000-0002-6710-8476}\inst{\ref{aff60},\ref{aff61}}
\and A.~Costille\inst{\ref{aff62}}
\and F.~Courbin\orcid{0000-0003-0758-6510}\inst{\ref{aff63},\ref{aff64}}
\and H.~M.~Courtois\orcid{0000-0003-0509-1776}\inst{\ref{aff65}}
\and M.~Cropper\orcid{0000-0003-4571-9468}\inst{\ref{aff27}}
\and A.~Da~Silva\orcid{0000-0002-6385-1609}\inst{\ref{aff66},\ref{aff67}}
\and H.~Degaudenzi\orcid{0000-0002-5887-6799}\inst{\ref{aff14}}
\and G.~De~Lucia\orcid{0000-0002-6220-9104}\inst{\ref{aff38}}
\and C.~Dolding\orcid{0009-0003-7199-6108}\inst{\ref{aff27}}
\and H.~Dole\orcid{0000-0002-9767-3839}\inst{\ref{aff31}}
\and F.~Dubath\orcid{0000-0002-6533-2810}\inst{\ref{aff14}}
\and C.~A.~J.~Duncan\orcid{0009-0003-3573-0791}\inst{\ref{aff33}}
\and X.~Dupac\inst{\ref{aff61}}
\and S.~Dusini\orcid{0000-0002-1128-0664}\inst{\ref{aff68}}
\and S.~Escoffier\orcid{0000-0002-2847-7498}\inst{\ref{aff69}}
\and M.~Fabricius\orcid{0000-0002-7025-6058}\inst{\ref{aff28},\ref{aff70}}
\and M.~Farina\orcid{0000-0002-3089-7846}\inst{\ref{aff24}}
\and R.~Farinelli\inst{\ref{aff4}}
\and S.~Ferriol\inst{\ref{aff29}}
\and F.~Finelli\orcid{0000-0002-6694-3269}\inst{\ref{aff4},\ref{aff71}}
\and P.~Fosalba\orcid{0000-0002-1510-5214}\inst{\ref{aff72},\ref{aff35}}
\and N.~Fourmanoit\orcid{0009-0005-6816-6925}\inst{\ref{aff69}}
\and M.~Frailis\orcid{0000-0002-7400-2135}\inst{\ref{aff38}}
\and E.~Franceschi\orcid{0000-0002-0585-6591}\inst{\ref{aff4}}
\and P.~Franzetti\inst{\ref{aff20}}
\and M.~Fumana\orcid{0000-0001-6787-5950}\inst{\ref{aff20}}
\and S.~Galeotta\orcid{0000-0002-3748-5115}\inst{\ref{aff38}}
\and K.~George\orcid{0000-0002-1734-8455}\inst{\ref{aff73}}
\and W.~Gillard\orcid{0000-0003-4744-9748}\inst{\ref{aff69}}
\and B.~Gillis\orcid{0000-0002-4478-1270}\inst{\ref{aff33}}
\and C.~Giocoli\orcid{0000-0002-9590-7961}\inst{\ref{aff4},\ref{aff41}}
\and J.~Gracia-Carpio\inst{\ref{aff28}}
\and A.~Grazian\orcid{0000-0002-5688-0663}\inst{\ref{aff42}}
\and F.~Grupp\inst{\ref{aff28},\ref{aff70}}
\and L.~Guzzo\orcid{0000-0001-8264-5192}\inst{\ref{aff74},\ref{aff36},\ref{aff75}}
\and S.~V.~H.~Haugan\orcid{0000-0001-9648-7260}\inst{\ref{aff76}}
\and H.~Hoekstra\orcid{0000-0002-0641-3231}\inst{\ref{aff1}}
\and W.~Holmes\inst{\ref{aff26}}
\and I.~M.~Hook\orcid{0000-0002-2960-978X}\inst{\ref{aff77}}
\and F.~Hormuth\inst{\ref{aff78}}
\and A.~Hornstrup\orcid{0000-0002-3363-0936}\inst{\ref{aff79},\ref{aff80}}
\and K.~Jahnke\orcid{0000-0003-3804-2137}\inst{\ref{aff10}}
\and M.~Jhabvala\inst{\ref{aff81}}
\and B.~Joachimi\orcid{0000-0001-7494-1303}\inst{\ref{aff82}}
\and E.~Keih\"anen\orcid{0000-0003-1804-7715}\inst{\ref{aff83}}
\and S.~Kermiche\orcid{0000-0002-0302-5735}\inst{\ref{aff69}}
\and A.~Kiessling\orcid{0000-0002-2590-1273}\inst{\ref{aff26}}
\and B.~Kubik\orcid{0009-0006-5823-4880}\inst{\ref{aff29}}
\and M.~K\"ummel\orcid{0000-0003-2791-2117}\inst{\ref{aff70}}
\and M.~Kunz\orcid{0000-0002-3052-7394}\inst{\ref{aff84}}
\and H.~Kurki-Suonio\orcid{0000-0002-4618-3063}\inst{\ref{aff85},\ref{aff86}}
\and R.~Laureijs\inst{\ref{aff2}}
\and A.~M.~C.~Le~Brun\orcid{0000-0002-0936-4594}\inst{\ref{aff87}}
\and S.~Ligori\orcid{0000-0003-4172-4606}\inst{\ref{aff49}}
\and P.~B.~Lilje\orcid{0000-0003-4324-7794}\inst{\ref{aff76}}
\and V.~Lindholm\orcid{0000-0003-2317-5471}\inst{\ref{aff85},\ref{aff86}}
\and I.~Lloro\orcid{0000-0001-5966-1434}\inst{\ref{aff88}}
\and G.~Mainetti\orcid{0000-0003-2384-2377}\inst{\ref{aff89}}
\and D.~Maino\inst{\ref{aff74},\ref{aff20},\ref{aff75}}
\and E.~Maiorano\orcid{0000-0003-2593-4355}\inst{\ref{aff4}}
\and O.~Mansutti\orcid{0000-0001-5758-4658}\inst{\ref{aff38}}
\and S.~Marcin\inst{\ref{aff90}}
\and O.~Marggraf\orcid{0000-0001-7242-3852}\inst{\ref{aff91}}
\and K.~Markovic\orcid{0000-0001-6764-073X}\inst{\ref{aff26}}
\and M.~Martinelli\orcid{0000-0002-6943-7732}\inst{\ref{aff22},\ref{aff92}}
\and N.~Martinet\orcid{0000-0003-2786-7790}\inst{\ref{aff62}}
\and F.~Marulli\orcid{0000-0002-8850-0303}\inst{\ref{aff93},\ref{aff4},\ref{aff41}}
\and R.~J.~Massey\orcid{0000-0002-6085-3780}\inst{\ref{aff94}}
\and E.~Medinaceli\orcid{0000-0002-4040-7783}\inst{\ref{aff4}}
\and S.~Mei\orcid{0000-0002-2849-559X}\inst{\ref{aff95},\ref{aff96}}
\and M.~Melchior\inst{\ref{aff97}}
\and Y.~Mellier\thanks{Deceased}\inst{\ref{aff98},\ref{aff99}}
\and M.~Meneghetti\orcid{0000-0003-1225-7084}\inst{\ref{aff4},\ref{aff41}}
\and E.~Merlin\orcid{0000-0001-6870-8900}\inst{\ref{aff22}}
\and G.~Meylan\inst{\ref{aff100}}
\and A.~Mora\orcid{0000-0002-1922-8529}\inst{\ref{aff101}}
\and M.~Moresco\orcid{0000-0002-7616-7136}\inst{\ref{aff93},\ref{aff4}}
\and L.~Moscardini\orcid{0000-0002-3473-6716}\inst{\ref{aff93},\ref{aff4},\ref{aff41}}
\and R.~Nakajima\orcid{0009-0009-1213-7040}\inst{\ref{aff91}}
\and C.~Neissner\orcid{0000-0001-8524-4968}\inst{\ref{aff102},\ref{aff52}}
\and R.~C.~Nichol\orcid{0000-0003-0939-6518}\inst{\ref{aff103}}
\and S.-M.~Niemi\orcid{0009-0005-0247-0086}\inst{\ref{aff50}}
\and C.~Padilla\orcid{0000-0001-7951-0166}\inst{\ref{aff102}}
\and S.~Paltani\orcid{0000-0002-8108-9179}\inst{\ref{aff14}}
\and F.~Pasian\orcid{0000-0002-4869-3227}\inst{\ref{aff38}}
\and K.~Pedersen\inst{\ref{aff104}}
\and W.~J.~Percival\orcid{0000-0002-0644-5727}\inst{\ref{aff105},\ref{aff106},\ref{aff107}}
\and V.~Pettorino\orcid{0000-0002-4203-9320}\inst{\ref{aff50}}
\and S.~Pires\orcid{0000-0002-0249-2104}\inst{\ref{aff108}}
\and G.~Polenta\orcid{0000-0003-4067-9196}\inst{\ref{aff109}}
\and M.~Poncet\inst{\ref{aff110}}
\and L.~A.~Popa\inst{\ref{aff111}}
\and L.~Pozzetti\orcid{0000-0001-7085-0412}\inst{\ref{aff4}}
\and F.~Raison\orcid{0000-0002-7819-6918}\inst{\ref{aff28}}
\and R.~Rebolo\orcid{0000-0003-3767-7085}\inst{\ref{aff58},\ref{aff112},\ref{aff113}}
\and A.~Renzi\orcid{0000-0001-9856-1970}\inst{\ref{aff114},\ref{aff68}}
\and J.~Rhodes\orcid{0000-0002-4485-8549}\inst{\ref{aff26}}
\and G.~Riccio\inst{\ref{aff13}}
\and E.~Romelli\orcid{0000-0003-3069-9222}\inst{\ref{aff38}}
\and M.~Roncarelli\orcid{0000-0001-9587-7822}\inst{\ref{aff4}}
\and E.~Rossetti\orcid{0000-0003-0238-4047}\inst{\ref{aff16}}
\and R.~Saglia\orcid{0000-0003-0378-7032}\inst{\ref{aff70},\ref{aff28}}
\and Z.~Sakr\orcid{0000-0002-4823-3757}\inst{\ref{aff115},\ref{aff116},\ref{aff117}}
\and D.~Sapone\orcid{0000-0001-7089-4503}\inst{\ref{aff118}}
\and B.~Sartoris\orcid{0000-0003-1337-5269}\inst{\ref{aff70},\ref{aff38}}
\and M.~Schirmer\orcid{0000-0003-2568-9994}\inst{\ref{aff10}}
\and P.~Schneider\orcid{0000-0001-8561-2679}\inst{\ref{aff91}}
\and T.~Schrabback\orcid{0000-0002-6987-7834}\inst{\ref{aff119}}
\and M.~Scodeggio\inst{\ref{aff20}}
\and A.~Secroun\orcid{0000-0003-0505-3710}\inst{\ref{aff69}}
\and E.~Sefusatti\orcid{0000-0003-0473-1567}\inst{\ref{aff38},\ref{aff37},\ref{aff39}}
\and G.~Seidel\orcid{0000-0003-2907-353X}\inst{\ref{aff10}}
\and S.~Serrano\orcid{0000-0002-0211-2861}\inst{\ref{aff72},\ref{aff120},\ref{aff35}}
\and P.~Simon\inst{\ref{aff91}}
\and C.~Sirignano\orcid{0000-0002-0995-7146}\inst{\ref{aff114},\ref{aff68}}
\and G.~Sirri\orcid{0000-0003-2626-2853}\inst{\ref{aff41}}
\and L.~Stanco\orcid{0000-0002-9706-5104}\inst{\ref{aff68}}
\and J.-L.~Starck\orcid{0000-0003-2177-7794}\inst{\ref{aff108}}
\and J.~Steinwagner\orcid{0000-0001-7443-1047}\inst{\ref{aff28}}
\and C.~Surace\orcid{0000-0003-2592-0113}\inst{\ref{aff62}}
\and P.~Tallada-Cresp\'{i}\orcid{0000-0002-1336-8328}\inst{\ref{aff51},\ref{aff52}}
\and D.~Tavagnacco\orcid{0000-0001-7475-9894}\inst{\ref{aff38}}
\and A.~N.~Taylor\inst{\ref{aff33}}
\and H.~I.~Teplitz\orcid{0000-0002-7064-5424}\inst{\ref{aff121}}
\and I.~Tereno\orcid{0000-0002-4537-6218}\inst{\ref{aff66},\ref{aff122}}
\and N.~Tessore\orcid{0000-0002-9696-7931}\inst{\ref{aff82}}
\and S.~Toft\orcid{0000-0003-3631-7176}\inst{\ref{aff3},\ref{aff123}}
\and R.~Toledo-Moreo\orcid{0000-0002-2997-4859}\inst{\ref{aff124}}
\and F.~Torradeflot\orcid{0000-0003-1160-1517}\inst{\ref{aff52},\ref{aff51}}
\and I.~Tutusaus\orcid{0000-0002-3199-0399}\inst{\ref{aff35},\ref{aff72},\ref{aff116}}
\and L.~Valenziano\orcid{0000-0002-1170-0104}\inst{\ref{aff4},\ref{aff71}}
\and J.~Valiviita\orcid{0000-0001-6225-3693}\inst{\ref{aff85},\ref{aff86}}
\and T.~Vassallo\orcid{0000-0001-6512-6358}\inst{\ref{aff38}}
\and A.~Veropalumbo\orcid{0000-0003-2387-1194}\inst{\ref{aff36},\ref{aff44},\ref{aff43}}
\and D.~Vibert\orcid{0009-0008-0607-631X}\inst{\ref{aff62}}
\and Y.~Wang\orcid{0000-0002-4749-2984}\inst{\ref{aff121}}
\and J.~Weller\orcid{0000-0002-8282-2010}\inst{\ref{aff70},\ref{aff28}}
\and A.~Zacchei\orcid{0000-0003-0396-1192}\inst{\ref{aff38},\ref{aff37}}
\and E.~Zucca\orcid{0000-0002-5845-8132}\inst{\ref{aff4}}
\and M.~Ballardini\orcid{0000-0003-4481-3559}\inst{\ref{aff125},\ref{aff126},\ref{aff4}}
\and E.~Bozzo\orcid{0000-0002-8201-1525}\inst{\ref{aff14}}
\and C.~Burigana\orcid{0000-0002-3005-5796}\inst{\ref{aff127},\ref{aff71}}
\and R.~Cabanac\orcid{0000-0001-6679-2600}\inst{\ref{aff116}}
\and M.~Calabrese\orcid{0000-0002-2637-2422}\inst{\ref{aff128},\ref{aff20}}
\and A.~Cappi\inst{\ref{aff4},\ref{aff129}}
\and D.~Di~Ferdinando\inst{\ref{aff41}}
\and J.~A.~Escartin~Vigo\inst{\ref{aff28}}
\and L.~Gabarra\orcid{0000-0002-8486-8856}\inst{\ref{aff130}}
\and W.~G.~Hartley\inst{\ref{aff14}}
\and M.~Huertas-Company\orcid{0000-0002-1416-8483}\inst{\ref{aff58},\ref{aff34},\ref{aff131},\ref{aff132}}
\and J.~Mart\'{i}n-Fleitas\orcid{0000-0002-8594-569X}\inst{\ref{aff133}}
\and S.~Matthew\orcid{0000-0001-8448-1697}\inst{\ref{aff33}}
\and N.~Mauri\orcid{0000-0001-8196-1548}\inst{\ref{aff57},\ref{aff41}}
\and R.~B.~Metcalf\orcid{0000-0003-3167-2574}\inst{\ref{aff93},\ref{aff4}}
\and A.~A.~Nucita\inst{\ref{aff134},\ref{aff135},\ref{aff136}}
\and A.~Pezzotta\orcid{0000-0003-0726-2268}\inst{\ref{aff36}}
\and M.~P\"ontinen\orcid{0000-0001-5442-2530}\inst{\ref{aff85}}
\and C.~Porciani\orcid{0000-0002-7797-2508}\inst{\ref{aff91}}
\and I.~Risso\orcid{0000-0003-2525-7761}\inst{\ref{aff36},\ref{aff44}}
\and V.~Scottez\orcid{0009-0008-3864-940X}\inst{\ref{aff98},\ref{aff137}}
\and M.~Sereno\orcid{0000-0003-0302-0325}\inst{\ref{aff4},\ref{aff41}}
\and M.~Tenti\orcid{0000-0002-4254-5901}\inst{\ref{aff41}}
\and M.~Viel\orcid{0000-0002-2642-5707}\inst{\ref{aff37},\ref{aff38},\ref{aff40},\ref{aff39},\ref{aff138}}
\and M.~Wiesmann\orcid{0009-0000-8199-5860}\inst{\ref{aff76}}
\and Y.~Akrami\orcid{0000-0002-2407-7956}\inst{\ref{aff139},\ref{aff140}}
\and S.~Alvi\orcid{0000-0001-5779-8568}\inst{\ref{aff125}}
\and I.~T.~Andika\orcid{0000-0001-6102-9526}\inst{\ref{aff141},\ref{aff142}}
\and S.~Anselmi\orcid{0000-0002-3579-9583}\inst{\ref{aff68},\ref{aff114},\ref{aff143}}
\and M.~Archidiacono\orcid{0000-0003-4952-9012}\inst{\ref{aff74},\ref{aff75}}
\and F.~Atrio-Barandela\orcid{0000-0002-2130-2513}\inst{\ref{aff144}}
\and E.~Aubourg\orcid{0000-0002-5592-023X}\inst{\ref{aff95},\ref{aff145}}
\and D.~Bertacca\orcid{0000-0002-2490-7139}\inst{\ref{aff114},\ref{aff42},\ref{aff68}}
\and M.~Bethermin\orcid{0000-0002-3915-2015}\inst{\ref{aff146}}
\and L.~Bisigello\orcid{0000-0003-0492-4924}\inst{\ref{aff42}}
\and A.~Blanchard\orcid{0000-0001-8555-9003}\inst{\ref{aff116}}
\and L.~Blot\orcid{0000-0002-9622-7167}\inst{\ref{aff147},\ref{aff87}}
\and M.~Bonici\orcid{0000-0002-8430-126X}\inst{\ref{aff105},\ref{aff20}}
\and S.~Borgani\orcid{0000-0001-6151-6439}\inst{\ref{aff148},\ref{aff37},\ref{aff38},\ref{aff39},\ref{aff138}}
\and M.~L.~Brown\orcid{0000-0002-0370-8077}\inst{\ref{aff59}}
\and S.~Bruton\orcid{0000-0002-6503-5218}\inst{\ref{aff149}}
\and A.~Calabro\orcid{0000-0003-2536-1614}\inst{\ref{aff22}}
\and B.~Camacho~Quevedo\orcid{0000-0002-8789-4232}\inst{\ref{aff37},\ref{aff40},\ref{aff38}}
\and F.~Caro\inst{\ref{aff22}}
\and C.~S.~Carvalho\inst{\ref{aff122}}
\and T.~Castro\orcid{0000-0002-6292-3228}\inst{\ref{aff38},\ref{aff39},\ref{aff37},\ref{aff138}}
\and F.~Cogato\orcid{0000-0003-4632-6113}\inst{\ref{aff93},\ref{aff4}}
\and S.~Conseil\orcid{0000-0002-3657-4191}\inst{\ref{aff29}}
\and A.~R.~Cooray\orcid{0000-0002-3892-0190}\inst{\ref{aff150}}
\and O.~Cucciati\orcid{0000-0002-9336-7551}\inst{\ref{aff4}}
\and G.~Daste\inst{\ref{aff62}}
\and F.~De~Paolis\orcid{0000-0001-6460-7563}\inst{\ref{aff134},\ref{aff135},\ref{aff136}}
\and G.~Desprez\orcid{0000-0001-8325-1742}\inst{\ref{aff2}}
\and A.~D\'iaz-S\'anchez\orcid{0000-0003-0748-4768}\inst{\ref{aff151}}
\and J.~J.~Diaz\orcid{0000-0003-2101-1078}\inst{\ref{aff58}}
\and S.~Di~Domizio\orcid{0000-0003-2863-5895}\inst{\ref{aff43},\ref{aff44}}
\and J.~M.~Diego\orcid{0000-0001-9065-3926}\inst{\ref{aff152}}
\and P.~Dimauro\orcid{0000-0001-7399-2854}\inst{\ref{aff153},\ref{aff22}}
\and P.-A.~Duc\orcid{0000-0003-3343-6284}\inst{\ref{aff146}}
\and M.~Y.~Elkhashab\orcid{0000-0001-9306-2603}\inst{\ref{aff38},\ref{aff39},\ref{aff148},\ref{aff37}}
\and A.~Enia\orcid{0000-0002-0200-2857}\inst{\ref{aff16},\ref{aff4}}
\and Y.~Fang\orcid{0000-0002-0334-6950}\inst{\ref{aff70}}
\and A.~G.~Ferrari\orcid{0009-0005-5266-4110}\inst{\ref{aff41}}
\and A.~Finoguenov\orcid{0000-0002-4606-5403}\inst{\ref{aff85}}
\and F.~Fontanot\orcid{0000-0003-4744-0188}\inst{\ref{aff38},\ref{aff37}}
\and A.~Franco\orcid{0000-0002-4761-366X}\inst{\ref{aff135},\ref{aff134},\ref{aff136}}
\and K.~Ganga\orcid{0000-0001-8159-8208}\inst{\ref{aff95}}
\and J.~Garc\'ia-Bellido\orcid{0000-0002-9370-8360}\inst{\ref{aff139}}
\and T.~Gasparetto\orcid{0000-0002-7913-4866}\inst{\ref{aff22}}
\and V.~Gautard\inst{\ref{aff154}}
\and E.~Gaztanaga\orcid{0000-0001-9632-0815}\inst{\ref{aff35},\ref{aff72},\ref{aff155}}
\and F.~Giacomini\orcid{0000-0002-3129-2814}\inst{\ref{aff41}}
\and F.~Gianotti\orcid{0000-0003-4666-119X}\inst{\ref{aff4}}
\and G.~Gozaliasl\orcid{0000-0002-0236-919X}\inst{\ref{aff156},\ref{aff85}}
\and M.~Gray\inst{\ref{aff62}}
\and M.~Guidi\orcid{0000-0001-9408-1101}\inst{\ref{aff16},\ref{aff4}}
\and C.~M.~Gutierrez\orcid{0000-0001-7854-783X}\inst{\ref{aff34}}
\and A.~Hall\orcid{0000-0002-3139-8651}\inst{\ref{aff33}}
\and C.~Hern\'andez-Monteagudo\orcid{0000-0001-5471-9166}\inst{\ref{aff113},\ref{aff58}}
\and H.~Hildebrandt\orcid{0000-0002-9814-3338}\inst{\ref{aff157}}
\and J.~Hjorth\orcid{0000-0002-4571-2306}\inst{\ref{aff104}}
\and J.~J.~E.~Kajava\orcid{0000-0002-3010-8333}\inst{\ref{aff158},\ref{aff159}}
\and Y.~Kang\orcid{0009-0000-8588-7250}\inst{\ref{aff14}}
\and V.~Kansal\orcid{0000-0002-4008-6078}\inst{\ref{aff160},\ref{aff161}}
\and D.~Karagiannis\orcid{0000-0002-4927-0816}\inst{\ref{aff125},\ref{aff162}}
\and K.~Kiiveri\inst{\ref{aff83}}
\and J.~Kim\orcid{0000-0003-2776-2761}\inst{\ref{aff130}}
\and C.~C.~Kirkpatrick\inst{\ref{aff83}}
\and S.~Kruk\orcid{0000-0001-8010-8879}\inst{\ref{aff61}}
\and V.~Le~Brun\orcid{0000-0002-5027-1939}\inst{\ref{aff62}}
\and J.~Le~Graet\orcid{0000-0001-6523-7971}\inst{\ref{aff69}}
\and L.~Legrand\orcid{0000-0003-0610-5252}\inst{\ref{aff163},\ref{aff164}}
\and M.~Lembo\orcid{0000-0002-5271-5070}\inst{\ref{aff99},\ref{aff125},\ref{aff126}}
\and F.~Lepori\orcid{0009-0000-5061-7138}\inst{\ref{aff165}}
\and G.~Leroy\orcid{0009-0004-2523-4425}\inst{\ref{aff23},\ref{aff94}}
\and G.~F.~Lesci\orcid{0000-0002-4607-2830}\inst{\ref{aff93},\ref{aff4}}
\and J.~Lesgourgues\orcid{0000-0001-7627-353X}\inst{\ref{aff53}}
\and T.~I.~Liaudat\orcid{0000-0002-9104-314X}\inst{\ref{aff145}}
\and A.~Loureiro\orcid{0000-0002-4371-0876}\inst{\ref{aff166},\ref{aff167}}
\and J.~Macias-Perez\orcid{0000-0002-5385-2763}\inst{\ref{aff168}}
\and M.~Magliocchetti\orcid{0000-0001-9158-4838}\inst{\ref{aff24}}
\and C.~Mancini\orcid{0000-0002-4297-0561}\inst{\ref{aff20}}
\and R.~Maoli\orcid{0000-0002-6065-3025}\inst{\ref{aff169},\ref{aff22}}
\and C.~J.~A.~P.~Martins\orcid{0000-0002-4886-9261}\inst{\ref{aff170},\ref{aff18}}
\and L.~Maurin\orcid{0000-0002-8406-0857}\inst{\ref{aff31}}
\and M.~Miluzio\inst{\ref{aff61},\ref{aff171}}
\and P.~Monaco\orcid{0000-0003-2083-7564}\inst{\ref{aff148},\ref{aff38},\ref{aff39},\ref{aff37},\ref{aff138}}
\and C.~Moretti\orcid{0000-0003-3314-8936}\inst{\ref{aff38},\ref{aff37},\ref{aff39},\ref{aff40}}
\and G.~Morgante\inst{\ref{aff4}}
\and S.~Nadathur\orcid{0000-0001-9070-3102}\inst{\ref{aff155}}
\and K.~Naidoo\orcid{0000-0002-9182-1802}\inst{\ref{aff155},\ref{aff82}}
\and P.~Natoli\orcid{0000-0003-0126-9100}\inst{\ref{aff125},\ref{aff126}}
\and A.~Navarro-Alsina\orcid{0000-0002-3173-2592}\inst{\ref{aff91}}
\and S.~Nesseris\orcid{0000-0002-0567-0324}\inst{\ref{aff139}}
\and D.~Paoletti\orcid{0000-0003-4761-6147}\inst{\ref{aff4},\ref{aff71}}
\and F.~Passalacqua\orcid{0000-0002-8606-4093}\inst{\ref{aff114},\ref{aff68}}
\and K.~Paterson\orcid{0000-0001-8340-3486}\inst{\ref{aff10}}
\and L.~Patrizii\inst{\ref{aff41}}
\and A.~Pisani\orcid{0000-0002-6146-4437}\inst{\ref{aff69}}
\and D.~Potter\orcid{0000-0002-0757-5195}\inst{\ref{aff165}}
\and S.~Quai\orcid{0000-0002-0449-8163}\inst{\ref{aff93},\ref{aff4}}
\and M.~Radovich\orcid{0000-0002-3585-866X}\inst{\ref{aff42}}
\and P.-F.~Rocci\inst{\ref{aff31}}
\and G.~Rodighiero\orcid{0000-0002-9415-2296}\inst{\ref{aff114},\ref{aff42}}
\and S.~Sacquegna\orcid{0000-0002-8433-6630}\inst{\ref{aff172},\ref{aff134},\ref{aff135}}
\and M.~Sahl\'en\orcid{0000-0003-0973-4804}\inst{\ref{aff173}}
\and D.~B.~Sanders\orcid{0000-0002-1233-9998}\inst{\ref{aff56}}
\and E.~Sarpa\orcid{0000-0002-1256-655X}\inst{\ref{aff40},\ref{aff138},\ref{aff39}}
\and C.~Scarlata\orcid{0000-0002-9136-8876}\inst{\ref{aff174}}
\and A.~Schneider\orcid{0000-0001-7055-8104}\inst{\ref{aff165}}
\and D.~Sciotti\orcid{0009-0008-4519-2620}\inst{\ref{aff22},\ref{aff92}}
\and E.~Sellentin\inst{\ref{aff175},\ref{aff1}}
\and F.~Shankar\orcid{0000-0001-8973-5051}\inst{\ref{aff9}}
\and L.~C.~Smith\orcid{0000-0002-3259-2771}\inst{\ref{aff176}}
\and E.~Soubrie\orcid{0000-0001-9295-1863}\inst{\ref{aff31}}
\and K.~Tanidis\orcid{0000-0001-9843-5130}\inst{\ref{aff130}}
\and C.~Tao\orcid{0000-0001-7961-8177}\inst{\ref{aff69}}
\and G.~Testera\inst{\ref{aff44}}
\and R.~Teyssier\orcid{0000-0001-7689-0933}\inst{\ref{aff177}}
\and S.~Tosi\orcid{0000-0002-7275-9193}\inst{\ref{aff43},\ref{aff44},\ref{aff36}}
\and A.~Troja\orcid{0000-0003-0239-4595}\inst{\ref{aff114},\ref{aff68}}
\and M.~Tucci\inst{\ref{aff14}}
\and C.~Valieri\inst{\ref{aff41}}
\and A.~Venhola\orcid{0000-0001-6071-4564}\inst{\ref{aff178}}
\and G.~Verza\orcid{0000-0002-1886-8348}\inst{\ref{aff179}}
\and P.~Vielzeuf\orcid{0000-0003-2035-9339}\inst{\ref{aff69}}
\and A.~Viitanen\orcid{0000-0001-9383-786X}\inst{\ref{aff83},\ref{aff14},\ref{aff22}}
\and N.~A.~Walton\orcid{0000-0003-3983-8778}\inst{\ref{aff176}}
\and J.~R.~Weaver\orcid{0000-0003-1614-196X}\inst{\ref{aff180},\ref{aff181}}}
										   
\institute{Leiden Observatory, Leiden University, Einsteinweg 55, 2333 CC Leiden, The Netherlands\label{aff1}
\and
Kapteyn Astronomical Institute, University of Groningen, PO Box 800, 9700 AV Groningen, The Netherlands\label{aff2}
\and
Cosmic Dawn Center (DAWN)\label{aff3}
\and
INAF-Osservatorio di Astrofisica e Scienza dello Spazio di Bologna, Via Piero Gobetti 93/3, 40129 Bologna, Italy\label{aff4}
\and
Dipartimento di Fisica e Astronomia, Universit\`{a} di Firenze, via G. Sansone 1, 50019 Sesto Fiorentino, Firenze, Italy\label{aff5}
\and
University of Trento, Via Sommarive 14, I-38123 Trento, Italy\label{aff6}
\and
INAF-Osservatorio Astrofisico di Arcetri, Largo E. Fermi 5, 50125, Firenze, Italy\label{aff7}
\and
SRON Netherlands Institute for Space Research, Landleven 12, 9747 AD, Groningen, The Netherlands\label{aff8}
\and
School of Physics \& Astronomy, University of Southampton, Highfield Campus, Southampton SO17 1BJ, UK\label{aff9}
\and
Max-Planck-Institut f\"ur Astronomie, K\"onigstuhl 17, 69117 Heidelberg, Germany\label{aff10}
\and
Department of Physical Sciences, Ritsumeikan University, Kusatsu, Shiga 525-8577, Japan\label{aff11}
\and
Academia Sinica Institute of Astronomy and Astrophysics (ASIAA), 11F of ASMAB, No.~1, Section 4, Roosevelt Road, Taipei 10617, Taiwan\label{aff12}
\and
INAF-Osservatorio Astronomico di Capodimonte, Via Moiariello 16, 80131 Napoli, Italy\label{aff13}
\and
Department of Astronomy, University of Geneva, ch. d'Ecogia 16, 1290 Versoix, Switzerland\label{aff14}
\and
Institute for Particle Physics and Astrophysics, Dept. of Physics, ETH Zurich, Wolfgang-Pauli-Strasse 27, 8093 Zurich, Switzerland\label{aff15}
\and
Dipartimento di Fisica e Astronomia, Universit\`a di Bologna, Via Gobetti 93/2, 40129 Bologna, Italy\label{aff16}
\and
Faculdade de Ci\^encias da Universidade do Porto, Rua do Campo de Alegre, 4150-007 Porto, Portugal\label{aff17}
\and
Instituto de Astrof\'isica e Ci\^encias do Espa\c{c}o, Universidade do Porto, CAUP, Rua das Estrelas, PT4150-762 Porto, Portugal\label{aff18}
\and
DTx -- Digital Transformation CoLAB, Building 1, Azur\'em Campus, University of Minho, 4800-058 Guimar\~aes, Portugal\label{aff19}
\and
INAF-IASF Milano, Via Alfonso Corti 12, 20133 Milano, Italy\label{aff20}
\and
Department of Mathematics and Physics, Roma Tre University, Via della Vasca Navale 84, 00146 Rome, Italy\label{aff21}
\and
INAF-Osservatorio Astronomico di Roma, Via Frascati 33, 00078 Monteporzio Catone, Italy\label{aff22}
\and
Department of Physics, Centre for Extragalactic Astronomy, Durham University, South Road, Durham, DH1 3LE, UK\label{aff23}
\and
INAF-Istituto di Astrofisica e Planetologia Spaziali, via del Fosso del Cavaliere, 100, 00100 Roma, Italy\label{aff24}
\and
School of Physics, HH Wills Physics Laboratory, University of Bristol, Tyndall Avenue, Bristol, BS8 1TL, UK\label{aff25}
\and
Jet Propulsion Laboratory, California Institute of Technology, 4800 Oak Grove Drive, Pasadena, CA, 91109, USA\label{aff26}
\and
Mullard Space Science Laboratory, University College London, Holmbury St Mary, Dorking, Surrey RH5 6NT, UK\label{aff27}
\and
Max Planck Institute for Extraterrestrial Physics, Giessenbachstr. 1, 85748 Garching, Germany\label{aff28}
\and
Universit\'e Claude Bernard Lyon 1, CNRS/IN2P3, IP2I Lyon, UMR 5822, Villeurbanne, F-69100, France\label{aff29}
\and
Univ. Lille, CNRS, Centrale Lille, UMR 9189 CRIStAL, 59000 Lille, France\label{aff30}
\and
Universit\'e Paris-Saclay, CNRS, Institut d'astrophysique spatiale, 91405, Orsay, France\label{aff31}
\and
Department of Physics and Astronomy, University of British Columbia, Vancouver, BC V6T 1Z1, Canada\label{aff32}
\and
Institute for Astronomy, University of Edinburgh, Royal Observatory, Blackford Hill, Edinburgh EH9 3HJ, UK\label{aff33}
\and
 Instituto de Astrof\'{\i}sica de Canarias, E-38205 La Laguna; Universidad de La Laguna, Dpto. Astrof\'\i sica, E-38206 La Laguna, Tenerife, Spain\label{aff34}
\and
Institute of Space Sciences (ICE, CSIC), Campus UAB, Carrer de Can Magrans, s/n, 08193 Barcelona, Spain\label{aff35}
\and
INAF-Osservatorio Astronomico di Brera, Via Brera 28, 20122 Milano, Italy\label{aff36}
\and
IFPU, Institute for Fundamental Physics of the Universe, via Beirut 2, 34151 Trieste, Italy\label{aff37}
\and
INAF-Osservatorio Astronomico di Trieste, Via G. B. Tiepolo 11, 34143 Trieste, Italy\label{aff38}
\and
INFN, Sezione di Trieste, Via Valerio 2, 34127 Trieste TS, Italy\label{aff39}
\and
SISSA, International School for Advanced Studies, Via Bonomea 265, 34136 Trieste TS, Italy\label{aff40}
\and
INFN-Sezione di Bologna, Viale Berti Pichat 6/2, 40127 Bologna, Italy\label{aff41}
\and
INAF-Osservatorio Astronomico di Padova, Via dell'Osservatorio 5, 35122 Padova, Italy\label{aff42}
\and
Dipartimento di Fisica, Universit\`a di Genova, Via Dodecaneso 33, 16146, Genova, Italy\label{aff43}
\and
INFN-Sezione di Genova, Via Dodecaneso 33, 16146, Genova, Italy\label{aff44}
\and
Department of Physics "E. Pancini", University Federico II, Via Cinthia 6, 80126, Napoli, Italy\label{aff45}
\and
European Southern Observatory, Karl-Schwarzschild-Str.~2, 85748 Garching, Germany\label{aff46}
\and
Dipartimento di Fisica, Universit\`a degli Studi di Torino, Via P. Giuria 1, 10125 Torino, Italy\label{aff47}
\and
INFN-Sezione di Torino, Via P. Giuria 1, 10125 Torino, Italy\label{aff48}
\and
INAF-Osservatorio Astrofisico di Torino, Via Osservatorio 20, 10025 Pino Torinese (TO), Italy\label{aff49}
\and
European Space Agency/ESTEC, Keplerlaan 1, 2201 AZ Noordwijk, The Netherlands\label{aff50}
\and
Centro de Investigaciones Energ\'eticas, Medioambientales y Tecnol\'ogicas (CIEMAT), Avenida Complutense 40, 28040 Madrid, Spain\label{aff51}
\and
Port d'Informaci\'{o} Cient\'{i}fica, Campus UAB, C. Albareda s/n, 08193 Bellaterra (Barcelona), Spain\label{aff52}
\and
Institute for Theoretical Particle Physics and Cosmology (TTK), RWTH Aachen University, 52056 Aachen, Germany\label{aff53}
\and
Deutsches Zentrum f\"ur Luft- und Raumfahrt e. V. (DLR), Linder H\"ohe, 51147 K\"oln, Germany\label{aff54}
\and
INFN section of Naples, Via Cinthia 6, 80126, Napoli, Italy\label{aff55}
\and
Institute for Astronomy, University of Hawaii, 2680 Woodlawn Drive, Honolulu, HI 96822, USA\label{aff56}
\and
Dipartimento di Fisica e Astronomia "Augusto Righi" - Alma Mater Studiorum Universit\`a di Bologna, Viale Berti Pichat 6/2, 40127 Bologna, Italy\label{aff57}
\and
Instituto de Astrof\'{\i}sica de Canarias, E-38205 La Laguna, Tenerife, Spain\label{aff58}
\and
Jodrell Bank Centre for Astrophysics, Department of Physics and Astronomy, University of Manchester, Oxford Road, Manchester M13 9PL, UK\label{aff59}
\and
European Space Agency/ESRIN, Largo Galileo Galilei 1, 00044 Frascati, Roma, Italy\label{aff60}
\and
ESAC/ESA, Camino Bajo del Castillo, s/n., Urb. Villafranca del Castillo, 28692 Villanueva de la Ca\~nada, Madrid, Spain\label{aff61}
\and
Aix-Marseille Universit\'e, CNRS, CNES, LAM, Marseille, France\label{aff62}
\and
Institut de Ci\`{e}ncies del Cosmos (ICCUB), Universitat de Barcelona (IEEC-UB), Mart\'{i} i Franqu\`{e}s 1, 08028 Barcelona, Spain\label{aff63}
\and
Instituci\'o Catalana de Recerca i Estudis Avan\c{c}ats (ICREA), Passeig de Llu\'{\i}s Companys 23, 08010 Barcelona, Spain\label{aff64}
\and
UCB Lyon 1, CNRS/IN2P3, IUF, IP2I Lyon, 4 rue Enrico Fermi, 69622 Villeurbanne, France\label{aff65}
\and
Departamento de F\'isica, Faculdade de Ci\^encias, Universidade de Lisboa, Edif\'icio C8, Campo Grande, PT1749-016 Lisboa, Portugal\label{aff66}
\and
Instituto de Astrof\'isica e Ci\^encias do Espa\c{c}o, Faculdade de Ci\^encias, Universidade de Lisboa, Campo Grande, 1749-016 Lisboa, Portugal\label{aff67}
\and
INFN-Padova, Via Marzolo 8, 35131 Padova, Italy\label{aff68}
\and
Aix-Marseille Universit\'e, CNRS/IN2P3, CPPM, Marseille, France\label{aff69}
\and
Universit\"ats-Sternwarte M\"unchen, Fakult\"at f\"ur Physik, Ludwig-Maximilians-Universit\"at M\"unchen, Scheinerstr.~1, 81679 M\"unchen, Germany\label{aff70}
\and
INFN-Bologna, Via Irnerio 46, 40126 Bologna, Italy\label{aff71}
\and
Institut d'Estudis Espacials de Catalunya (IEEC),  Edifici RDIT, Campus UPC, 08860 Castelldefels, Barcelona, Spain\label{aff72}
\and
University Observatory, LMU Faculty of Physics, Scheinerstr.~1, 81679 Munich, Germany\label{aff73}
\and
Dipartimento di Fisica "Aldo Pontremoli", Universit\`a degli Studi di Milano, Via Celoria 16, 20133 Milano, Italy\label{aff74}
\and
INFN-Sezione di Milano, Via Celoria 16, 20133 Milano, Italy\label{aff75}
\and
Institute of Theoretical Astrophysics, University of Oslo, P.O. Box 1029 Blindern, 0315 Oslo, Norway\label{aff76}
\and
Department of Physics, Lancaster University, Lancaster, LA1 4YB, UK\label{aff77}
\and
Felix Hormuth Engineering, Goethestr. 17, 69181 Leimen, Germany\label{aff78}
\and
Technical University of Denmark, Elektrovej 327, 2800 Kgs. Lyngby, Denmark\label{aff79}
\and
Cosmic Dawn Center (DAWN), Denmark\label{aff80}
\and
NASA Goddard Space Flight Center, Greenbelt, MD 20771, USA\label{aff81}
\and
Department of Physics and Astronomy, University College London, Gower Street, London WC1E 6BT, UK\label{aff82}
\and
Department of Physics and Helsinki Institute of Physics, Gustaf H\"allstr\"omin katu 2, University of Helsinki, 00014 Helsinki, Finland\label{aff83}
\and
Universit\'e de Gen\`eve, D\'epartement de Physique Th\'eorique and Centre for Astroparticle Physics, 24 quai Ernest-Ansermet, CH-1211 Gen\`eve 4, Switzerland\label{aff84}
\and
Department of Physics, P.O. Box 64, University of Helsinki, 00014 Helsinki, Finland\label{aff85}
\and
Helsinki Institute of Physics, Gustaf H{\"a}llstr{\"o}min katu 2, University of Helsinki, 00014 Helsinki, Finland\label{aff86}
\and
Laboratoire d'etude de l'Univers et des phenomenes eXtremes, Observatoire de Paris, Universit\'e PSL, Sorbonne Universit\'e, CNRS, 92190 Meudon, France\label{aff87}
\and
SKAO, Jodrell Bank, Lower Withington, Macclesfield SK11 9FT, UK\label{aff88}
\and
Centre de Calcul de l'IN2P3/CNRS, 21 avenue Pierre de Coubertin 69627 Villeurbanne Cedex, France\label{aff89}
\and
University of Applied Sciences and Arts of Northwestern Switzerland, School of Computer Science, 5210 Windisch, Switzerland\label{aff90}
\and
Universit\"at Bonn, Argelander-Institut f\"ur Astronomie, Auf dem H\"ugel 71, 53121 Bonn, Germany\label{aff91}
\and
INFN-Sezione di Roma, Piazzale Aldo Moro, 2 - c/o Dipartimento di Fisica, Edificio G. Marconi, 00185 Roma, Italy\label{aff92}
\and
Dipartimento di Fisica e Astronomia "Augusto Righi" - Alma Mater Studiorum Universit\`a di Bologna, via Piero Gobetti 93/2, 40129 Bologna, Italy\label{aff93}
\and
Department of Physics, Institute for Computational Cosmology, Durham University, South Road, Durham, DH1 3LE, UK\label{aff94}
\and
Universit\'e Paris Cit\'e, CNRS, Astroparticule et Cosmologie, 75013 Paris, France\label{aff95}
\and
CNRS-UCB International Research Laboratory, Centre Pierre Bin\'etruy, IRL2007, CPB-IN2P3, Berkeley, USA\label{aff96}
\and
University of Applied Sciences and Arts of Northwestern Switzerland, School of Engineering, 5210 Windisch, Switzerland\label{aff97}
\and
Institut d'Astrophysique de Paris, 98bis Boulevard Arago, 75014, Paris, France\label{aff98}
\and
Institut d'Astrophysique de Paris, UMR 7095, CNRS, and Sorbonne Universit\'e, 98 bis boulevard Arago, 75014 Paris, France\label{aff99}
\and
Institute of Physics, Laboratory of Astrophysics, Ecole Polytechnique F\'ed\'erale de Lausanne (EPFL), Observatoire de Sauverny, 1290 Versoix, Switzerland\label{aff100}
\and
Telespazio UK S.L. for European Space Agency (ESA), Camino bajo del Castillo, s/n, Urbanizacion Villafranca del Castillo, Villanueva de la Ca\~nada, 28692 Madrid, Spain\label{aff101}
\and
Institut de F\'{i}sica d'Altes Energies (IFAE), The Barcelona Institute of Science and Technology, Campus UAB, 08193 Bellaterra (Barcelona), Spain\label{aff102}
\and
School of Mathematics and Physics, University of Surrey, Guildford, Surrey, GU2 7XH, UK\label{aff103}
\and
DARK, Niels Bohr Institute, University of Copenhagen, Jagtvej 155, 2200 Copenhagen, Denmark\label{aff104}
\and
Waterloo Centre for Astrophysics, University of Waterloo, Waterloo, Ontario N2L 3G1, Canada\label{aff105}
\and
Department of Physics and Astronomy, University of Waterloo, Waterloo, Ontario N2L 3G1, Canada\label{aff106}
\and
Perimeter Institute for Theoretical Physics, Waterloo, Ontario N2L 2Y5, Canada\label{aff107}
\and
Universit\'e Paris-Saclay, Universit\'e Paris Cit\'e, CEA, CNRS, AIM, 91191, Gif-sur-Yvette, France\label{aff108}
\and
Space Science Data Center, Italian Space Agency, via del Politecnico snc, 00133 Roma, Italy\label{aff109}
\and
Centre National d'Etudes Spatiales -- Centre spatial de Toulouse, 18 avenue Edouard Belin, 31401 Toulouse Cedex 9, France\label{aff110}
\and
Institute of Space Science, Str. Atomistilor, nr. 409 M\u{a}gurele, Ilfov, 077125, Romania\label{aff111}
\and
Consejo Superior de Investigaciones Cientificas, Calle Serrano 117, 28006 Madrid, Spain\label{aff112}
\and
Universidad de La Laguna, Dpto. Astrof\'\i sica, E-38206 La Laguna, Tenerife, Spain\label{aff113}
\and
Dipartimento di Fisica e Astronomia "G. Galilei", Universit\`a di Padova, Via Marzolo 8, 35131 Padova, Italy\label{aff114}
\and
Institut f\"ur Theoretische Physik, University of Heidelberg, Philosophenweg 16, 69120 Heidelberg, Germany\label{aff115}
\and
Institut de Recherche en Astrophysique et Plan\'etologie (IRAP), Universit\'e de Toulouse, CNRS, UPS, CNES, 14 Av. Edouard Belin, 31400 Toulouse, France\label{aff116}
\and
Universit\'e St Joseph; Faculty of Sciences, Beirut, Lebanon\label{aff117}
\and
Departamento de F\'isica, FCFM, Universidad de Chile, Blanco Encalada 2008, Santiago, Chile\label{aff118}
\and
Universit\"at Innsbruck, Institut f\"ur Astro- und Teilchenphysik, Technikerstr. 25/8, 6020 Innsbruck, Austria\label{aff119}
\and
Satlantis, University Science Park, Sede Bld 48940, Leioa-Bilbao, Spain\label{aff120}
\and
Infrared Processing and Analysis Center, California Institute of Technology, Pasadena, CA 91125, USA\label{aff121}
\and
Instituto de Astrof\'isica e Ci\^encias do Espa\c{c}o, Faculdade de Ci\^encias, Universidade de Lisboa, Tapada da Ajuda, 1349-018 Lisboa, Portugal\label{aff122}
\and
Niels Bohr Institute, University of Copenhagen, Jagtvej 128, 2200 Copenhagen, Denmark\label{aff123}
\and
Universidad Polit\'ecnica de Cartagena, Departamento de Electr\'onica y Tecnolog\'ia de Computadoras,  Plaza del Hospital 1, 30202 Cartagena, Spain\label{aff124}
\and
Dipartimento di Fisica e Scienze della Terra, Universit\`a degli Studi di Ferrara, Via Giuseppe Saragat 1, 44122 Ferrara, Italy\label{aff125}
\and
Istituto Nazionale di Fisica Nucleare, Sezione di Ferrara, Via Giuseppe Saragat 1, 44122 Ferrara, Italy\label{aff126}
\and
INAF, Istituto di Radioastronomia, Via Piero Gobetti 101, 40129 Bologna, Italy\label{aff127}
\and
Astronomical Observatory of the Autonomous Region of the Aosta Valley (OAVdA), Loc. Lignan 39, I-11020, Nus (Aosta Valley), Italy\label{aff128}
\and
Universit\'e C\^{o}te d'Azur, Observatoire de la C\^{o}te d'Azur, CNRS, Laboratoire Lagrange, Bd de l'Observatoire, CS 34229, 06304 Nice cedex 4, France\label{aff129}
\and
Department of Physics, Oxford University, Keble Road, Oxford OX1 3RH, UK\label{aff130}
\and
Universit\'e PSL, Observatoire de Paris, Sorbonne Universit\'e, CNRS, LERMA, 75014, Paris, France\label{aff131}
\and
Universit\'e Paris-Cit\'e, 5 Rue Thomas Mann, 75013, Paris, France\label{aff132}
\and
Aurora Technology for European Space Agency (ESA), Camino bajo del Castillo, s/n, Urbanizacion Villafranca del Castillo, Villanueva de la Ca\~nada, 28692 Madrid, Spain\label{aff133}
\and
Department of Mathematics and Physics E. De Giorgi, University of Salento, Via per Arnesano, CP-I93, 73100, Lecce, Italy\label{aff134}
\and
INFN, Sezione di Lecce, Via per Arnesano, CP-193, 73100, Lecce, Italy\label{aff135}
\and
INAF-Sezione di Lecce, c/o Dipartimento Matematica e Fisica, Via per Arnesano, 73100, Lecce, Italy\label{aff136}
\and
ICL, Junia, Universit\'e Catholique de Lille, LITL, 59000 Lille, France\label{aff137}
\and
ICSC - Centro Nazionale di Ricerca in High Performance Computing, Big Data e Quantum Computing, Via Magnanelli 2, Bologna, Italy\label{aff138}
\and
Instituto de F\'isica Te\'orica UAM-CSIC, Campus de Cantoblanco, 28049 Madrid, Spain\label{aff139}
\and
CERCA/ISO, Department of Physics, Case Western Reserve University, 10900 Euclid Avenue, Cleveland, OH 44106, USA\label{aff140}
\and
Technical University of Munich, TUM School of Natural Sciences, Physics Department, James-Franck-Str.~1, 85748 Garching, Germany\label{aff141}
\and
Max-Planck-Institut f\"ur Astrophysik, Karl-Schwarzschild-Str.~1, 85748 Garching, Germany\label{aff142}
\and
Laboratoire Univers et Th\'eorie, Observatoire de Paris, Universit\'e PSL, Universit\'e Paris Cit\'e, CNRS, 92190 Meudon, France\label{aff143}
\and
Departamento de F{\'\i}sica Fundamental. Universidad de Salamanca. Plaza de la Merced s/n. 37008 Salamanca, Spain\label{aff144}
\and
IRFU, CEA, Universit\'e Paris-Saclay 91191 Gif-sur-Yvette Cedex, France\label{aff145}
\and
Universit\'e de Strasbourg, CNRS, Observatoire astronomique de Strasbourg, UMR 7550, 67000 Strasbourg, France\label{aff146}
\and
Center for Data-Driven Discovery, Kavli IPMU (WPI), UTIAS, The University of Tokyo, Kashiwa, Chiba 277-8583, Japan\label{aff147}
\and
Dipartimento di Fisica - Sezione di Astronomia, Universit\`a di Trieste, Via Tiepolo 11, 34131 Trieste, Italy\label{aff148}
\and
California Institute of Technology, 1200 E California Blvd, Pasadena, CA 91125, USA\label{aff149}
\and
Department of Physics \& Astronomy, University of California Irvine, Irvine CA 92697, USA\label{aff150}
\and
Departamento F\'isica Aplicada, Universidad Polit\'ecnica de Cartagena, Campus Muralla del Mar, 30202 Cartagena, Murcia, Spain\label{aff151}
\and
Instituto de F\'isica de Cantabria, Edificio Juan Jord\'a, Avenida de los Castros, 39005 Santander, Spain\label{aff152}
\and
Observatorio Nacional, Rua General Jose Cristino, 77-Bairro Imperial de Sao Cristovao, Rio de Janeiro, 20921-400, Brazil\label{aff153}
\and
CEA Saclay, DFR/IRFU, Service d'Astrophysique, Bat. 709, 91191 Gif-sur-Yvette, France\label{aff154}
\and
Institute of Cosmology and Gravitation, University of Portsmouth, Portsmouth PO1 3FX, UK\label{aff155}
\and
Department of Computer Science, Aalto University, PO Box 15400, Espoo, FI-00 076, Finland\label{aff156}
\and
Ruhr University Bochum, Faculty of Physics and Astronomy, Astronomical Institute (AIRUB), German Centre for Cosmological Lensing (GCCL), 44780 Bochum, Germany\label{aff157}
\and
Department of Physics and Astronomy, Vesilinnantie 5, University of Turku, 20014 Turku, Finland\label{aff158}
\and
Serco for European Space Agency (ESA), Camino bajo del Castillo, s/n, Urbanizacion Villafranca del Castillo, Villanueva de la Ca\~nada, 28692 Madrid, Spain\label{aff159}
\and
ARC Centre of Excellence for Dark Matter Particle Physics, Melbourne, Australia\label{aff160}
\and
Centre for Astrophysics \& Supercomputing, Swinburne University of Technology,  Hawthorn, Victoria 3122, Australia\label{aff161}
\and
Department of Physics and Astronomy, University of the Western Cape, Bellville, Cape Town, 7535, South Africa\label{aff162}
\and
DAMTP, Centre for Mathematical Sciences, Wilberforce Road, Cambridge CB3 0WA, UK\label{aff163}
\and
Kavli Institute for Cosmology Cambridge, Madingley Road, Cambridge, CB3 0HA, UK\label{aff164}
\and
Department of Astrophysics, University of Zurich, Winterthurerstrasse 190, 8057 Zurich, Switzerland\label{aff165}
\and
Oskar Klein Centre for Cosmoparticle Physics, Department of Physics, Stockholm University, Stockholm, SE-106 91, Sweden\label{aff166}
\and
Astrophysics Group, Blackett Laboratory, Imperial College London, London SW7 2AZ, UK\label{aff167}
\and
Univ. Grenoble Alpes, CNRS, Grenoble INP, LPSC-IN2P3, 53, Avenue des Martyrs, 38000, Grenoble, France\label{aff168}
\and
Dipartimento di Fisica, Sapienza Universit\`a di Roma, Piazzale Aldo Moro 2, 00185 Roma, Italy\label{aff169}
\and
Centro de Astrof\'{\i}sica da Universidade do Porto, Rua das Estrelas, 4150-762 Porto, Portugal\label{aff170}
\and
HE Space for European Space Agency (ESA), Camino bajo del Castillo, s/n, Urbanizacion Villafranca del Castillo, Villanueva de la Ca\~nada, 28692 Madrid, Spain\label{aff171}
\and
INAF - Osservatorio Astronomico d'Abruzzo, Via Maggini, 64100, Teramo, Italy\label{aff172}
\and
Theoretical astrophysics, Department of Physics and Astronomy, Uppsala University, Box 516, 751 37 Uppsala, Sweden\label{aff173}
\and
Minnesota Institute for Astrophysics, University of Minnesota, 116 Church St SE, Minneapolis, MN 55455, USA\label{aff174}
\and
Mathematical Institute, University of Leiden, Einsteinweg 55, 2333 CA Leiden, The Netherlands\label{aff175}
\and
Institute of Astronomy, University of Cambridge, Madingley Road, Cambridge CB3 0HA, UK\label{aff176}
\and
Department of Astrophysical Sciences, Peyton Hall, Princeton University, Princeton, NJ 08544, USA\label{aff177}
\and
Space physics and astronomy research unit, University of Oulu, Pentti Kaiteran katu 1, FI-90014 Oulu, Finland\label{aff178}
\and
Center for Computational Astrophysics, Flatiron Institute, 162 5th Avenue, 10010, New York, NY, USA\label{aff179}
\and
Department of Physics, Massachusetts Institute of Technology, Cambridge, MA 02139, USA\label{aff180}
\and
MIT Kavli Institute for Astrophysics and Space Research, Massachusetts Institute of Technology, Cambridge, MA 02139, USA\label{aff181}}     

\abstract{The slitless spectroscopy mode of the \ac{nisp} on board the \Euclid telescope has enabled efficient spectroscopy of objects within a large field of view. Nevertheless, the relatively low spectral resolution, overlapping spectra, and contamination pose challenges to source classification and redshift determination using the \ac{nisp} spectra alone. 
In this work, we present a large and homogeneous sample of bright quasars identified from the Euclid Quick Data Release (Q1), constructed by combining high-purity candidate selections from \gaia and WISE with the new spectroscopic capabilities of \Euclid. 
Through visual inspection of the \Euclid spectra of these quasar candidates, we identify approximately 3500 quasars and determine reliable redshifts in the range of $0< z \lesssim4.8$. Of these, 2686 are new spectroscopic identifications relative to existing public compilations. We generated the first \Euclid composite spectrum of quasars covering rest-frame \ac{nuv} to \ac{nir} wavelengths without telluric lines, which will be pivotal to \ac{nir} quasar spectral analysis. We obtained an empirical spectroscopic depth of $\JE \lesssim 21.5$ and $\HE \lesssim 21.3$ at the sensitivity of the Wide Field Survey, beyond which the number of securely identified quasars declines sharply. Accordingly, the sample presented in this paper comprises spectroscopically confirmed quasars brighter than these limits. 
We analysed morphological parameters from the Visible Camera (VIS) using Sérsic and model-independent (CAS) metrics, and a deep-learning point spread function fraction to track nuclear dominance. The VIS morphologies show a clear redshift dependence: at low redshift ($z<0.5$), obvious host structures are common and a single Sérsic model fits about half of the sources; at intermediate redshift ($0.5<z<2$), the nuclear component dominates, with 90\% of the Sérsic fits saturating at the upper index limit. In this intermediate redshift regime, $f_{\sfont{PSF}}$ is available, and we use it as a more reliable compactness measure than the single-Sérsic and CAS parameters to quantify nuclear versus host emission. We also explore the novel \Euclid \ac{nir} colour space and discuss the role of these quasars in refining active galactic nucleus selection techniques for future \Euclid data releases. }

   \titlerunning{Bright quasars with \Euclid spectroscopy}
   \authorrunning{Euclid Collaboration: Y.~Fu et al.}
   
   \maketitle
\nolinenumbers

\section{\label{sec:Intro}Introduction}

Powered by accreting supermassive black holes, quasars are among the most luminous and distant objects in the Universe. Quasars act as beacons that allow us to probe recent to early cosmic epochs and trace the large‐scale structure of the cosmos \citep[e.g.][]{2017AJ....154...28B,2020MNRAS.499..210N,2023ARA&A..61..373F}. Traditionally, quasar surveys have primarily relied on multi-colour selection techniques to isolate candidates, with subsequent slit or multi-fibre spectroscopy to confirm their nature and determine precise redshifts \citep[e.g.][]{2002AJ....123.2945R,2004MNRAS.349.1397C,2015ApJS..221...27M,2023ApJ...944..107C}. Ground-based quasar surveys have adopted slitless spectroscopy to find quasars in a cost-effective way. However, such slitless campaigns suffer from low spectral resolution, overlapping spectra, and contamination, often necessitating follow-up slit spectroscopy for confirmation \citep[e.g.][]{1986ApJ...306..411S,1991ApJS...75..273O,1999AJ....117...40S}.

The advent of space-based observations and improvements in data reduction have substantially improved the quality of slitless spectroscopic data. For example, the \HST's (HST) Advanced Camera for Surveys (ACS) and Wide Field Camera 3 (WFC3) have produced high-quality grism data \citep[e.g.][]{Momcheva_2016ApJS_3dhst_survey,2019ApJ...870..133E} that are processed with dedicated pipelines \citep[e.g.][]{2009PASP..121...59K,2019ascl.soft05001B}. More recently, with the low background, high sensitivity, and high spatial resolution of the \textit{James Webb} Space Telescope (JWST), the Near Infrared Camera \citep[NIRCam;][]{2005SPIE.5904....1R,2023PASP..135b8001R} and the Near Infrared Imager and Slitless Spectrograph \citep[NIRISS;][]{2022PASP..134b5002W,2023PASP..135i8001D} on board JWST have enabled unprecedented studies of distant galaxies \citep[e.g.][]{2022ApJ...938L..13R,2023ApJ...953...53S,2023MNRAS.525.2864O,2024MNRAS.535.1067M}. 

The European Space Agency's (ESA) \Euclid\ mission \citep{EuclidSkyOverview} is designed to probe the dark matter and dark energy of the Universe by studying weak lensing and galaxy clustering over approximately one third of the sky in both the optical and near-infrared using the Visible Camera \citep[VIS;][]{EuclidSkyVIS} and the \acl{nisp} \citep[\acs{nisp};][]{EuclidSkyNISP}. A key feature of \Euclid\ is the slitless spectroscopic mode of \ac{nisp}, which can simultaneously capture spectra of sources in a large field of view of 0.57\,$\deg^2$. Three red grisms covering the same \RGE band (1206--1892 nm) are adopted in the Euclid Wide Survey to provide spectra with different dispersion directions of \ang{0}, \ang{180}, and \ang{270} with respect to the detector columns. Dispersed slitless images of the three grims are combined to disentangle overlapping spectra from multiple sources, and generate the final clean spectra. The red grims have a resolving power of ${{\cal R}_{{\rm{R}}{{\rm{G}}_{\rm{E}}}}} > 480$ for a source with a \ang{;;0.5} diameter \citep{EuclidSkyNISP,EuclidSkyOverview}. This spectroscopic capability enables a census of bright quasars and enhances the efficiency of quasar discovery by leveraging the high sensitivity and spatial resolution from space. For example, \citet{Banados25} have recently discovered a $z=5.404$ quasar, EUCL J181530.01+652054.0, using \ac{nisp} spectroscopy.

\citet{EP-Lusso} provide a detailed prediction of the \ac{nisp} spectroscopic mode for \acp{agn} using mock spectra. They demonstrate that redshift measurements are robust when the H\,$\alpha$ emission line is visible within the spectral coverage of \RGE ($0.89 < z < 1.83$) at a line flux greater than $2 \times 10^{-16}$~erg\,s$^{-1}$\,cm$^{-2}$. Outside this redshift range, however, redshift measurements are inefficient due to a low \ac{snr} or a lack of prominent emission lines or both.

Early investigations of \acp{agn} in the first Euclid Quick Data Release \citep[Q1;][]{Q1-TP001} already demonstrate the potential of \Euclid for \ac{agn} science. For example, \citet{Q1-SP027} present an \ac{agn} candidate catalogue with a total of 229\,779 objects selected with multi-wavelength data, while \citet{Q1-SP003} identify \Euclid counterparts to X-ray sources in the Deep Fields, most of which are \acp{agn}, using catalogues from \textit{eROSITA} \citep{2024A&A...682A..34M}, \XMMN \citep{2020A&A...641A.136W}, and \Chandra \citep{2024ApJS..274...22E}. 

In this work, we present a homogenous bright quasar sample identified with spectroscopic data of Q1, complementary to the efforts of \citet{Q1-SP027} and \citet{Q1-SP003}. We select quasar candidates from all-sky optical and mid-infrared databases, and identify the objects based on prominent emission lines in the \Euclid slitless spectra. Our analysis demonstrates that slitless spectroscopy, as implemented in the \Euclid\ mission, not only overcomes some of the limitations of colour-based quasar selection techniques, but also provides a valuable dataset for studying the large-scale structure of the Universe through quasar clustering and cross-correlation with galaxy and weak lensing maps \citep[e.g.][]{2003MNRAS.342..467M,2015MNRAS.449.4326P,2023ApJ...946...27P,2023JCAP...11..043A}. Spectral properties of these quasars obtained with multi-component fitting, including spectral indices, line widths, and black hole masses,  will be reported in Euclid Collaboration: Calhau et al. (in preparation).  

The paper is organised as follows. \Cref{sec:sample} describes the sample selections of quasar candidates and the crossmatch to the \Euclid spectral catalogue. \Cref{sec:spec_ident} describes the identification and redshift determination procedure. We present our results in \cref{sec:results} and discuss their implications in \cref{sec:discussion}. Finally, \cref{sec:conclusions} concludes with a summary of our findings and prospects for future studies. All magnitudes are in the AB system \citep{1983ApJ...266..713O} unless stated otherwise.

\section{Data\label{sec:sample}}

In this work, we use external quasar candidates from \gaia and AllWISE as the input sample for the identification with \Euclid spectroscopy. Below, we briefly introduce the Q1 data we use, and describe the contents of the input quasar candidate sample and the matched Q1 spectroscopic sample.

\subsection{Q1 photometry and spectroscopy \label{sec:data-q1}}

The Q1 \citep[][]{Q1-TP001,Q1cite} dataset contains images and photometric catalogues from both VIS (\IE band) and \ac{nisp} (\YE, \JE, and \HE bands), and \ac{1d} spectra of the \ac{nisp} spectroscopic mode. 
In this work, we use the main photometric catalogue (\verb|catalogue.mer_catalogue| in the Euclid Science Archive\footnote{\url{https://eas.esac.esa.int/sas/}}) generated by the MERge Processing Function \citep[MER;][]{Q1-TP004}, which include aperture fluxes, template-fit and Sérsic-fit fluxes, and quality flags in each band, as well as morphological information for all sources detected in the Euclid Deep Fields. We also use the main morphology catalogue \citep[\texttt{catalogue.mer\_morphology};][]{Q1-SP040} to assess the morphological properties of the final sample.

Combined \ac{1d} slitless spectroscopic data of sources brighter than $\HE=22.5$ have been generated by the dedicated SIR spectroscopic processing function \citep[][]{Q1-TP006}. The slitless spectra used in this work are retrieved from ESA Datalabs\footnote{\url{https://datalabs.esa.int/}} \citep{Datalabscite}. 

Because the observed slitless spectrum of an object is the convolution of its intrinsic spectrum with its wavelength‑dependent spatial light profile along the dispersion direction, sources with larger extents will have lower spectral resolution. In addition, the blue and red ends of the observed spectrum tend to show upturns due to the convolution nature of the observed spectrum, and the lower \ac{snr} at both ends \citep[see e.g.][]{2004ApJS..154..501P}. Each of the original \ac{1d} spectra contains 531 data points, covering 11\,900--19\,002 \AA\ with a wavelength interval of 13.4 \AA. To keep only the useful data from the slitless spectra, we trim the blue and red ends, retaining 12\,047--18\,734 \AA, which yields 500 data points per spectrum.

\subsection{Reliable quasar candidates from \gaia and AllWISE \label{sec:input_quasar}}

\gdr{3} announced a sample of 6.6 million quasar candidates \citep[the \texttt{qso\_candidates} table, hereafter the GDR3 QSO candidate catalogue;][]{GaiaCollaboration_2023_2023A&A...674A...1G,GaiaCollaboration_2023_2023A&A...674A..41G}, which has high completeness thanks to the combination of several different modules, including the Discrete Source Classifier (DSC), the Quasar Classifier (QSOC), the variability classification module, the surface brightness profile module, and the \gdr{3} Celestial Reference Frame source table. The DSC uses the \gaia Blue (BP) and Red (RP) Photometer \citep{2023A&A...674A...2D} spectrum together with the mean $G$-band magnitude, the variability in this band, the parallax, and the proper motion to classify each \gaia source probabilistically into five classes: quasars, galaxies, stars, white dwarfs, and physical binary stars. The QSOC determines the redshifts using BP and RP spectra of the sources classified as quasars by the DSC \citep{2023A&A...674A..31D}. The variability classification module identifies 25 classes of variable sources (including \ac{agn} candidates) from the variability of the \gaia light curves using supervised machine learning \citep{2023A&A...674A..14R,2023A&A...674A..24C}. Both the surface-brightness profile module and the \gdr{3} Celestial Reference Frame source table are based on external catalogues of quasars and quasar candidates \citep[see][for a complete list]{2023A&A...674A..11D,2022A&A...667A.148G}. 
Despite its high completeness, the GDR3 QSO candidate catalogue has an estimated low purity of quasars (52\,\%) and a large scatter of redshift estimates \citep{GaiaCollaboration_2023_2023A&A...674A..41G}.

Instead of using the original GDR3 QSO candidate catalogue, we take its purified subsets to find \Euclid counterparts of the sources. These purified catalogues include the following. 

\begin{enumerate}
    \item Quaia \citep{Storey-Fisher2024} with nearly 1.3 million sources at $G<20.5$. This sample is selected using a set of cuts involving proper motion, \gaia and UnWISE \citep{2019ApJS..240...30S} colours and magnitudes, which are designed to remove stellar contaminants of the Milky Way and the Large and Small Magellanic Clouds. 
    \item CatNorth \citep{Fu2024_CatNorth} with more than 1.5 million sources down to the \gaia limiting magnitude ($G<21.0$) in the 3\,$\pi$ sky of the Pan-STARRS1 \citep[PS1;][]{Chambers2016} footprint ($\delta>-30\degr$). This catalogue is primarily built with a machine learning classification model trained on colour and morphological features from \gaia, PS1, and CatWISE2020 \citep{2021ApJS..253....8M}. An additional probabilistic cut on proper motion \citep[probability density of zero proper motion; see][for the definition]{2021ApJS..254....6F,Fu2024_CatNorth} is applied to further purify the candidates. The CatNorth catalogue has a purity of approximately 90\,\%.
    \item CatSouth \citep{Fu2025_catsouth} with 0.9 million sources with $G<21.0$ covered by the fourth data release (DR4) of the SkyMapper Southern Survey \citep[SMSS; $\delta\lesssim 16\degr$;][]{Onken_2024_2024PASA...41...61O}. This catalogue is built with the same method as CatNorth, while based on data from \gaia, SMSS DR4, and VISTA (Visible and Infrared Survey Telescope for Astronomy) surveys \citep{2006Msngr.126...41E,2010NewA...15..433M,2011A&A...527A.116C,2013Msngr.154...35M,2013Msngr.154...32E}, as well as CatWISE2020. 
\end{enumerate}

All three purified catalogues propagate the \gdr{3} QSOC template-matching redshift and combine it with multi-band photometry to derive improved photometric redshifts. These photometric redshifts reduce the fractions of catastrophic outliers by more than 15\,\% compared to the original QSOC redshift, at the cost of a modest decrease in precision \citep{Storey-Fisher2024,Fu2024_CatNorth}. The compilation of the three catalogues above contains more than 1.9 million unique (with unique \gaia \verb|source_id|) quasar candidates in the entire sky. This purified GDR3 QSO candidate catalogue will be referred to as GDR3-QSOs hereafter for simplicity. Together, 4467 sources are matched to the Q1 main photometric catalogue (see \cref{sec:data-q1} for details) using \gaia \verb|source_id|, which is listed in both GDR3-QSOs and \verb|catalogue.mer_catalogue| of Q1.

In addition to GDR3-QSOs, which are mainly optically bright quasars, we also include \ac{agn} candidates selected with data from the Wide-field Infrared Survey Explorer \citep[WISE;][]{2010AJ....140.1868W}, a NASA mission that has surveyed the entire sky in the 3.4-, 4.6-, 12-, and 22-$\mu$m mid-infrared bands (W1, W2, W3, and W4). The AllWISE source catalogue was built by combining data from the WISE cryogenic and NEOWISE \citep{2011ApJ...731...53M} post-cryogenic survey phases, providing positions, proper motions, four-band fluxes, and flux variability statistics for over 747 million objects. Using $W1-W2$ colour and $W2$ magnitude from AllWISE, \citet{2018ApJS..234...23A} constructed two large catalogues of \ac{agn} candidates across 75\,\% of the sky: the R90 catalogue with 90\,\% reliability, and the C75 catalogue with 75\,\% completeness. In total, 8202 sources from the R90 \ac{agn} candidate catalogue are matched to the Q1 main photometric catalogue using a radius of \ang{;;1.5}.

Combining GDR3-QSOs and the R90 \ac{agn} candidates yields 10\,201 unique Q1 sources. Among them, 5083 are detected by \gaia (including 616 sources from R90), and 5118 are not in \gaia. We refer to these two subsets as \gaia and non-\gaia subsets, the latter only selected using the R90 \ac{agn} candidate catalogue. Matching the input quasar candidates with the Q1 spectra source table (\texttt{sedm.spectra\_source} in the Euclid Science Archive) gives 9214 sources. The numbers of input sources of different subsets are summarised in \cref{tab:gaia_non_gaia_summary}.

\begin{table}[h]
\centering
\caption{Summary of numbers of input sources and identified quasars.}
\label{tab:gaia_non_gaia_summary}
\begin{tabular}{lccc}
\toprule
Sample & \gaia & Non-\gaia & All \\
\midrule
Photometric (MER) sources            & 5083 & 5118 & 10\,201 \\
Spectroscopic (SIR) sources            & 4965 & 4249 &  9214 \\
Identified quasars     & 2753 &  715 &  3468 \\
\midrule
Success rate      & 55.4\,\% & 16.8\,\% &  37.6\,\% \\
\bottomrule
\end{tabular}
\tablefoot{The success rate is the number of identified quasars divided by that of SIR sources.}
\end{table}

\section{Source classification and spectral redshift determination\label{sec:spec_ident}}

\subsection{Template-matching redshifts for the candidates}
For each observed quasar spectrum, we first estimated the redshift by comparing it to a rest-frame template via a Pearson correlation function (PCF, also known as the normalised cross-correlation) method \citep{2018A&A...616A...6S}. We began with the template from \citet{Glikman_2006}, based on 27 bright quasars observed with the NASA \ac{irtf}. After constructing a new composite spectrum in \cref{subsec:spec_comps}, we re-ran the PCF with the new composite as a template to refine the redshift estimates; for the final measurements we adopted a piecewise template that combines \citet{2001AJ....122..549V}, the mean composite from this work, and \citet{Glikman_2006}, see \cref{app:piecewise_template}. The PCF algorithm proceeds as follows:

\begin{enumerate}
    \item The template wavelengths are shifted into the observed frame using the relation
    \begin{equation}
    \lambda_{\mathrm{obs}} = \lambda_{\mathrm{rest}} \, (1+z)\;,
    \end{equation}
    where $z$ is the trial redshift.
    
    \item The shifted template is then interpolated onto the observed wavelength grid. Both the observed flux, $F_{\mathrm{obs}}(\lambda)$, and the interpolated template flux, $F_{\mathrm{temp}}(\lambda; z)$, are median-normalised as 
    \begin{gather}
            \tilde{F}_{\mathrm{obs}}(\lambda) = \frac{F_{\mathrm{obs}}(\lambda)}{\mathrm{median}[F_{\mathrm{obs}}(\lambda)]}\,, \\
    \tilde{F}_{\mathrm{temp}}(\lambda; z) = \frac{F_{\mathrm{temp}}(\lambda; z)}{\mathrm{median}[F_{\mathrm{temp}}(\lambda; z)]}\,,
    \end{gather}
    to minimise the effect of continuum differences.
    
    \item The PCF is computed over the set, $\mathcal{I}$, of overlapping wavelength bins using the median-normalised spectra above:
    \begin{equation}
    r(z) = \frac{\sum_{i\in\mathcal{I}} \left[\tilde{F}_{\mathrm{obs},i} - \langle \tilde{F}_{\mathrm{obs}} \rangle\right]\left[\tilde{F}_{\mathrm{temp},i}(z) - \langle \tilde{F}_{\mathrm{temp}}(z) \rangle\right]}{\sqrt{\sum_{i\in\mathcal{I}} \left[\tilde{F}_{\mathrm{obs},i} - \langle \tilde{F}_{\mathrm{obs}} \rangle\right]^2 \, \sum_{i\in\mathcal{I}} \left[\tilde{F}_{\mathrm{temp},i}(z) - \langle \tilde{F}_{\mathrm{temp}}(z) \rangle\right]^2}}\;,
    \end{equation}
    where $\langle \tilde{F} \rangle$ denotes the mean of the median-normalised spectrum.
    
    \item To account for the logarithmic nature of wavelength shifts, the redshift grid is sampled with a step size that scales with $1+z$:
    \begin{equation}
    \Delta z = \delta \, (1+z)\;,
    \end{equation}
    where $\delta$ is a base step (e.g. $\delta = 0.001$ in this work).
    
    \item The best-fit template redshift (hereafter \ztemp) is then adopted as the value of $z$ that maximises $r(z)$:
    \begin{equation}
    \ztemp = \underset{z}{\mathrm{arg\max}}\; r(z)\;.
    \end{equation}
    
\end{enumerate}

A primary advantage of using the PCF is that it constrains the correlation to be in the range between $-1$ and $+1$, thereby removing dependence on the absolute flux scale or any additive offsets between the observed spectrum and the template. This normalisation ensures that the metric is driven solely by the relative shapes and positions of spectral features. By employing the PCF method, our analysis emphasises the similarity in spectral features rather than overall flux levels.

\subsection{Visual inspection}

An interactive visual inspection tool \verb|PGSpecPlot|, part of the Python package \verb|specbox| \citep{fu_2026_18642758}, was used to check the spectra of the quasar candidates. During visual inspection, each spectrum is displayed sequentially with an overplotted quasar template adjusted to the estimated template redshift (\ztemp). Using interactive controls, including a slider with non-linear $(1+z)$ scaling and a corresponding spin box (a text box with an up-down control), the user can verify the template redshift and adjust it as necessary. Keyboard shortcuts facilitate the rapid classification of each spectrum (e.g. flagging non-quasar objects or uncertain cases). A history of inspected spectra is maintained in a comma-separated value file so that previously processed spectra can be automatically loaded.

In addition to the initial \ztemp, a \gaia redshift is displayed in the window when available. The \gaia redshift (hereafter $z_{\gaia}$) is primarily based on the \gdr{3} low-resolution spectral template-matching redshift from QSOC \citep[\texttt{redshift\_qsoc};][]{2023A&A...674A..31D,GaiaCollaboration_2023_2023A&A...674A..41G}, supplemented by the photometric redshift from CatNorth and CatSouth when \verb|redshift_qsoc| is not available. Because \zgaia and \ztemp are independent spectroscopic redshift estimates obtained from different wavelength ranges (\zgaia from the BP or RP bands within 330--1050 nm, and \ztemp from \RGE in the \ac{nir}), a \ztemp that is close to \zgaia (with $|\ztemp-\zgaia|/[1+\ztemp]<0.15$) is taken as a secure visual redshift (\zvi) in most cases, even when only one emission line is present in the wavelength range of \RGE. Nevertheless, when \zgaia is unavailable and only one emission line is seen, the redshift can be highly uncertain. Such single-line spectra are labelled as `uncertain-redshift' objects unless the emission line and continuum fitted the template with high confidence during visual inspection.

\begin{figure*}[htb]
\centering
\includegraphics[angle=0,width=1.0\hsize]{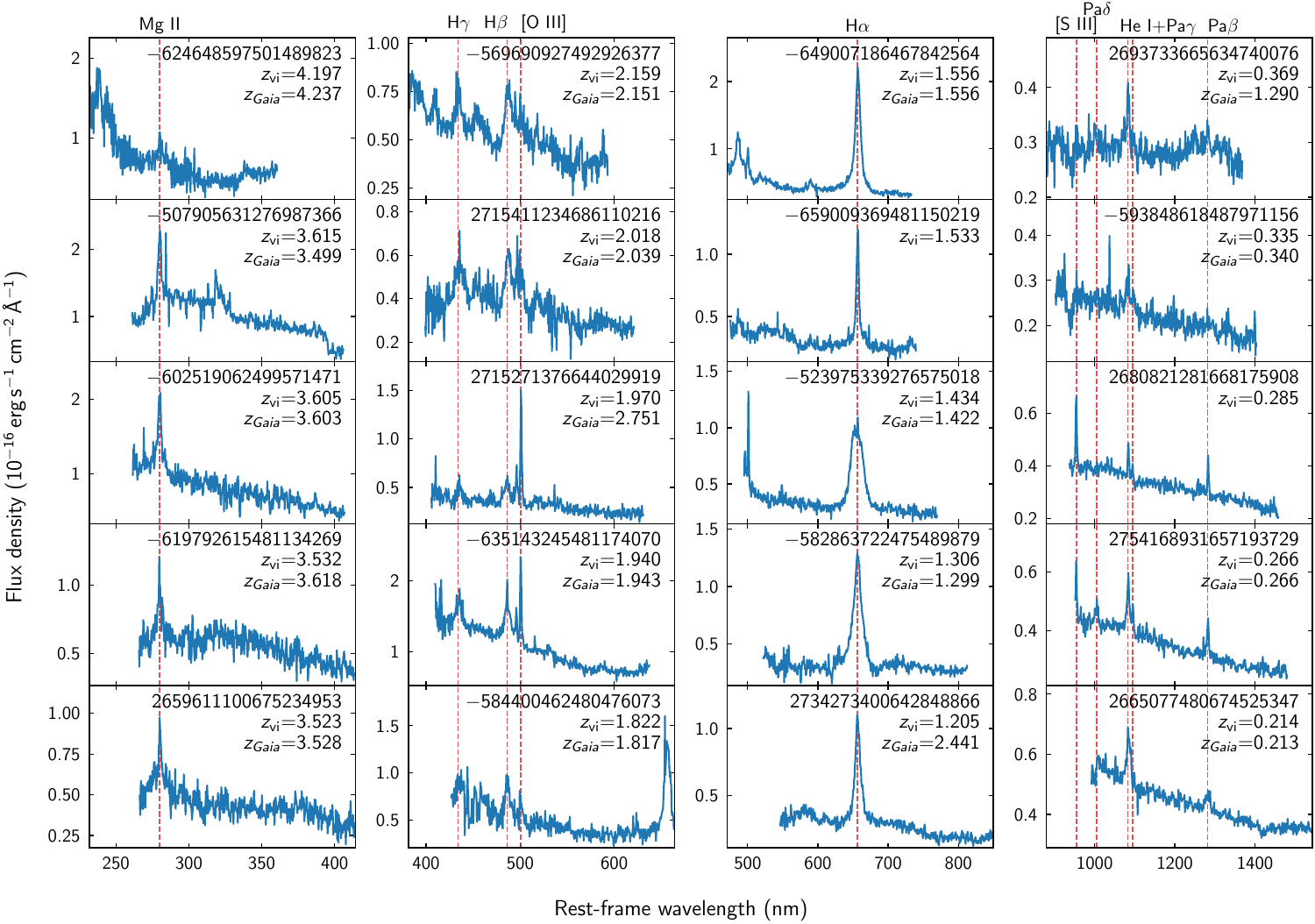}
\caption{Example \Euclid spectra of the visually identified quasars in the rest frame. The four columns from left to right show prominent emission lines used for the visual inspection in descending order of redshift: \ion{Mg}{ii} in the first column; H\,$\beta$, [\ion{O}{iii}], and H\,$\gamma$ in the second column; H\,$\alpha$ in the third column; and \ion{He}{i}\,+\,Pa\,$\gamma$, Pa\,$\beta$, Pa\,$\delta$, and [\ion{S}{iii}] in the last column. }
\label{fig:spectra_examples}
\end{figure*}

As expected from the predictions \citep{EP-Lusso}, quasars with $0.89<z<1.83$ are most easily identified with the H\,$\alpha$ emission line, which is the strongest and broadest line among all emission lines detected by \RGE. Quasars at lower redshift ($z<0.89$) are mainly identified with the combination of several rest-frame \ac{nir} emission lines, i.e.  [\ion{S}{iii}] $\lambda$9071, Pa\,$\delta$, \ion{He}{i}\,+\,Pa\,$\gamma$ (blended), and Pa\,$\beta$. At $1.83<z<2.85$, quasars are identified with H\,$\beta$, [\ion{O}{iii}], and H\,$\gamma$. At $2.85<z<3.3$, the quasar spectra lack strong features except for H\,$\gamma$ and the pseudo-continuum (`small blue bump'), and they are mainly identified with the agreement between \ztemp and \zgaia. At $z>3.3$, \ion{Mg}{ii} enters the wavelength range, and the redshift determination is secured with the combination of \ztemp (from the \ion{Mg}{ii} line) and \zgaia. \Cref{fig:spectra_examples} shows example \Euclid spectra of identified quasars at different rest-frame wavelength ranges, which correspond to the different emission line features described above.

Approximately 2300 sources are labelled as `uncertain-redshift' objects, making up 25\,\% of the entire sample (9214 spectra). Given the complexity of the data reduction and limited depth of the slitless spectroscopy, there are also approximately 3400 unidentifiable spectra with: (i) corrupted data with too many invalid or anomalously high flux values; or (ii) featureless spectra due to either weak lines or low \ac{snr}. These spectra are currently classified as unknown sources. However, given the high reliability of our input catalogues, it is likely that many of these objects are indeed quasars that could be confirmed with deeper observations or improved data processing in the future.

\subsection{Spurious redshift rejection, consistency check with DESI, and the final redshift}

To evaluate the performance of our visual redshift $z_{\mathrm{vi}}$, we first examined the source concentration parameter \mumaxmag (\verb|mumax_minus_mag| in \verb|catalogue.mer_catalogue|). Here, $\mu_{\mathrm{max}}$ is the source peak surface brightness above the background, and $\mathrm{mag}$ is the magnitude used to compute point-like probability, both given by SourceXtractor++\footnote{\url{https://github.com/astrorama/SourceXtractorPlusPlus}} \citep[][]{2020ASPC..527..461B,2022arXiv221202428K} during the MER data reduction. The estimator \mumaxmag is related to the concentration of light at the peak versus the total
magnitude; at a given magnitude, sources with smaller \mumaxmag are more point-like \citep{Q1-TP004}. This parameter has also been used in \citet{2012MNRAS.426.3369J}, \citet{2022arXiv220705709S}, \citet{2023A&A...671A.146E}, \citet{Q1-SP027}, and \citet{Q1-SP003} as input for point/extended source classification. 

\begin{figure}[htb]
\centering
\includegraphics[angle=0,width=1.0\hsize]{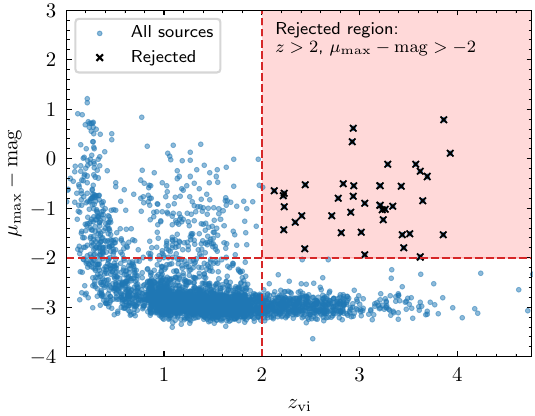}
\caption{Concentration parameter \mumaxmag as a function of the visual redshift \zvi. The rejected region is shaded in pink, and is defined with $\mumaxmag > -2$ at $z_{\mathrm{vi}} > 2$, as indicated by the dashed red lines. Sources inside the rejected region are considered to have spurious redshifts and are marked with black crosses.}
\label{fig:mumax_rejection}
\end{figure}

\Cref{fig:mumax_rejection} shows the distribution of visually identified quasars in the \mumaxmag-\zvi plane. In general, \mumaxmag decreases as \zvi becomes higher. 
Most sources exhibit compact morphologies (low \mumaxmag values) at $\zvi \gtrsim 0.8$, consistent with unresolved point-like sources. However, a distinct subset of sources at $z_{\mathrm{vi}} > 2$ shows significantly higher \mumaxmag values, indicative of extended or low-concentration profiles that are atypical for high-redshift quasars. These sources are flagged as spurious redshifts and excluded from the final sample using a conservative empirical cut: $\mumaxmag > -2$ at $z_{\mathrm{vi}} > 2$, as indicated by the dashed red lines. This selection removes sources whose morphology and redshift are inconsistent with expectations for high-redshift quasars, improving the redshift accuracy of the final catalogue.

As an external consistency check on our visually confirmed redshifts, we crossmatched our quasar sample with Data Release~1 of the Dark Energy Spectroscopic Instrument \citep[DESI DR1;][]{DESI_DR1} using a radius of \ang{;;1.0} and obtained 454 objects in common. We flag discrepant cases using the criterion $\left|\zvi-\zdesi\right|/(1+\zdesi)>0.15$, yielding 16 outliers. We then visually reinspect both the DESI and \Euclid spectra for all 16 sources. In 11 cases, the DESI spectra support the same redshift solution as our \Euclid-based identification, while the DESI catalogue redshift is inconsistent with the spectral features. In the remaining five cases, we revise our visual redshifts; the updated $\zvi$ values are now consistent with $\zdesi$ and are adopted in the released catalogue. The spectra of the 16 initially discrepant sources are shown in Appendix~\ref{app:desi_check}.

The visually identified quasar sample contains 3468 sources (38\,\% of the entire sample) covering the redshift range of $0< z \lesssim4.8$. In addition to the crossmatch to DESI DR1 above, we also crossmatch the full sample with the Million Quasar Catalogue \citep[Milliquas v8;][]{2023OJAp....6E..49F} using a radius of \ang{;;1.0} and find 710 sources in common. The union of the DESI DR1 matches (454 objects) and the Milliquas matches (710 objects) contain 782 sources, implying that 2686 quasars in our sample are new spectroscopic identifications.

\section{The bright quasar sample identified with Q1 spectroscopy\label{sec:results}}
\subsection{Photometric properties}

The 3468 sources of the visually identified quasar sample are compiled into a catalogue detailed in \cref{tab:main-metadata}. This catalogue includes source IDs and coordinates, redshifts, spectroscopic quality indicators, \ac{psf} fraction measurements from VIS imaging \citep{Q1-SP015} for a redshift-limited subsample ($0.5<z<2$), and magnitudes across the \Euclid bands. Among them, 2753 are \gaia sources and 715 are non-\gaia sources.

\begin{figure}[htb]
\centering
\includegraphics[width=1.0\hsize]{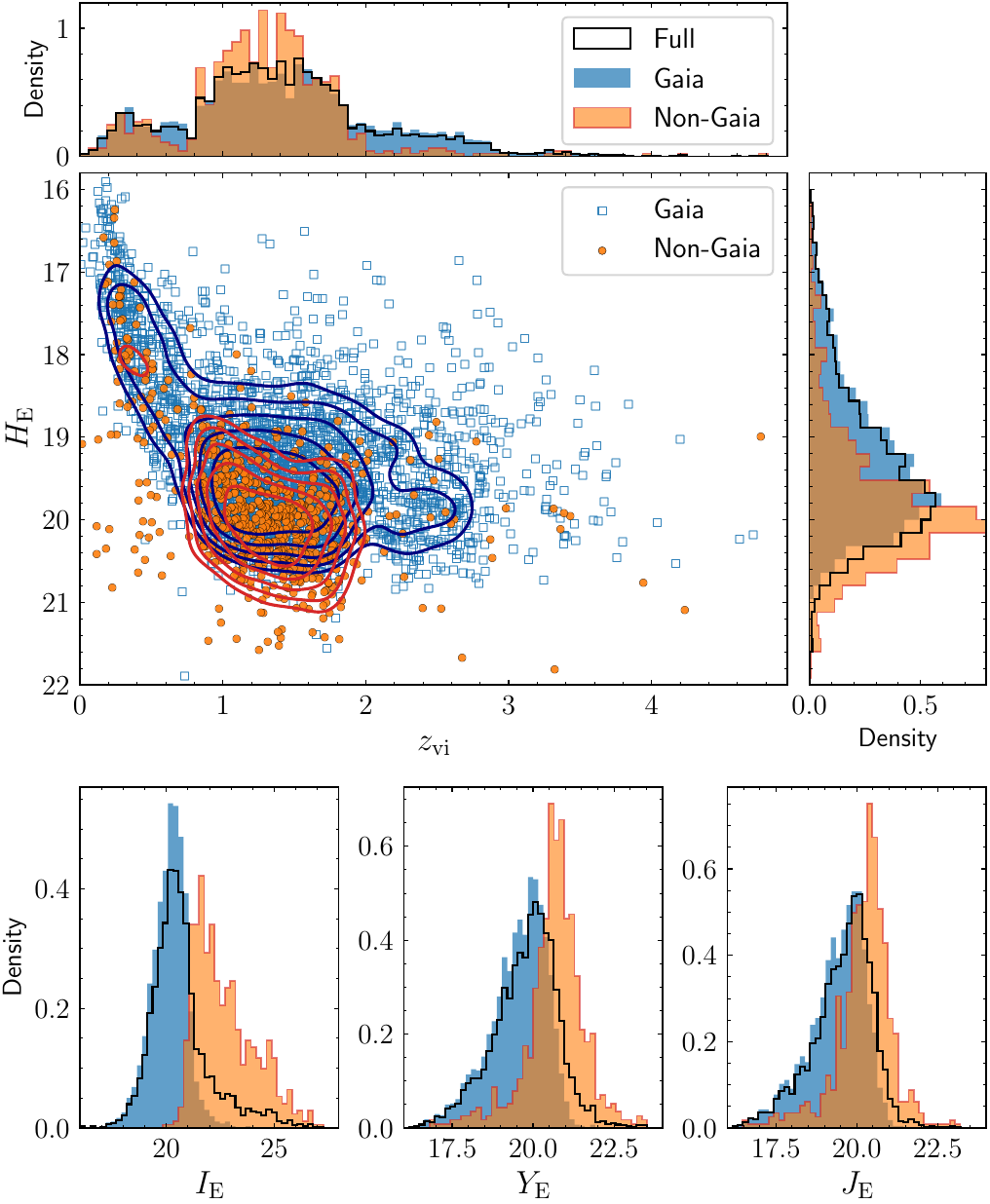}
\caption{
Redshift-magnitude distributions for the visually identified quasar sample using \Euclid Q1 data. 
\emph{Top:} Two-dimensional distribution of \HE versus $z_{\mathrm{vi}}$. 
\gaia-detected sources are shown as blue open squares, and sources not in \gdr{3} are shown as orange circles. 
Blue contours trace the density of the \gaia subset, and red contours trace the density of the non-\gaia subset. 
The marginal histograms for $z_{\mathrm{vi}}$ (above) and \HE (right) show the full sample (black steps), the \gaia subset (blue steps), and the non-\gaia subset (orange steps). 
\emph{Bottom:} One-dimensional histograms of \IE, \YE, and \JE for the same three samples. 
All histograms are normalised to unit area.
}

\label{fig:z_mag_hist}
\end{figure}

The redshift and magnitude distributions of the full sample, and \gaia or non-\gaia subsets, are shown in \cref{fig:z_mag_hist}. The redshift density distributions peak in the range $0.89\lesssim z\lesssim 1.83$ when the H\,$\alpha$ emission line is in the observed wavelength range, and drop sharply at $z\approx0.89$ and $z\approx1.83$ when H\,$\alpha$ moves out of the wavelength range. As shown in the histograms of \Euclid magnitudes, these visually identified quasars represent a bright sample with a median \IE of 20.5, and median magnitudes of 19.9 in \YE, 19.7 in \JE, and 19.5 in \HE. The non-\gaia subset is overall 2 magnitudes fainter than the \gaia subset in \IE, 0.8 magnitudes fainter in \JE and 0.6 magnitudes fainter in \HE. The faintest identified quasars in this work have $\IE \approx 27$, $\YE \approx 23$, and $\JE \approx \HE \approx 22.5$. 

\begin{figure*}
\centering
\includegraphics[width=\textwidth]{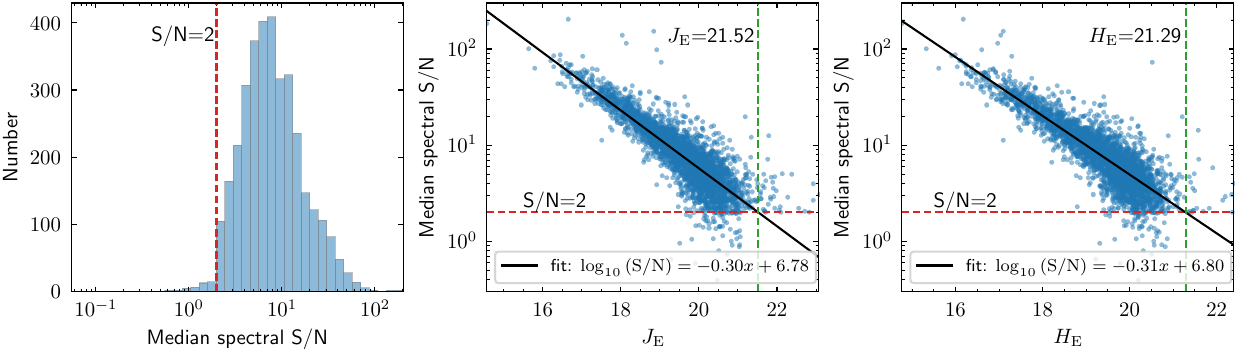}
\caption{
\emph{Left:} Distribution of the median spectral S/N within [12\,047, 18\,734] \AA\ for visually identified quasars. The dashed red line at $\mathrm{S/N} = 2$ marks an empirical threshold below which the number of successful identifications declines rapidly.  
\emph{Middle and Right:} Median S/N versus \JE and \HE magnitudes, respectively. Blue points show individual quasars, and the black lines indicate linear fits to $\log_{10}(\mathrm{S/N})$, with the fitted relations annotated. The dashed red line again marks $\mathrm{S/N} = 2$, and its intersection with the fit defines empirical limiting magnitudes of $\JE \approx 21.5$ and $\HE \approx 21.3$, shown by dashed green lines. These values represent practical limits for reliable spectral identification of quasars in Q1.}
\label{fig:snr_vs_mag}
\end{figure*}

To assess the effective depth of reliable spectral identification in our sample, we examined the relationship between the median \ac{snr} of the \ac{nisp} spectra within [12\,047, 18\,734] \AA~and the \JE and \HE magnitudes. As shown in \cref{fig:snr_vs_mag}, the number of visually confirmed quasars declines steeply below $\mathrm{S/N} = 2$, indicating an empirical limit where spectral identifications become increasingly difficult. By fitting a linear relation to $\log_{10}(\ac{snr})$ as a function of magnitude, we find that this empirical transition occurs at approximately $\JE = 21.5$ and $\HE = 21.3$. These values define the practical magnitude limits beyond which reliable redshift determination becomes rare in Q1 slitless spectroscopy for quasars.

We also compare the colour distributions of the \gaia and non-\gaia quasar subsets on the AllWISE and \Euclid colour-colour diagrams. To illustrate colour spaces occupied by stars, we show in the diagrams \Euclid point-like sources that are selected from \verb|catalogue.mer_catalogue| using 
\begin{verbatim}
    mumax_minus_mag<-2.6 AND spurious_flag=0
    AND flux_vis_psf>0 AND flux_y_templfit>0
    AND flux_h_templfit>0 AND flux_j_templfit>0.
\end{verbatim}
These point-like criteria are similar to the definition of \Euclid point-like sources in \citet{Q1-SP027}. 

\begin{figure}[htb]
\centering
\includegraphics[width=0.5\textwidth]{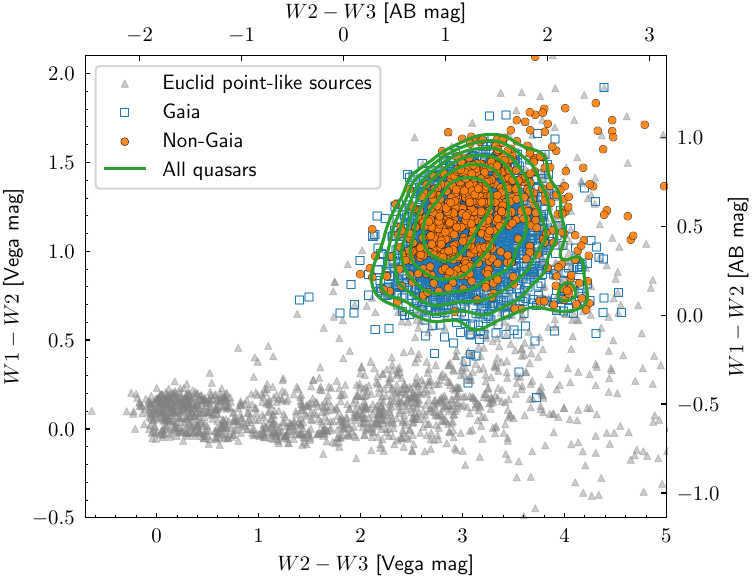}
\caption{$W1-W2$ versus $W2-W3$ colour-colour diagram of spectroscopically identified quasars in this work, where \gaia-detected sources are shown as blue open squares, and sources not in \gdr{3} are shown as orange circles. The density distribution of the full identified quasar sample is indicated with green contour lines. \Euclid point-like sources are shown as grey triangles. To ensure the reliability of the colours, only sources with adequate \ac{snr} ($\texttt{w1snr}>5$, $\texttt{w2snr}>5$, and $\texttt{w3snr}>3$) are shown.}
\label{fig:wise_ccd}
\end{figure}

\begin{figure*}[htb]
\centering
\includegraphics[width=1.0\hsize]{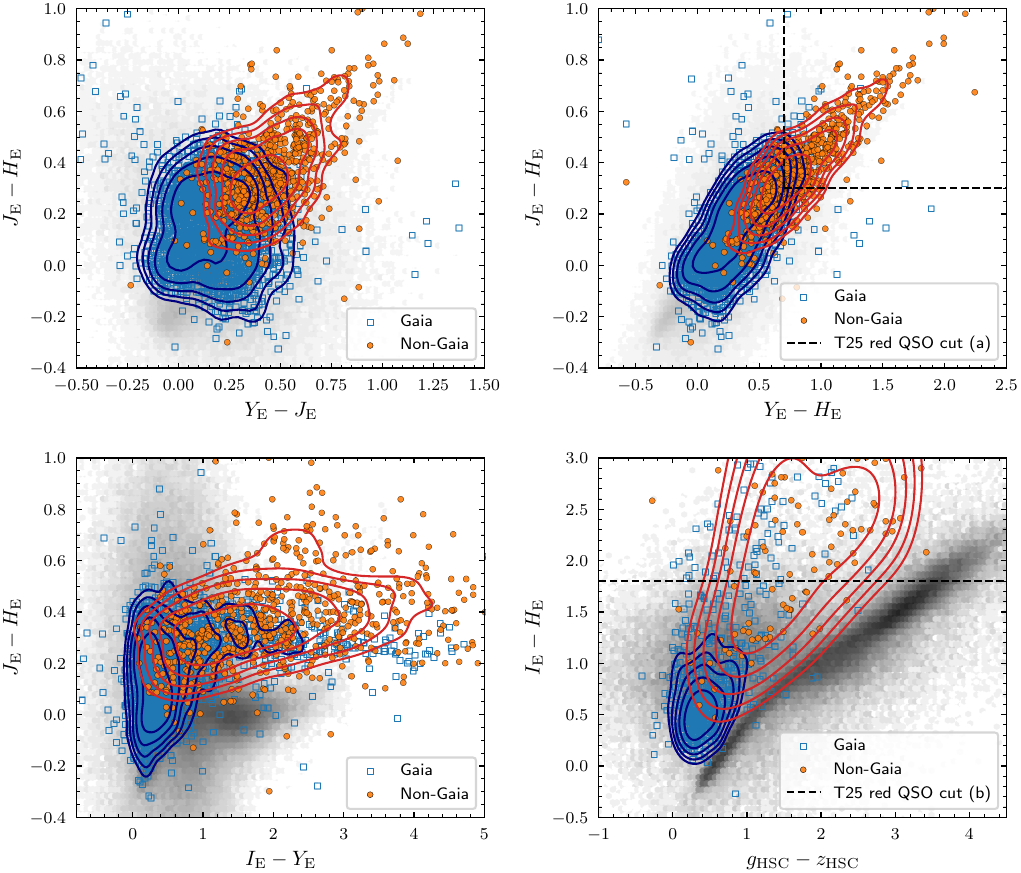}
\caption{\Euclid and external \citep[Hyper Suprime-Cam, HSC;][]{2018PASJ...70S...1M,2018PASJ...70S...4A} colour-colour diagrams of spectroscopically identified quasars in this work, where \gaia-detected sources are shown as blue open squares, and sources not in \gdr{3} are shown as orange circles. The density distribution of the \gaia-detected sample is indicated with blue contour lines, and that of the non-\gaia sample is indicated with red contour lines. The density of \Euclid point-like sources is shown as grey hexagonal binning plots in the background.}
\label{fig:euclid_ccds}
\end{figure*}

The \gaia and non-\gaia quasar subsets share nearly identical colour spaces on the $W1-W2$ versus $W2-W3$ diagram (\cref{fig:wise_ccd}). These have been described as \ac{agn} regions by many previous studies \citep[e.g.][]{2012ApJ...753...30S,2012AJ....144...49W,2012MNRAS.426.3271M,2018ApJS..234...23A}. On \Euclid colour planes (\cref{fig:euclid_ccds}), however, the \gaia and non-\gaia quasar samples become more separable. The non-\gaia subset occupies colour spaces redder than the \gaia subset, while partly overlapping. The optical-faint and infrared-bright selection of the non-\gaia subset makes it a representative sample of red quasars. As shown in \cref{fig:euclid_ccds}, the \Euclid colour cuts for red quasars ($\YE-\HE>0.7$, $\JE-\HE>0.3$, and $\IE-\HE>1.8$) from \citet[][hereafter \citetalias{Q1-SP023}]{Q1-SP023} select the redder half of non-\gaia quasars, which only have a small overlap with the \gaia quasars.

\subsection{A stacked emission-line map and the redshift challenge} \label{sec:emission-line-map}

To ensure robust composite building and follow-up spectral analysis, we selected a golden sample of 2868 visually identified quasars with the following constraints: (i) median \ac{snr} of the spectrum higher than 3 within [12\,047, 18\,734] \AA; and (ii) containing $\leq 15$ invalid pixels (NaN, or zero flux values). 

\begin{figure*}[htb]
\sidecaption
\includegraphics[width=12cm]{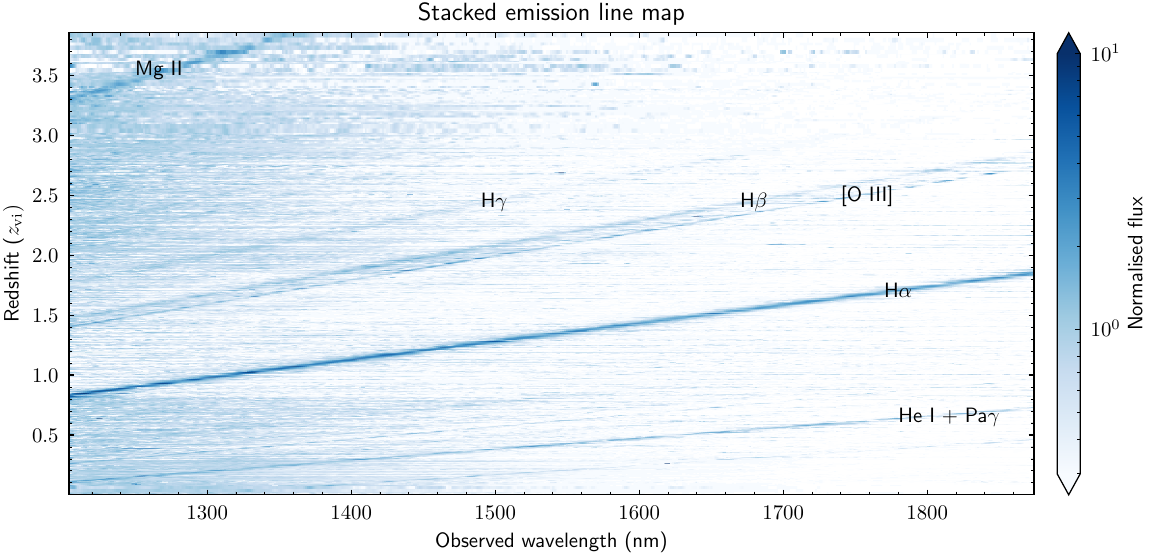}
\caption{Stacked emission-line map of the golden sample of 2868 visually identified quasars. Each row of the pixels represents a spectrum normalised using its 25th and 95th percentiles. Prominent emission lines are marked with text labels.}
\label{fig:stacked_em_map}
\end{figure*}

To better understand the systematic effects of the redshift determination using \RGE spectra, we constructed a stacked emission-line map of the golden sample on the redshift--wavelength plane (\cref{fig:stacked_em_map}) after normalising each spectrum with a percentile-based scaling,
\begin{equation}
    F_{\mathrm{norm}}(\lambda) = \frac{F_{\mathrm{obs}}(\lambda)-P_{25}(F_{\mathrm{obs}}(\lambda))}{P_{95}(F_{\mathrm{obs}}(\lambda))-P_{25}(F_{\mathrm{obs}}(\lambda))}\;,
\end{equation}
\noindent where $F_{\mathrm{obs}}(\lambda)$ is the original flux density array, and $P_{25}$ and $P_{95}$ denote the 25th and 95th percentiles of $F_{\mathrm{obs}}(\lambda)$, respectively. This normalisation reduces the continuum level and enhances the emission lines on the stacked plot. 

From \cref{fig:stacked_em_map}, the most prominent emission line among all spectra is H\,$\alpha$, which lies in the wavelength range of [12\,047, 18\,734] \AA~at redshift $0.83\lesssim z\lesssim 1.85$. Other prominent emission lines include \ion{He}{i}\,+\,Pa\,$\gamma$ at $0.11<z<0.73$, H\,$\beta$\,+\,[\ion{O}{iii}] at $1.5\lesssim z \lesssim 2.8$, and \ion{Mg}{ii} at $z>3.3$. While present in the emission-line map, H\,$\gamma$ becomes faint at high redshift, especially when H\,$\beta$\,+\,[\ion{O}{iii}] move out of the observed wavelength range. The low \ac{ew} and \ac{snr} of H\,$\gamma$ pose a challenge to the redshift determination at $2.8<z<3.3$, which can only be alleviated by combining with data at other wavelengths. In future data releases of \Euclid, the blue-grism data of \ac{nisp}, covering 926--1366 nm, will be available for Euclid Deep and Auxiliary fields \citep{EuclidSkyOverview}. Inclusion of the blue-grism data will increase the efficiency of source identification and redshift determination in these deep fields.

\subsection{Composite \Euclid spectra of the golden sample of bright quasars \label{subsec:spec_comps}} 

A mean or median composite spectrum of a sample of quasars is useful for understanding the average spectral properties across a wide wavelength range, and for providing a representative spectrum template \citep[e.g.][]{2001AJ....122..549V,Glikman_2006}. To generate the mean and median composite quasar spectra spanning the rest-frame optical to \ac{nir} wavelengths, we adopted a procedure similar to that described by \citet{2001AJ....122..549V}, with a \ac{1d} `drizzle' technique to improve the sampling of the low-resolution spectra. We briefly introduce the procedure below, and refer to \cref{app:composite_method} for technical details. We first corrected the Milky Way dust extinction of the spectra using $E(B-V)$ values from the Galactic dust map produced by \citet{2016A&A...596A.109P}, and the extinction law of \citet{2023ApJ...950...86G} assuming $R_{V}=3.1$. The packages \texttt{dustmaps} \citep{2018JOSS....3..695G} and \texttt{dust\_extinction} \citep{2024JOSS....9.7023G} are used for the correction. The spectra are then sorted in ascending order of redshift. The spectrum with the lowest redshift is normalised arbitrarily to have a unit mean flux density. Each subsequent spectrum is shifted to the rest-frame based on its measured redshift and normalised to match the mean flux density of the mean composite built from all lower-redshift spectra, within their overlapping wavelength range.

All spectra were resampled onto a common rest-frame wavelength grid with a constant bin size of $\Delta \lambda = 4$\,\AA, using a flux-conserving re-binning method to preserve the integrated flux density in each bin. This process is mathematically equivalent to a \ac{1d} version of the Drizzle algorithm \citep{2002PASP..114..144F}, which has been widely applied to image reconstructions by combining dithered, undersampled images in HST and JWST surveys \citep[e.g.][]{2011ApJS..197...36K,2023ApJS..268...64W,2023ApJ...946L..12B}. We set $\Delta\lambda=4$\,\AA\ to match the smallest native rest-frame pixel size near the blue end while maintaining higher per-pixel \ac{snr} at the highest redshifts. For reference, the observed sampling of 13.4\,\AA\ at $z=2.35$ corresponds to $13.4\,\AA/(1+2.35)=4$\,\AA\ in the rest frame, and the blue-end wavelength at this redshift is $12047\,\AA/(1+2.35)=3596$\,\AA. Consequently, the common 4-\AA\ grid oversamples the data at longer rest wavelengths and applies mild downsampling at the shortest rest wavelengths contributed by the highest-redshift quasars. By stacking the oversampled rest-frame spectra of different redshifts through drizzle, the composite recovers some information that is lost in the individual spectra at rest-frame $\lambda>3596\,\AA$, giving finer emission line features than individual ones.

The arithmetic mean composite (mean composite for short) is calculated as the equal-weight mean of the sigma-clipped normalised flux densities in each bin. The uncertainties of the mean composite are computed through error propagation, assuming uncorrelated input pixels. In parallel, the \ac{rms} flux that quantifies the object-to-object dispersion is recorded. A median and a geometric mean composite spectrum are also generated with the same sigma-clipped data as used for the mean composite. The mean, median, and geometric mean composite spectra, along with the uncertainties of the arithmetic and geometric mean composites, \ac{rms} flux, \ac{snr}, and number of spectra in each wavelength bin ($N_{\mathrm{spec}}$), are tabulated in \cref{tab:comp_spec_data}.

\begin{table}[htb]
\centering
\caption{Mean, geometric mean, and median \Euclid Q1 quasar composite spectra, along with \ac{rms}, \ac{snr} and the number of spectra in each wavelength bin ($N_{\mathrm{spec}}$).}
\label{tab:comp_spec_data}
\resizebox{\columnwidth}{!}{
\begin{tabular}{lcccccc}
\toprule
Wavelength & Mean $F_\lambda$ & Geometric mean $F_\lambda$ & Median $F_\lambda$ & RMS & S/N & $N_{\mathrm{spec}}$ \\
(nm) & (arb.) & (arb.) & (arb.) & (arb.) &  &  \\
\midrule
240.0 & $15.00 \pm 1.08$ & $14.71 \pm 1.13$ & 14.35 & 2.97 & 13.83 & 5 \\
240.4 & $13.11 \pm 1.03$ & $12.51 \pm 0.96$ & 13.44 & 3.70 & 12.74 & 5 \\
240.8 & $15.90 \pm 1.22$ & $15.65 \pm 1.18$ & 16.19 & 2.83 & 12.99 & 5 \\
241.2 & $14.82 \pm 1.16$ & $14.13 \pm 1.26$ & 14.39 & 4.59 & 12.73 & 5 \\
241.6 & $14.93 \pm 1.06$ & $14.87 \pm 1.05$ & 14.59 & 1.35 & 14.12 & 5 \\
... & ... & ... & ... & ... & ... & ... \\
1698.4 & $0.83 \pm 0.01$ & $0.83 \pm 0.02$ & 0.84 & 0.06 & 55.20 & 9 \\
1698.8 & $0.83 \pm 0.02$ & $0.82 \pm 0.02$ & 0.83 & 0.06 & 50.31 & 9 \\
1699.2 & $0.77 \pm 0.02$ & $0.72 \pm 0.02$ & 0.83 & 0.20 & 46.35 & 9 \\
1699.6 & $0.84 \pm 0.02$ & $0.84 \pm 0.02$ & 0.83 & 0.05 & 48.51 & 8 \\
1700.0 & $0.86 \pm 0.02$ & $0.86 \pm 0.02$ & 0.86 & 0.08 & 50.59 & 8 \\
\bottomrule
\end{tabular}
}
\tablefoot{\cref{tab:comp_spec_data} is available in its entirety at the CDS. A portion is shown here for guidance regarding its form and content.}
\end{table}

\begin{figure*}[htb]
\centering
\includegraphics[angle=0,width=1.0\hsize]{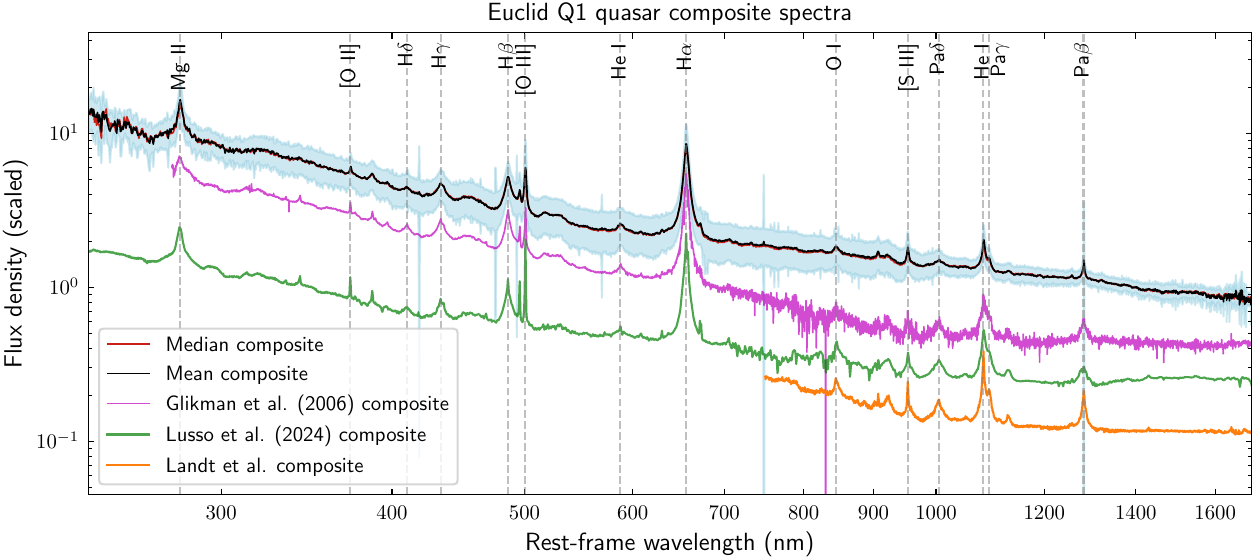}
\caption{Mean and median composite quasar spectra constructed from our golden sample (black and red curves, respectively), with the \ac{rms} scatter around the mean composite indicated by the shaded blue region. The quasar composite spectrum from \citet{Glikman_2006} is plotted in magenta, the type 1 \ac{agn} composite from \citet{EP-Lusso} is plotted in green, and the mean composite constructed using data from \citet{2008ApJS..174..282L,2011MNRAS.414..218L,2013MNRAS.432..113L} is plotted in orange. Prominent emission lines are marked for reference.}
\label{fig:spec_comps}
\end{figure*}

\Cref{fig:spec_comps} presents the mean and median composite spectra derived from our sample, together with the \ac{rms} scatter around the mean composite. The mean quasar composite spectrum from \citet{Glikman_2006}, the mean composite of type 1 \acp{agn} from \citet{EP-Lusso}, and a mean composite spectrum constructed by \citet{EP-Lusso} using datasets from \citet{2008ApJS..174..282L,2011MNRAS.414..218L,2013MNRAS.432..113L}, are also shown for comparison. Prominent emission lines are marked in the composite spectra for reference, including \ion{Mg}{ii} $\lambda2800$, [\ion{O}{ii}] $\lambda3728$, \ion{H}{$\gamma$}, \ion{H}{$\beta$}, [\ion{O}{iii}] $\lambda\lambda4960,5008$, \ion{H}{$\alpha$}, \ion{He}{i}, \ion{Pa}{$\gamma$}, and \ion{Pa}{$\beta$}. The mean and median composite spectra are consistent with each other.

\begin{figure}[htb]
\centering
\includegraphics[width=\hsize]{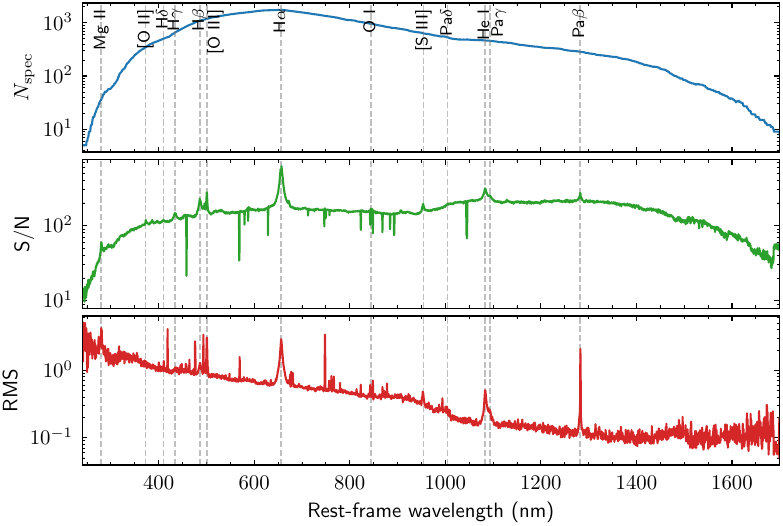}
\caption{
Diagnostics of the \Euclid Q1 mean quasar composite as a function of rest-frame wavelength for a bin size of $\Delta\lambda = 4$\,\AA. 
\emph{Top}: number of spectra contributing to each wavelength bin. 
\emph{Middle}: \ac{snr} of the mean composite per bin. 
\emph{Bottom}: \ac{rms} dispersion of the contributing spectra about the mean. 
Vertical dashed lines mark the rest wavelengths of prominent emission lines, which are labelled in the top panel.
}
\label{fig:nspec_snr}
\end{figure}

The number of contributing spectra ($N_{\mathrm{spec}}$), the \ac{snr}, and the \ac{rms} flux of the mean composite as functions of rest-frame wavelength are shown in \cref{fig:nspec_snr}. Between 100 and more than 1000 spectra contribute to each wavelength bin over the range $0.32$--$1.48\,\micron$, which yields \ac{snr} values above 100 in the majority of bins. The \ac{rms} about the mean is typically between 0.1 and 1 in the continuum (in units of the mean flux), and usually increases by less than an order of magnitude at the positions of emission lines, except around \ion{Pa}{$\beta$} where the \ac{rms} rises by about an order of magnitude. This behaviour indicates a moderate level of intrinsic quasar diversity that dominates the variance near strong lines.

The \citet{Glikman_2006} composite (up to $3.52\,\micron$) is constructed using data from 27 bright quasars observed with the NASA \ac{irtf}, and the \citet{EP-Lusso} type 1 composite (up to $3.61\,\micron$) is built with 23 quasars from \citet{Glikman_2006} and nine additional hard-X-ray-selected \acp{agn} observed with the folded-port infrared echellette \citep[FIRE;][]{2008SPIE.7014E..0US} from \citet{2022ApJS..261....8R}. The \citet{2008ApJS..174..282L,2011MNRAS.414..218L,2013MNRAS.432..113L} composite covering $0.75$--$2.3\,\micron$ is based on 29 well-known local type 1 \ac{agn} observed at the NASA \ac{irtf} and the Gemini North observatory. At $\lambda<0.7\,\micron$, the mean composite spectrum from this work (hereafter the \Euclid composite) shows a consistent continuum slope to the composites from \citet{Glikman_2006} and \citet{EP-Lusso}. At longer wavelengths, both \citet{Glikman_2006} and \citet{EP-Lusso} composites display residual telluric absorption features, while the \Euclid composite shows a clean continuum, free of telluric absorption. The \citet{2008ApJS..174..282L,2011MNRAS.414..218L,2013MNRAS.432..113L} composite shows a smoother continuum than the other two ground-based composite spectra, and a steeper slope at $0.75<\lambda<0.98\,\micron$. 

\begin{figure*}[htb]
\centering
\includegraphics[angle=0,width=1\hsize]{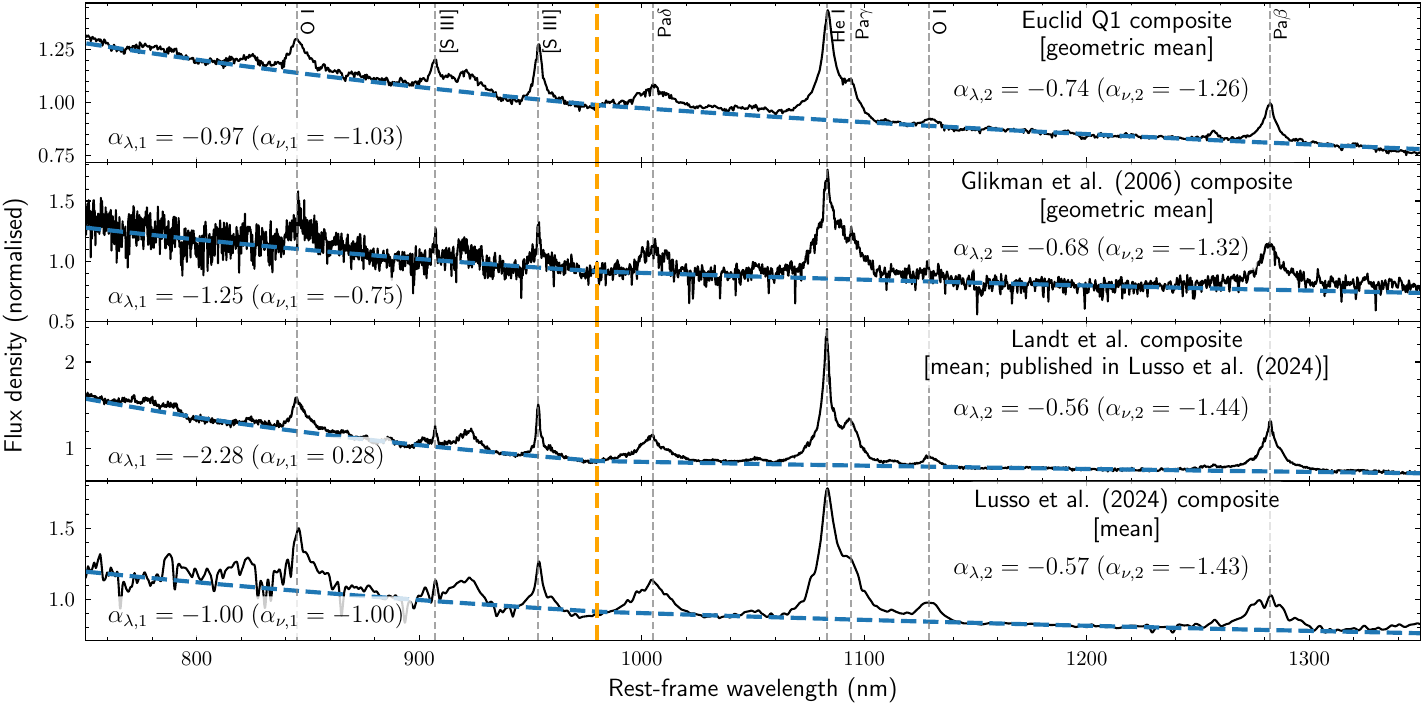}
\caption{
\Euclid Q1 geometric mean composite and three literature quasar composites, with \ac{nir} continua fitted with broken power laws over $0.75$--$1.35~\mu\mathrm{m}$. From top to bottom, the panels show the \Euclid Q1 composite (geometric mean), the geometric mean composite from \citet{Glikman_2006}, the mean Landt et al.\ composite as published by \citet{EP-Lusso}, and the type~1 \ac{agn} composite from \citet{EP-Lusso}. The black curves show the composite spectra, the dashed blue lines show the best-fitting broken power laws in $F_\lambda$, and the vertical dashed orange lines indicate the break wavelength of 980 nm. The annotated indices $\alpha_{\lambda,1}$ ($\alpha_{\nu,1}$) and $\alpha_{\lambda,2}$ ($\alpha_{\nu,2}$) are the corresponding spectral slopes in $F_\lambda$ ($F_\nu$) blueward and redward of the break. Selected emission lines are labelled in the top panel for reference.
}
\label{fig:nir_pl_fit}
\end{figure*}

To quantitatively compare the \ac{nir} continuum shape with previous work, we fitted a broken power law in $F_\lambda$ to the \Euclid Q1 geometric mean composite and to three published quasar composites over the wavelength range $0.75$--$1.35\,\micron$ (\cref{fig:nir_pl_fit}), using the spectral fitting package \texttt{QSOFITMORE} \citep{fu_2025_15571037}. The break was fixed at $0.98\,\micron$ based on visual inspection. The upper limit of $1.35\,\micron$ was chosen to exclude the strong long-wavelength upturn of the \citet{EP-Lusso} composite and to reduce the impact of small-number statistics in the Q1 composite. We parameterised the power-law model as $F_\lambda \propto \lambda^{\alpha_{\lambda}}$ and describe the broken power law with indices $\alpha_{\lambda,1}$ and $\alpha_{\lambda,2}$ blueward and redward of the break, respectively. 

For the \Euclid Q1 composite, we obtain $\alpha_{\lambda,1} = -0.97$ and $\alpha_{\lambda,2} = -0.73$. The type~1 composite from \citet{EP-Lusso} has similar slopes, $\alpha_{\lambda,1} = -1.00$ and $\alpha_{\lambda,2} = -0.57$, while the \citet{Glikman_2006} geometric mean composite yields $\alpha_{\lambda,1} = -1.25$ and $\alpha_{\lambda,2} = -0.68$. The mean Landt et al.\ composite shows a rather steep blue segment with $\alpha_{\lambda,1} = -2.28$ over $0.75$--$0.98\,\micron$, then turns over to a much flatter slope, $\alpha_{\lambda,2} = -0.56$, over $0.98$--$1.35\,\micron$. In the $F_\lambda$ representation, all four composites show a mild flattening of the \ac{nir} continuum toward longer wavelengths. For comparison with the literature, the corresponding $\alpha_\nu$ values (computed via $\alpha_\nu = -[\alpha_\lambda + 2]$) are listed in parentheses in \cref{fig:nir_pl_fit}. 
Our \Euclid composite has the flattest continuum in the \ac{nir} (with the smallest change of slopes at the break), while the Landt et al.\ mean composite exhibits the largest curvature (change of slopes). As shown by \citet{2013MNRAS.432..113L}, the \ac{nir} \ac{sed} of \acp{agn} affected by strong host galaxy light are much flatter than those of \acp{agn} with low host contribution. The flatness of the \ac{nir} continuum of our \Euclid composite is therefore most likely due to the host galaxy light contribution of the low-redshift \acp{agn}. 

The observed central wavelengths of selected emission lines (\cref{tab:line_ids}) were measured from the mean composite with spectral fitting using \texttt{QSOFITMORE} \citep{fu_2025_15571037}. For broad emission lines, including \ion{Mg}{ii}, \ion{H}{$\beta$}, \ion{H}{$\alpha$}, \ion{O}{i}, \ion{He}{i}, \ion{Pa}{$\delta$}, \ion{Pa}{$\gamma$}, and \ion{Pa}{$\beta$}, we fitted a narrow Gaussian component ($\mathrm{FWHM}\leq 1200\, \kms$) and 2--3 broad Gaussian components ($\mathrm{FWHM}> 1200\, \kms$) to each line profile, and report the central wavelengths of the narrow components. For narrow (forbidden) lines, including [\ion{O}{ii}], [\ion{Ne}{iii}], [\ion{O}{iii}], and [\ion{S}{iii}], we fitted only one narrow component to each profile. The uncertainties are the standard deviations of a Monte Carlo simulation of 50 fits. As shown in \cref{tab:line_ids}, the line centres of narrow lines typically show lower uncertainties than those of the broad lines ([\ion{O}{iii}]$\lambda5008$ has the lowest wavelength uncertainty), indicating the significance of narrow lines in redshift determination. 

\begin{table}[htb]
\centering
\caption{Major emission lines identified from the mean composite quasar spectrum.}
\label{tab:line_ids}
\resizebox{\columnwidth}{!}{
\begin{tabular}{@{}lcccrr@{}}
\toprule
Line ID & $\lambda_{\mathrm{vac}}$ & $\lambda_{\mathrm{obs}}$ & Intensity & EW & FWHM \\
 & (\AA) & (\AA) & [100$F/F({\rm H}\alpha)$] & (\AA) & ($\mathrm{km~s^{-1}}$) \\ \midrule
{\ion{Mg}{ii}} & 2798.75 & $2799.37\pm0.8$ & $16.0 \pm 8.7$ & 7.5 & 6444.9 \\
{[\ion{O}{ii}]} & 3728.48 & $3731.54\pm0.45$ & $2.2 \pm 0.2$ & 1.9 & 1200.0 \\
{[\ion{Ne}{iii}]} & 3868.58 & $3868.73\pm0.85$ & $2.4 \pm 0.3$ & 2.3 & 1200.0 \\
{H$\gamma$} & 4341.68 & $4338.57\pm1.33$ & $10.3 \pm 1.0$ & 12.5 & 4195.5 \\
{H$\beta$} & 4862.68 & $4863.73\pm0.86$ & $38.6 \pm 3.9$ & 61.3 & 8241.9 \\
{[\ion{O}{iii}]} & 4960.3 & $4961.25\pm0.3$ & $4.5 \pm 0.6$ & 7.2 & 1200.0 \\
{[\ion{O}{iii}]} & 5008.24 & $5007.79\pm0.07$ & $13.6 \pm 1.4$ & 21.7 & 1200.0 \\
{\ion{He}{i}} & 5877.29 & $5873.04\pm5.88$ & $0.1 \pm 0.1$ & 0.2 & 316.2 \\
{H$\alpha$} & 6564.61 & $6564.63\pm1.34$ & $100.0$ & 219.6 & 5871.2 \\
{\ion{O}{i}} & 8448.8 & $8445.11\pm3.37$ & $1.0 \pm 1.0$ & 2.9 & 3209.3 \\
{[\ion{S}{iii}]} & 9071.1 & $9068.76\pm0.66$ & $1.2 \pm 0.1$ & 3.7 & 1200.0 \\
{[\ion{S}{iii}]} & 9533.2 & $9535.53\pm0.35$ & $3.0 \pm 0.3$ & 9.7 & 1185.9 \\
{Pa$\delta$} & 10052.1 & $10055.74\pm1.55$ & $4.4 \pm 0.5$ & 15.1 & 5378.1 \\
{\ion{He}{i}} & 10833.2 & $10836.36\pm0.21$ & $3.0 \pm 0.3$ & 10.9 & 1016.5 \\
{Pa$\gamma$} & 10941.1 & $10941.91\pm0.75$ & $0.6 \pm 0.2$ & 2.3 & 972.8 \\
{\ion{O}{i}} & 11290.0 & $11297.54\pm2.38$ & $0.4 \pm 0.1$ & 1.7 & 2825.8 \\
{Pa$\beta$} & 12821.6 & $12825.47\pm0.23$ & $5.2 \pm 0.5$ & 22.3 & 2825.8 \\
\bottomrule
\end{tabular}
}
\tablefoot{For each line, its vacuum wavelength ($\lambda_{\mathrm{vac}}$) is listed, along with observed wavelength ($\lambda_{\mathrm{obs}}$) with 1$\sigma$ uncertainty, relative intensity to H\,$\alpha$, \ac{ew}, and \ac{fwhm}.}
\end{table}

\section{Discussion\label{sec:discussion}}

\subsection{Morphological properties of the bright quasar sample \label{sec:morphology}}

{With a pixel scale of \ang{;;0.1} and stable \ac{psf} of VIS imaging, \Euclid resolves obvious galaxy structure from the local Universe to at least $z=1.5$ \citep[with particularly rich statistics at $0.3<z<0.7$; see e.g.][]{Q1-SP047}. These data enable the detailed structural characterisation of quasar hosts through uniform measurements of Sérsic-based and model-independent parameters \citep{Q1-SP040,Q1-TP004}, as well as the separation of compact nuclear light from extended galaxy emission \citep{Q1-SP015}.

To take advantage of the detailed structural information provided by the VIS images, we examined two redshift regimes separately: (i) a low-redshift ($z<0.5$) subset with 341 sources, and (ii) an intermediate-redshift ($0.5<z<2$) subset with 2361 sources. Both subsets are selected to have valid morphological parameters measured from VIS imaging, including parameters from the single-component Sérsic fit \citep{Q1-SP040}: Sérsic index ($n_{\sfont{VIS}}$), effective radius ($R_{\mathrm{e,}\sfont{VIS}}$), and axis ratio ($\rho_{\sfont{VIS}}$); and the model-independent CAS parameters \citep{2003ApJS..147....1C,Q1-TP004}: concentration, asymmetry, and clumpiness (smoothness). The intermediate-redshift subset is further supplemented with a VIS \ac{psf} fraction ($f_{\sfont{PSF}}$), derived by the deep-learning-based $f_{\sfont{PSF}}$ prediction model trained on simulated galaxy images with different levels of $f_{\sfont{PSF}}$ added to them \citep{Q1-SP015}. 
We choose the redshift range $0.5<z<2$ because the model was trained and validated in this range.

As shown in \cref{fig:morph_vis_corner_lowz} and \Cref{tab:vis_morph_stats}, the low-redshift population shows a large source extent with median $\mumaxmag\ =-1.67$, significantly higher than the median $\mumaxmag\ =-2.87$ of the intermediate-redshift sample. The \mumaxmag value anticorrelates with the Sérsic index and concentration. In particular, the most compact sources in this population ($\mumaxmag \approx -2.5$) have a Sérsic index around 5.5, which is the upper limit set by \citet{Q1-SP040}. They have suggested that fits with Sérsic indices above 5.45 should be removed from any Sérsic-based analysis. In this low-redshift population, 50\,\% of the sources have $n_{\sfont{VIS}}$ higher than 5.45. Such near-boundary Sérsic indices indicate that the single-component Sérsic model cannot describe the light profiles of \acp{agn} with bright cores. When requiring $n_{\sfont{VIS}}<5.45$, the median $n_{\sfont{VIS}}$ is 2.31, which is still significantly higher than the peak value of 0.8 among all galaxies in \citet{Q1-SP040}. The median $R_{\mathrm{e,}\sfont{VIS}}$ of $n_{\sfont{VIS}}<5.45$ sources is \ang{;;1.0}, identical to the peak value of $R_{\mathrm{e,}\sfont{VIS}}$ in \citet{Q1-SP040}.

The CAS parameters further characterise the resolved structure of the low-redshift quasars. 
The median concentration ($C=4.27$) is high compared to the general galaxy population \citep[e.g. $C\approx2.5$, see][]{Q1-SP040}
but with significant scatter, reflecting the coexistence of bright nuclear light and extended hosts. 
Asymmetry values are moderate (median $A=0.36$) with a tail to $A>1$, 
indicating disturbed morphologies or nearby companions. 
The clumpiness parameter is elevated (median $S = 0.19$), suggesting clumpy emission in many cases, possibly due to ongoing star formation \citep[e.g.][]{2003ApJS..147....1C, 2014ARA&A..52..291C}.
Overall, at $z < 0.5$, the host galaxies of bright quasars are frequently resolved and often display interacting features and star-forming clumps.

\begin{table}[t]
\centering
\caption{Median VIS morphology parameters and $f_{\sfont{PSF}}$.}
\label{tab:vis_morph_stats}
\small
\begin{tabular}{lcccc}
\hline\hline
 & \multicolumn{2}{c}{$z<0.5$} & \multicolumn{2}{c}{$0.5<z<2$} \\
\cline{2-3} \cline{4-5}
Parameter & all & $n_{\sfont{VIS}}<5.45$ & all & $n_{\sfont{VIS}}<5.45$ \\
\hline
$N$                & 341 & 171 & 2361 & 230 \\
$n_{\sfont{VIS}}$             & 5.35 & 2.31 & 5.50 & 2.25 \\
$\rho_{\sfont{VIS}}$          & 0.76 & 0.71 & 0.69 & 0.67 \\
$R_{\mathrm{e,}\sfont{VIS}}$ & 0.78 & 1.00 & 0.01 & 0.42 \\
$C$                           & 4.27 & 3.96 & 2.56 & 2.93 \\
$A$                           & 0.36 & 0.33 & 0.55 & 0.50 \\
$S$                           & 0.19 & 0.19 & 0.08 & 0.19 \\
\mumaxmag             & $-1.67$ & $-0.94$ & $-2.87$ & $-1.68$ \\
$f_{\sfont{PSF}}$             & --- & --- & 0.76 & 0.16 \\
\hline
\end{tabular}
\tablefoot{$f_{\sfont{PSF}}$ is available only for $0.5<z<2$.}
\end{table}

\begin{figure}[h]
\centering
\includegraphics[angle=0,width=1.0\hsize]{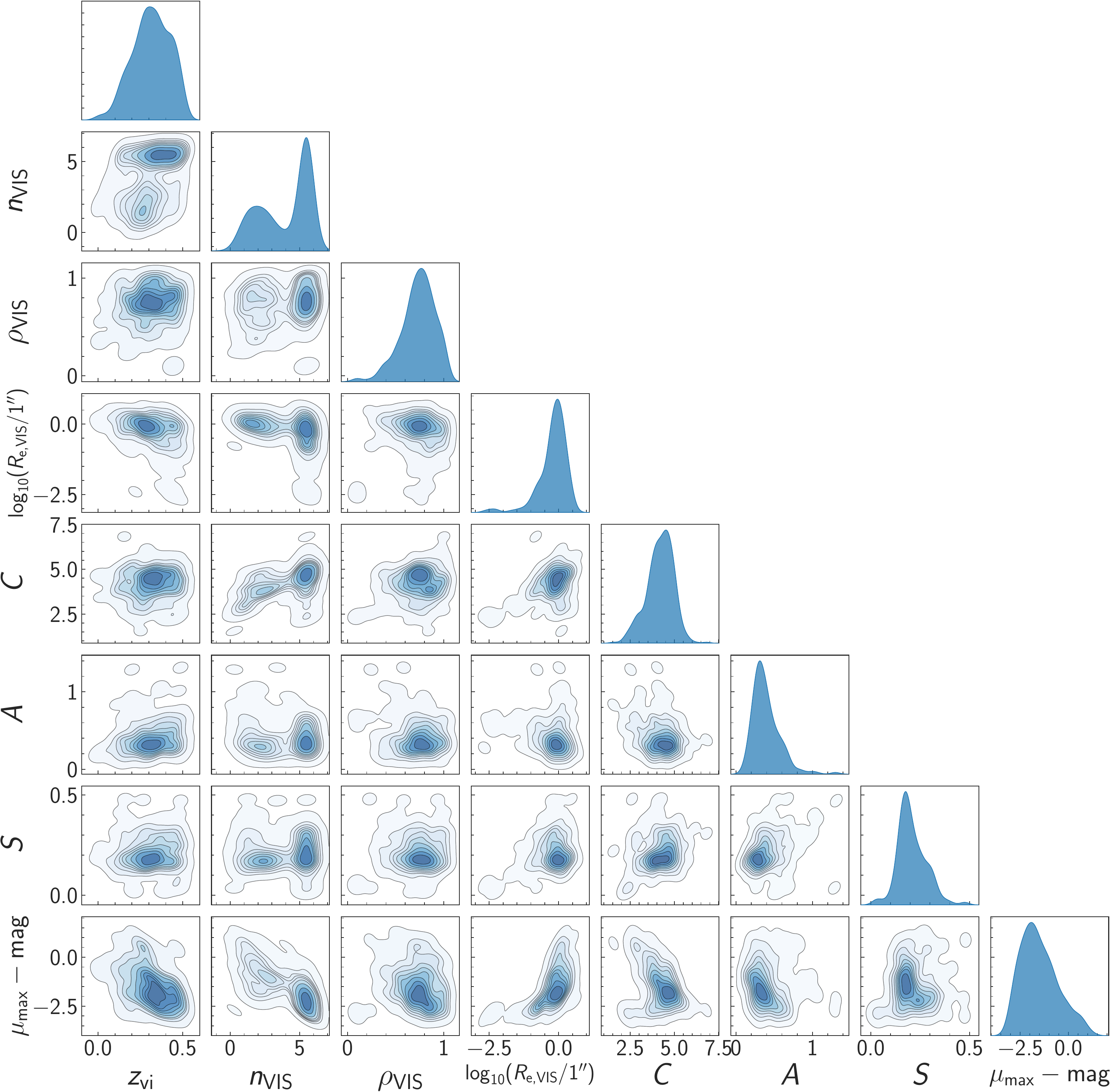}
\caption{
Corner plot showing the joint distributions of morphological parameters measured from \Euclid VIS imaging for 341 sources at $z<0.5$. Displayed parameters include visual-inspection redshift ($z_{\mathrm{vi}}$), Sérsic index ($n_{\sfont{VIS}}$), axis ratio ($\rho_{\sfont{VIS}}$), logarithmic effective radius ($\log_{10}(R_{\mathrm{e,}\sfont{VIS}}/1^{\prime\prime})$), concentration ($C$), asymmetry ($A$), clumpiness ($S$), and \mumaxmag.
}
\label{fig:morph_vis_corner_lowz}
\end{figure}

\begin{figure}[h]
\centering
\includegraphics[angle=0,width=1.0\hsize]{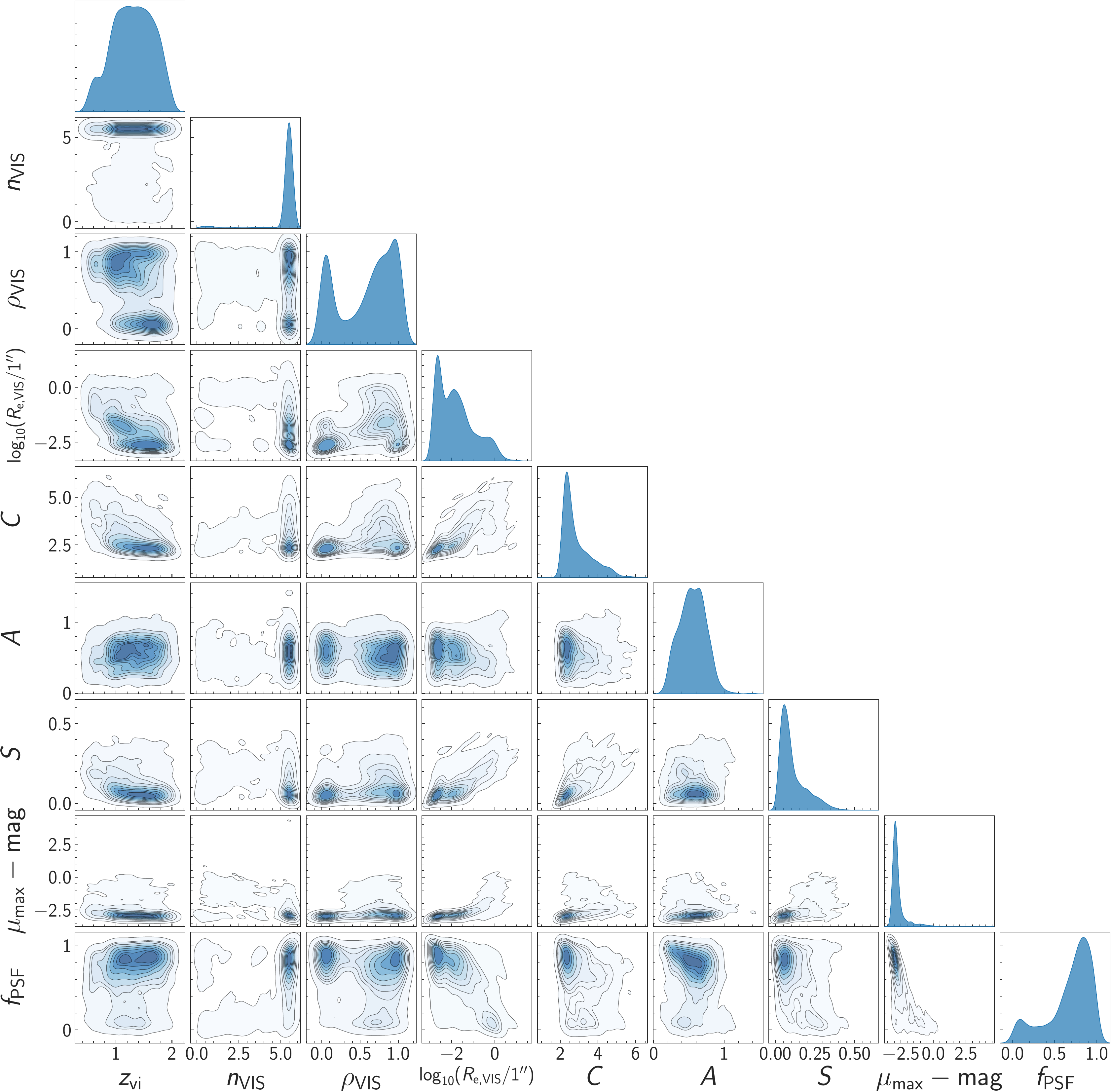}
\caption{
Same as \cref{fig:morph_vis_corner_lowz}, but for 2361 sources at $0.5<z<2$ with measurements of \ac{agn} \ac{psf} contribution fraction ($f_{\sfont{PSF}}$).
}
\label{fig:morph_vis_corner_fpsf}
\end{figure}

Compared to the low-redshift subset, the intermediate-redshift quasars at $0.5<z<2$ are markedly more compact (see \cref{fig:morph_vis_corner_fpsf} and \Cref{tab:vis_morph_stats}). More than 90\,\% of the sources (2031) have near-boundary Sérsic indices of $n_{\sfont{VIS}}>5.45$, which results in a median $n_{\sfont{VIS}}=5.5$ and a median $R_{\mathrm{e,}\sfont{VIS}}=\ang{;;0.01}$ ($1/10$ of the VIS pixel size). Such a high fraction of saturated Sérsic fits indicates that the single-component Sérsic model cannot capture the unresolved nuclear dominance of these sources. When restricting to $n_{\sfont{VIS}}<5.45$, the median Sérsic index decreases to 2.25 and the median effective radius increases to \ang{;;0.42}, values more consistent with resolved galaxy light profiles.

The VIS PSF fraction ($f_{\sfont{PSF}}$) provides a more direct handle on this unresolved light. Its distribution peaks around 0.8 with another weaker peak around 0.1 (\cref{fig:morph_vis_corner_fpsf}), confirming that most sources are dominated by their cores, consistent with their Sérsic fits converging to extreme values. Sources with high $f_{\sfont{PSF}}$ values also correspond to the most point-like objects, characterised by $\mumaxmag \approx -2.9$. However, low $f_{\sfont{PSF}}$ values (e.g. $f_{\sfont{PSF}}<0.1$) do not necessarily imply the absence of \ac{agn} activity, because uncertainties of the measurements could dominate at such low levels.

As shown in \cref{fig:morph_vis_corner_fpsf} and \cref{tab:vis_morph_stats}, the CAS parameters of the intermediate-redshift quasars show very different distributions compared to those of the low-redshift quasars. Highly left-skewed distributions of concentration (median $C=2.56$) and clumpiness (median $S=0.08$) are seen in intermediate-redshift quasars, in contrast to the nearly symmetric distributions of the two parameters for the low-redshift quasars (median $C=4.27$ and median $S=0.19$). The concentration and clumpiness distributions of the intermediate-redshift quasars appear unphysical, because quasar hosts are expected to be more concentrated and clumpier towards higher redshift.This discrepancy likely arises because the nuclear \ac{psf} components were not removed before the measurements. Indeed, sources with low nuclear fractions ($f_{\mathrm{PSF}} \lesssim 0.3$) show more realistic Sérsic indices and CAS parameters, reinforcing that uncorrected nuclear light biases the structural measurements of the host galaxies.

Taken together, these results demonstrate that VIS morphology captures a strong redshift dependence in the apparent structure of bright quasars: (i) at $z<0.5$, obvious host galaxy structures are frequently resolved, and a single Sérsic model can describe the light profiles of half of the sources; (ii) at $0.5<z<2$, the nuclear component dominates, driving Sérsic indices of 90\,\% of the sources to their upper limits. In this regime ($0.5<z<2$), $f_{\sfont{PSF}}$ offers a more reliable compactness measure and light profile characterisation than the single Sérsic fits, allowing us to quantify the balance between nuclear and host emission across the bright quasar sample. Representative \Euclid \ac{1d} spectra and VIS cutouts illustrating these trends are shown in \cref{app:euclid_cutouts} for low-redshift sources ($z<0.5$; \cref{fig:cutouts-lowz}) and for $0.5<z<2$ quasars spanning different $f_{\sfont{PSF}}$ levels (\cref{fig:cutouts-fpsf}). A more detailed analysis of \ac{agn} host galaxy morphology, including re-measurement of morphological parameters with methods tailored to \ac{agn} host galaxies, and dedicated tests to quantify measurement sensitivity as a function of host galaxy surface brightness, redshift, and nucleus-to-host contrast, will be presented in a follow-up study.}

\subsection{\Euclid colour space, multi-wavelength selection, and future improvements}

The advent of \textit{Euclid}'s high-quality, wide-area near-infrared photometry opens a novel colour space for quasar selection, providing new opportunities to discover dust-reddened or intrinsically red quasars that are missed by traditional optical or \ac{uv} selection. The extended wavelength coverage of the \Euclid NIR bands, combined with precise VIS photometry, enables more robust discrimination of quasars from stars and galaxies, particularly at redshifts and extinction levels where classical techniques are less effective. Our initial exploration demonstrates that red quasars occupy distinct loci in \Euclid colour-colour diagrams, highlighting the potential of the survey to identify populations that are heavily underrepresented in previous optical surveys.

Mid-infrared photometry from WISE has already proven highly effective for selecting luminous \acp{agn} and quasars via their characteristic red MIR colours. Our analysis confirms that WISE-based colour cuts efficiently select \ac{agn}-dominated sources, as reflected in the properties of our pre-selected sample. However, WISE alone cannot distinguish between the most dust-obscured and less obscured quasars. The synergy between \Euclid and WISE thus offers a powerful selection strategy: by combining WISE MIR colours with deep VIS and NIR photometry from \Euclid, future quasar searches can achieve high completeness, efficiently probing obscured \ac{agn} populations.

Moreover, incorporating multi-band photometric data into the spectroscopic identification process will substantially improve the completeness and reliability of the quasar sample. Photometric pre-selection not only increases the likelihood of identifying genuine quasars, especially those with atypical \ac{sed}s, but also enhances redshift determination by providing prior constraints and mitigating degeneracies in the spectral fitting. The inclusion of \Euclid and external photometry will be essential for robust identification and breaking redshift degeneracies from only one emission line.

A limitation of the present work is that the initial sample relies on pre-selection using \gaia and WISE catalogues, which may miss quasars outside the colour selection windows or at the limits of the \gaia and WISE sensitivity. In future \Euclid data releases, the construction of a more complete quasar sample will be enabled by combining traditional pre-selected candidates with new candidates identified from \Euclid photometry and by the SPE pipeline. This composite approach, leveraging the full capability of the \Euclid data, will allow for the recovery of quasars with a wider range of colours, luminosities, and host properties.

In addition, we note that measuring the intrinsic spectral properties of quasars requires careful treatment of the host galaxy contribution. As demonstrated in \cref{sec:morphology}, even in \ac{agn}-dominated sources, the host can provide a significant fraction of the observed flux, especially in the VIS band. Correctly separating nuclear and host emission is challenging and may introduce additional uncertainties in line and continuum measurements. Future analyses will benefit from improved morphological decomposition and \ac{sed} modelling to robustly isolate the quasar component and fully exploit \textit{Euclid}'s spectrophotometric capabilities. 

\section{Conclusions\label{sec:conclusions}}

In this work, we have presented a large sample of 3468 bright quasars covering the redshift range of $0< z \lesssim4.8$ identified with red-grism spectroscopy of the first \Euclid Quick Data Release, including 2686 sources with new spectroscopic identifications relative to existing public compilations. To ensure identification efficiency, we focused on a high-purity input quasar candidate sample of 9214 sources based on GDR3-QSOs (Quaia, CatNorth, and CatSouth) and the AllWISE R90 \ac{agn} table. Through a template matching process based on the Pearson correlation function and a visual inspection campaign, we labelled quasars and determine their spectroscopic redshifts. The success rate of the spectroscopic identification is 38\,\%. 

From the identified sample, we estimated an empirical spectroscopic depth of $\JE \lesssim 21.5$ and $\HE \lesssim 21.3$ at the sensitivity of the Wide Field Survey, beyond which the number of securely identified quasars declines sharply. Our investigation of the novel \Euclid colour space demonstrates its power in uncovering redder, dust-obscured quasars that may be missed by traditional optical and MIR selections. The synergy between \Euclid and WISE photometry promises even greater completeness and diversity in future quasar surveys. 

We constructed the first \Euclid composite spectrum of bright quasars, covering rest-frame \ac{nuv} to \ac{nir} wavelengths and free from telluric absorption, providing a valuable benchmark for future spectral studies. The spectroscopic bright quasar catalogue of this work, and the composite quasar spectrum, will be available at CDS. In addition, the spectral properties of the quasars will be derived by multi-component fitting (Euclid Collaboration: Calhau et al. in preparation).  

We characterised VIS morphologies using single-Sérsic and model-independent (CAS) parameters, supplemented by a deep-learning PSF fraction, $f_{\sfont{PSF}}$. At low redshift ($z<0.5$), obvious host structures are frequently resolved, and a single Sérsic model describes the light profiles of about half of the sources. At intermediate redshift ($0.5<z<2$), the nuclear component dominates, driving Sérsic indices of roughly 90\,\% of the sources to the upper bound; in this regime $f_{\sfont{PSF}}$ provides a more reliable compactness measure and is used to quantify the balance between nuclear and host emission across the sample.

Finally, we discuss current limitations, including the pre-selection bias and the challenges of host-nucleus decomposition, and outline pathways for improvement with upcoming \Euclid data releases and advanced selection methods. The results presented here demonstrate the capability of \Euclid slitless spectroscopy in building large quasar samples and provide a basis for extending this approach to the wide-area \ac{agn} censuses enabled by forthcoming \Euclid data releases.

\paragraph{Data availability.} The full Tables \ref{tab:comp_spec_data} and \ref{tab:main-metadata} are only available in electronic form at the CDS via anonymous ftp to \url{cdsarc.u-strasbg.fr} (\url{130.79.128.5}) or via \url{http://cdsweb.u-strasbg.fr/cgi-bin/qcat?J/A+A/}.

\begin{acknowledgements}
We thank the two anonymous referees for their constructive reports and helpful suggestions, which improved the clarity and quality of this paper.
\AckEC  
\AckQone
\AckDatalabs
This work has made use of data from the European Space Agency (ESA) mission
{\it Gaia} (\url{https://www.cosmos.esa.int/gaia}), processed by the {\it Gaia}
Data Processing and Analysis Consortium (DPAC,
\url{https://www.cosmos.esa.int/web/gaia/dpac/consortium}). Funding for the DPAC
has been provided by national institutions, in particular the institutions
participating in the {\it Gaia} Multilateral Agreement. 
This publication makes use of data products from the Wide-field Infrared Survey Explorer, which is a joint project of the University of California, Los Angeles, and the Jet Propulsion Laboratory/California Institute of Technology, and NEOWISE, which is a project of the Jet Propulsion Laboratory/California Institute of Technology. WISE and NEOWISE are funded by the National Aeronautics and Space Administration. This publication makes use of data from the Hyper Suprime-Cam (HSC). 

This research used data obtained with the Dark Energy Spectroscopic Instrument (DESI). DESI construction and operations is managed by the Lawrence Berkeley National Laboratory. This material is based upon work supported by the U.S. Department of Energy, Office of Science, Office of High-Energy Physics, under Contract No. DE–AC02–05CH11231, and by the National Energy Research Scientific Computing Center, a DOE Office of Science User Facility under the same contract. Additional support for DESI was provided by the U.S. National Science Foundation (NSF), Division of Astronomical Sciences under Contract No. AST-0950945 to the NSF’s National Optical-Infrared Astronomy Research Laboratory; the Science and Technology Facilities Council of the United Kingdom; the Gordon and Betty Moore Foundation; the Heising-Simons Foundation; the French Alternative Energies and Atomic Energy Commission (CEA); the National Council of Humanities, Science and Technology of Mexico (CONAHCYT); the Ministry of Science and Innovation of Spain (MICINN), and by the DESI Member Institutions: www.desi.lbl.gov/collaborating-institutions. The DESI collaboration is honored to be permitted to conduct scientific research on I’oligam Du’ag (Kitt Peak), a mountain with particular significance to the Tohono O’odham Nation. Any opinions, findings, and conclusions or recommendations expressed in this material are those of the author(s) and do not necessarily reflect the views of the U.S. National Science Foundation, the U.S. Department of Energy, or any of the listed funding agencies. This research uses services or data provided by the SPectra Analysis and Retrievable Catalog Lab (SPARCL) and the Astro Data Lab, which are both part of the Community Science and Data Center (CSDC) program at NSF National Optical-Infrared Astronomy Research Laboratory. NOIRLab is operated by the Association of Universities for Research in Astronomy (AURA), Inc. under a cooperative agreement with the National Science Foundation.

This research made use of hips2fits\footnote{\url{https://alasky.cds.unistra.fr/hips-image-services/hips2fits}}, a service provided by CDS.

YF is financed by the Dutch Research Council (NWO) grant number OCENW.XL21.XL21.025. KIC acknowledges funding from the Dutch Research Council (NWO) through the award of the Vici Grant
VI.C.212.036.

VA, JC and FR acknowledge the support from the INAF Large Grant "\ac{agn} \& \Euclid: a close entanglement" Ob. Fu. 01.05.23.01.14.
\end{acknowledgements}

\bibliography{references}

@software{fu_2026_18642758,
  author       = {Fu, Yuming},
  title        = {specbox: a simple tool to manipulate and visualize UV/optical/NIR spectra for astronomical research},
  month        = feb,
  year         = 2026,
  publisher    = {Zenodo},
  version      = {v1.0.0},
  doi          = {10.5281/zenodo.18642758},
  url          = {https://doi.org/10.5281/zenodo.18642758},
  howpublished =  {v1.0.0, Zenodo,
                   \url{https://doi.org/10.5281/zenodo.18642758}},
}

@article{2017AJ....154...28B,
  author =        {{Blanton}, Michael R. and {Bershady}, Matthew A. and
                   {Abolfathi}, Bela and {Albareti}, Franco D. and
                   {Allende Prieto}, Carlos and {Almeida}, Andres and
                   {Alonso-Garc{\'\i}a}, Javier and {Anders}, Friedrich and
                   et al.},
  journal =       {\aj},
  month =         jul,
  number =        {1},
  pages =         {28},
  title =         {{Sloan Digital Sky Survey IV: Mapping the Milky Way,
                   Nearby Galaxies, and the Distant Universe}},
  volume =        {154},
  year =          {2017},
  doi =           {10.3847/1538-3881/aa7567},
  eid =           {28},
}

@article{2020MNRAS.499..210N,
  author =        {{Neveux}, Richard and {Burtin}, Etienne and
                   {de Mattia}, Arnaud and {Smith}, Alex and
                   {Ross}, Ashley J. and {Hou}, Jiamin and
                   {Bautista}, Julian and {Brinkmann}, Jonathan and
                   {Chuang}, Chia-Hsun and {Dawson}, Kyle S. and
                   {Gil-Mar{\'\i}n}, H{\'e}ctor and {Lyke}, Brad W. and
                   {de la Macorra}, Axel and
                   {du Mas des Bourboux}, H{\'e}lion and
                   {Mohammad}, Faizan G. and {M{\"u}ller}, Eva-Maria and
                   {Myers}, Adam D. and {Newman}, Jeffrey A. and
                   {Percival}, Will J. and {Rossi}, Graziano and
                   {Schneider}, Donald and {Vivek}, M. and
                   {Zarrouk}, Pauline and {Zhao}, Cheng and
                   {Zhao}, Gong-Bo},
  journal =       {\mnras},
  month =         nov,
  number =        {1},
  pages =         {210-229},
  title =         {{The completed SDSS-IV extended Baryon Oscillation
                   Spectroscopic Survey: BAO and RSD measurements from
                   the anisotropic power spectrum of the quasar sample
                   between redshift 0.8 and 2.2}},
  volume =        {499},
  year =          {2020},
  doi =           {10.1093/mnras/staa2780},
}

@article{2023ARA&A..61..373F,
  author =        {{Fan}, Xiaohui and {Ba{\~n}ados}, Eduardo and
                   {Simcoe}, Robert A.},
  journal =       {\araa},
  month =         aug,
  pages =         {373-426},
  title =         {{Quasars and the Intergalactic Medium at Cosmic
                   Dawn}},
  volume =        {61},
  year =          {2023},
  doi =           {10.1146/annurev-astro-052920-102455},
}

@article{2002AJ....123.2945R,
  author =        {{Richards}, Gordon T. and {Fan}, Xiaohui and
                   {Newberg}, Heidi Jo and {Strauss}, Michael A. and
                   {Vanden Berk}, Daniel E. and {Schneider}, Donald P. and
                   {Yanny}, Brian and {Boucher}, Adam and
                   {Burles}, Scott and {Frieman}, Joshua A. and
                   {Gunn}, James E. and {Hall}, Patrick B. and
                   {Ivezi{\'c}}, {\v{Z}}eljko and {Kent}, Stephen and
                   {Loveday}, Jon and {Lupton}, Robert H. and
                   {Rockosi}, Constance M. and {Schlegel}, David J. and
                   {Stoughton}, Chris and {SubbaRao}, Mark and
                   {York}, Donald G.},
  journal =       {\aj},
  month =         jun,
  number =        {6},
  pages =         {2945-2975},
  title =         {{Spectroscopic Target Selection in the Sloan Digital
                   Sky Survey: The Quasar Sample}},
  volume =        {123},
  year =          {2002},
  doi =           {10.1086/340187},
}

@article{2004MNRAS.349.1397C,
  author =        {{Croom}, S.~M. and {Smith}, R.~J. and {Boyle}, B.~J. and
                   {Shanks}, T. and {Miller}, L. and {Outram}, P.~J. and
                   {Loaring}, N.~S.},
  journal =       {\mnras},
  month =         apr,
  number =        {4},
  pages =         {1397-1418},
  title =         {{The 2dF QSO Redshift Survey - XII. The spectroscopic
                   catalogue and luminosity function}},
  volume =        {349},
  year =          {2004},
  doi =           {10.1111/j.1365-2966.2004.07619.x},
}

@article{2015ApJS..221...27M,
  author =        {{Myers}, Adam D. and
                   {Palanque-Delabrouille}, Nathalie and
                   {Prakash}, Abhishek and {P{\^a}ris}, Isabelle and
                   {Yeche}, Christophe and {Dawson}, Kyle S. and
                   {Bovy}, Jo and {Lang}, Dustin and
                   {Schlegel}, David J. and {Newman}, Jeffrey A. and
                   {Petitjean}, Patrick and {Kneib}, Jean-Paul and
                   {Laurent}, Pierre and {Percival}, Will J. and
                   {Ross}, Ashley J. and {Seo}, Hee-Jong and
                   {Tinker}, Jeremy L. and {Armengaud}, Eric and
                   {Brownstein}, Joel and {Burtin}, Etienne and
                   {Cai}, Zheng and {Comparat}, Johan and
                   {Kasliwal}, Mansi and {Kulkarni}, Shrinivas R. and
                   {Laher}, Russ and {Levitan}, David and
                   {McBride}, Cameron K. and {McGreer}, Ian D. and
                   {Miller}, Adam A. and {Nugent}, Peter and
                   {Ofek}, Eran and {Rossi}, Graziano and {Ruan}, John and
                   {Schneider}, Donald P. and {Sesar}, Branimir and
                   {Streblyanska}, Alina and {Surace}, Jason},
  journal =       {\apjs},
  month =         dec,
  number =        {2},
  pages =         {27},
  title =         {{The SDSS-IV Extended Baryon Oscillation
                   Spectroscopic Survey: Quasar Target Selection}},
  volume =        {221},
  year =          {2015},
  doi =           {10.1088/0067-0049/221/2/27},
  eid =           {27},
}

@article{2023ApJ...944..107C,
  author =        {{Chaussidon}, Edmond and {Y{\`e}che}, Christophe and
                   {Palanque-Delabrouille}, Nathalie and
                   {Alexander}, David M. and {Yang}, Jinyi and
                   {Ahlen}, Steven and {Bailey}, Stephen and
                   {Brooks}, David and {Cai}, Zheng and
                   {Chabanier}, Sol{\`e}ne and {Davis}, Tamara M. and
                   {Dawson}, Kyle and {de laMacorra}, Axel and
                   {Dey}, Arjun and {Dey}, Biprateep and
                   {Eftekharzadeh}, Sarah and {Eisenstein}, Daniel J. and
                   {Fanning}, Kevin and {Font-Ribera}, Andreu and
                   {Gazta{\~n}aga}, Enrique and
                   {A Gontcho}, Satya Gontcho and
                   {Gonzalez-Morales}, Alma X. and {Guy}, Julien and
                   {Herrera-Alcantar}, Hiram K. and {Honscheid}, Klaus and
                   {Ishak}, Mustapha and {Jiang}, Linhua and
                   {Juneau}, Stephanie and {Kehoe}, Robert and
                   {Kisner}, Theodore and {Kov{\'a}cs}, Andras and
                   {Kremin}, Anthony and {Lan}, Ting-Wen and
                   {Landriau}, Martin and {Le Guillou}, Laurent and
                   {Levi}, Michael E. and {Magneville}, Christophe and
                   {Martini}, Paul and {Meisner}, Aaron M. and
                   {Moustakas}, John and
                   {Mu{\~n}oz-Guti{\'e}rrez}, Andrea and
                   {Myers}, Adam D. and {Newman}, Jeffrey A. and
                   {Nie}, Jundan and {Percival}, Will J. and
                   {Poppett}, Claire and {Prada}, Francisco and
                   {Raichoor}, Anand and {Ravoux}, Corentin and
                   {Ross}, Ashley J. and {Schlafly}, Edward and
                   {Schlegel}, David and {Tan}, Ting and
                   {Tarl{\'e}}, Gregory and {Zhou}, Rongpu and
                   {Zhou}, Zhimin and {Zou}, Hu},
  journal =       {\apj},
  month =         feb,
  number =        {1},
  pages =         {107},
  title =         {{Target Selection and Validation of DESI Quasars}},
  volume =        {944},
  year =          {2023},
  doi =           {10.3847/1538-4357/acb3c2},
  eid =           {107},
}

@article{1986ApJ...306..411S,
  author =        {{Schmidt}, M. and {Schneider}, D.~P. and
                   {Gunn}, J.~E.},
  journal =       {\apj},
  month =         jul,
  pages =         {411},
  title =         {{Spectroscopic CCD Surveys for Quasars at Large
                   Redshift. I. A Deep PFUEI Survey}},
  volume =        {306},
  year =          {1986},
  doi =           {10.1086/164354},
}

@article{1991ApJS...75..273O,
  author =        {{Osmer}, Patrick S. and {Hewett}, Paul C.},
  journal =       {\apjs},
  month =         feb,
  pages =         {273},
  title =         {{A New Survey for Quasar Clustering}},
  volume =        {75},
  year =          {1991},
  doi =           {10.1086/191532},
}

@article{1999AJ....117...40S,
  author =        {{Schneider}, D.~P. and {Schmidt}, Maarten and
                   {Gunn}, J.~E.},
  journal =       {\aj},
  month =         jan,
  number =        {1},
  pages =         {40-55},
  title =         {{Spectroscopic CCD Surveys for Quasars at Large
                   Redshift. V. The Palomar Scan GRISM Survey Catalog}},
  volume =        {117},
  year =          {1999},
  doi =           {10.1086/300703},
}

@article{Momcheva_2016ApJS_3dhst_survey,
  author =        {{Momcheva}, Ivelina G. and {Brammer}, Gabriel B. and
                   {van Dokkum}, Pieter G. and {Skelton}, Rosalind E. and
                   {Whitaker}, Katherine E. and {Nelson}, Erica J. and
                   {Fumagalli}, Mattia and {Maseda}, Michael V. and
                   {Leja}, Joel and {Franx}, Marijn and
                   {Rix}, Hans-Walter and {Bezanson}, Rachel and
                   {Da Cunha}, Elisabete and {Dickey}, Claire and
                   {F{\"o}rster Schreiber}, Natascha M. and
                   {Illingworth}, Garth and {Kriek}, Mariska and
                   {Labb{\'e}}, Ivo and {Ulf Lange}, Johannes and
                   {Lundgren}, Britt F. and {Magee}, Daniel and
                   {Marchesini}, Danilo and {Oesch}, Pascal and
                   {Pacifici}, Camilla and {Patel}, Shannon G. and
                   {Price}, Sedona and {Tal}, Tomer and {Wake}, David A. and
                   {van der Wel}, Arjen and {Wuyts}, Stijn},
  journal =       {\apjs},
  month =         aug,
  number =        {2},
  pages =         {27},
  title =         {{The 3D-HST Survey: Hubble Space Telescope WFC3/G141
                   Grism Spectra, Redshifts, and Emission Line
                   Measurements for \raisebox{-0.5ex}\textasciitilde
                   100,000 Galaxies}},
  volume =        {225},
  year =          {2016},
  doi =           {10.3847/0067-0049/225/2/27},
  eid =           {27},
}

@article{2019ApJ...870..133E,
  author =        {{Estrada-Carpenter}, Vicente and {Papovich}, Casey and
                   {Momcheva}, Ivelina and {Brammer}, Gabriel and
                   {Long}, James and {Quadri}, Ryan F. and
                   {Bridge}, Joanna and {Dickinson}, Mark and
                   {Ferguson}, Henry and {Finkelstein}, Steven and
                   {Giavalisco}, Mauro and {Gosmeyer}, Catherine M. and
                   {Lotz}, Jennifer and {Salmon}, Brett and
                   {Skelton}, Rosalind E. and {Trump}, Jonathan R. and
                   {Weiner}, Benjamin},
  journal =       {\apj},
  month =         jan,
  number =        {2},
  pages =         {133},
  title =         {{CLEAR. I. Ages and Metallicities of Quiescent
                   Galaxies at 1.0 < z < 1.8 Derived from Deep Hubble
                   Space Telescope Grism Data}},
  volume =        {870},
  year =          {2019},
  doi =           {10.3847/1538-4357/aaf22e},
  eid =           {133},
}

@article{2009PASP..121...59K,
  author =        {{K{\"u}mmel}, M. and {Walsh}, J.~R. and {Pirzkal}, N. and
                   {Kuntschner}, H. and {Pasquali}, A.},
  journal =       {\pasp},
  month =         jan,
  number =        {875},
  pages =         {59},
  title =         {{The Slitless Spectroscopy Data Extraction Software
                   aXe}},
  volume =        {121},
  year =          {2009},
  doi =           {10.1086/596715},
}

@misc{2019ascl.soft05001B,
  author =        {{Brammer}, Gabe},
  howpublished =  {Astrophysics Source Code Library, record
                   ascl:1905.001},
  month =         may,
  title =         {{Grizli: Grism redshift and line analysis software}},
  year =          {2019},
  eid =           {ascl:1905.001},
}

@inproceedings{2005SPIE.5904....1R,
  author =        {{Rieke}, Marcia J. and {Kelly}, Douglas and
                   {Horner}, Scott},
  booktitle =     {Cryogenic Optical Systems and Instruments XI},
  editor =        {{Heaney}, James B. and {Burriesci}, Lawrence G.},
  month =         aug,
  pages =         {1-8},
  series =        {Society of Photo-Optical Instrumentation Engineers
                   (SPIE) Conference Series},
  title =         {{Overview of James Webb Space Telescope and NIRCam's
                   Role}},
  volume =        {5904},
  year =          {2005},
  doi =           {10.1117/12.615554},
}

@article{2023PASP..135b8001R,
  author =        {{Rieke}, Marcia J. and {Kelly}, Douglas M. and
                   {Misselt}, Karl and {Stansberry}, John and
                   {Boyer}, Martha and {Beatty}, Thomas and
                   {Egami}, Eiichi and {Florian}, Michael and
                   {Greene}, Thomas P. and {Hainline}, Kevin and et al.},
  journal =       {\pasp},
  month =         feb,
  number =        {1044},
  pages =         {028001},
  title =         {{Performance of NIRCam on JWST in Flight}},
  volume =        {135},
  year =          {2023},
  doi =           {10.1088/1538-3873/acac53},
  eid =           {028001},
}

@article{2022PASP..134b5002W,
  author =        {{Willott}, Chris J. and {Doyon}, Ren{\'e} and
                   {Albert}, Loic and {Brammer}, Gabriel B. and
                   {Dixon}, William V. and {Muzic}, Koraljka and
                   {Ravindranath}, Swara and {Scholz}, Aleks and
                   {Abraham}, Roberto and {Artigau}, {\'E}tienne and
                   {Brada{\v{c}}}, Maru{\v{s}}a and {Goudfrooij}, Paul and
                   {Hutchings}, John B. and {Iyer}, Kartheik G. and
                   {Jayawardhana}, Ray and {LaMassa}, Stephanie and
                   {Martis}, Nicholas and {Meyer}, Michael R. and
                   {Morishita}, Takahiro and {Mowla}, Lamiya and
                   {Muzzin}, Adam and {Noirot}, Ga{\"e}l and
                   {Pacifici}, Camilla and {Rowlands}, Neil and
                   {Sarrouh}, Ghassan and {Sawicki}, Marcin and
                   {Taylor}, Joanna M. and {Volk}, Kevin and
                   {Zabl}, Johannes},
  journal =       {\pasp},
  month =         feb,
  number =        {1032},
  pages =         {025002},
  title =         {{The Near-infrared Imager and Slitless Spectrograph
                   for the James Webb Space Telescope. II. Wide Field
                   Slitless Spectroscopy}},
  volume =        {134},
  year =          {2022},
  doi =           {10.1088/1538-3873/ac5158},
  eid =           {025002},
}

@article{2023PASP..135i8001D,
  author =        {{Doyon}, Ren{\'e} and {Willott}, Chris J. and
                   {Hutchings}, John B. and {Sivaramakrishnan}, Anand and
                   {Albert}, Lo{\"\i}c and {Lafreni{\`e}re}, David and
                   {Rowlands}, Neil and {Bego{\~n}a Vila}, M. and
                   {Martel}, Andr{\'e} R. and {LaMassa}, Stephanie and
                   {Aldridge}, David and {Artigau}, {\'E}tienne and
                   {Cameron}, Peter and {Chayer}, Pierre and
                   {Cook}, Neil J. and {Cooper}, Rachel A. and
                   {Darveau-Bernier}, Antoine and {Dupuis}, Jean and
                   {Earnshaw}, Colin and {Espinoza}, N{\'e}stor and
                   {Filippazzo}, Joseph C. and {Fullerton}, Alexander W. and
                   {Gaudreau}, Daniel and {Gawlik}, Roman and
                   {Goudfrooij}, Paul and {Haley}, Craig and
                   {Kammerer}, Jens and {Kendall}, David and
                   {Lambros}, Scott D. and {Ignat}, Luminita Ilinca and
                   {Maszkiewicz}, Michael and {McColgan}, Ashley and
                   {Morishita}, Takahiro and
                   {Ouellette}, Nathalie N. -Q. and {Pacifici}, Camilla and
                   {Philippi}, Natasha and {Radica}, Michael and
                   {Ravindranath}, Swara and {Rowe}, Jason and
                   {Roy}, Arpita and {Roy}, Niladri and {Saad}, Karl and
                   {Sohn}, Sangmo Tony and {Talens}, Geert Jan and
                   {Touahri}, Driss and {Thatte}, Deepashri and
                   {Taylor}, Joanna M. and {Vandal}, Thomas and
                   {Volk}, Kevin and {Wander}, Michel and
                   {Warner}, Gerald and {Zheng}, Sheng-Hai and
                   {Zhou}, Julia and {Abraham}, Roberto and
                   {Beaulieu}, Mathilde and {Benneke}, Bj{\"o}rn and
                   {Ferrarese}, Laura and {Jayawardhana}, Ray and
                   {Johnstone}, Doug and {Kaltenegger}, Lisa and
                   {Meyer}, Michael R. and {Pipher}, Judy L. and
                   {Rameau}, Julien and {Rieke}, Marcia and
                   {Salhi}, Salma and {Sawicki}, Marcin},
  journal =       {\pasp},
  month =         sep,
  number =        {1051},
  pages =         {098001},
  title =         {{The Near Infrared Imager and Slitless Spectrograph
                   for the James Webb Space Telescope. I. Instrument
                   Overview and In-flight Performance}},
  volume =        {135},
  year =          {2023},
  doi =           {10.1088/1538-3873/acd41b},
  eid =           {098001},
}

@article{2022ApJ...938L..13R,
  author =        {{Roberts-Borsani}, Guido and {Morishita}, Takahiro and
                   {Treu}, Tommaso and {Brammer}, Gabriel and
                   {Strait}, Victoria and {Wang}, Xin and
                   {Bradac}, Marusa and {Acebron}, Ana and
                   {Bergamini}, Pietro and {Boyett}, Kristan and
                   {Calabr{\'o}}, Antonello and {Castellano}, Marco and
                   {Fontana}, Adriano and {Glazebrook}, Karl and
                   {Grillo}, Claudio and {Henry}, Alaina and
                   {Jones}, Tucker and {Malkan}, Matthew and
                   {Marchesini}, Danilo and {Mascia}, Sara and
                   {Mason}, Charlotte and {Mercurio}, Amata and
                   {Merlin}, Emiliano and {Nanayakkara}, Themiya and
                   {Pentericci}, Laura and {Rosati}, Piero and
                   {Santini}, Paola and {Scarlata}, Claudia and
                   {Trenti}, Michele and {Vanzella}, Eros and
                   {Vulcani}, Benedetta and {Willott}, Chris},
  journal =       {\apjl},
  month =         oct,
  number =        {2},
  pages =         {L13},
  title =         {{Early Results from GLASS-JWST. I: Confirmation of
                   Lensed z {\ensuremath{\geq}} 7 Lyman-break Galaxies
                   behind the Abell 2744 Cluster with NIRISS}},
  volume =        {938},
  year =          {2022},
  doi =           {10.3847/2041-8213/ac8e6e},
  eid =           {L13},
}

@article{2023ApJ...953...53S,
  author =        {{Sun}, Fengwu and {Egami}, Eiichi and {Pirzkal}, Nor and
                   {Rieke}, Marcia and {Baum}, Stefi and {Boyer}, Martha and
                   {Boyett}, Kristan and {Bunker}, Andrew J. and
                   {Cameron}, Alex J. and {Curti}, Mirko and
                   {Eisenstein}, Daniel J. and {Gennaro}, Mario and
                   {Greene}, Thomas P. and {Jaffe}, Daniel and
                   {Kelly}, Doug and {Koekemoer}, Anton M. and
                   {Kumari}, Nimisha and {Maiolino}, Roberto and
                   {Maseda}, Michael and {Perna}, Michele and
                   {Rest}, Armin and {Robertson}, Brant E. and
                   {Schlawin}, Everett and {Smit}, Renske and
                   {Stansberry}, John and {Sunnquist}, Ben and
                   {Tacchella}, Sandro and {Williams}, Christina C. and
                   {Willmer}, Christopher N.~A.},
  journal =       {\apj},
  month =         aug,
  number =        {1},
  pages =         {53},
  title =         {{First Sample of H{\ensuremath{\alpha}}+[O
                   III]{\ensuremath{\lambda}}5007 Line Emitters at z > 6
                   Through JWST/NIRCam Slitless Spectroscopy: Physical
                   Properties and Line-luminosity Functions}},
  volume =        {953},
  year =          {2023},
  doi =           {10.3847/1538-4357/acd53c},
  eid =           {53},
}

@article{2023MNRAS.525.2864O,
  author =        {{Oesch}, P.~A. and {Brammer}, G. and {Naidu}, R.~P. and
                   {Bouwens}, R.~J. and {Chisholm}, J. and
                   {Illingworth}, G.~D. and {Matthee}, J. and
                   {Nelson}, E. and {Qin}, Y. and {Reddy}, N. and
                   et al.},
  journal =       {\mnras},
  month =         oct,
  number =        {2},
  pages =         {2864-2874},
  title =         {{The JWST FRESCO survey: legacy NIRCam/grism
                   spectroscopy and imaging in the two GOODS fields}},
  volume =        {525},
  year =          {2023},
  doi =           {10.1093/mnras/stad2411},
}

@article{2024MNRAS.535.1067M,
  author =        {{Meyer}, R.~A. and {Oesch}, P.~A. and
                   {Giovinazzo}, E. and {Weibel}, A. and {Brammer}, G. and
                   {Matthee}, J. and {Naidu}, R.~P. and {Bouwens}, R.~J. and
                   {Chisholm}, J. and {Covelo-Paz}, A. and
                   {Fudamoto}, Y. and {Maseda}, M. and {Nelson}, E. and
                   {Shivaei}, I. and {Xiao}, M. and
                   {Herard-Demanche}, T. and {Illingworth}, G.~D. and
                   {Kerutt}, J. and {Kramarenko}, I. and {Labbe}, I. and
                   {Leonova}, E. and {Magee}, D. and {Matharu}, J. and
                   {Prieto Lyon}, G. and {Reddy}, N. and {Schaerer}, D. and
                   {Shapley}, A. and {Stefanon}, M. and {Wozniak}, M.~A. and
                   {Wuyts}, S.},
  journal =       {\mnras},
  month =         nov,
  number =        {1},
  pages =         {1067-1094},
  title =         {{JWST FRESCO: a comprehensive census of H
                   {\ensuremath{\beta}} + [O III] emitters at 6.8 < z <
                   9.0 in the GOODS fields}},
  volume =        {535},
  year =          {2024},
  doi =           {10.1093/mnras/stae2353},
}

@article{EuclidSkyOverview,
  author =        {{Euclid Collaboration: Mellier}, Y. and {Abdurro'uf} and
                   {Acevedo~Barroso}, J.A. and others},
  journal =       {A\&A},
  pages =         {A1},
  title =         {Euclid - I. Overview of the Euclid mission},
  volume =        {697},
  year =          {2025},
  doi =           {10.1051/0004-6361/202450810},
  url =           {https://doi.org/10.1051/0004-6361/202450810},
}

@article{EuclidSkyVIS,
  author =        {{Euclid Collaboration: Cropper}, M. and
                   {Al-Bahlawan}, A. and {Amiaux}, J. and others},
  journal =       {A\&A},
  pages =         {A2},
  title =         {Euclid - II. The VIS instrument},
  volume =        {697},
  year =          {2025},
  doi =           {10.1051/0004-6361/202450996},
  url =           {https://doi.org/10.1051/0004-6361/202450996},
}

@article{EuclidSkyNISP,
  author =        {{Euclid Collaboration: Jahnke}, K. and {Gillard}, W. and
                   {Schirmer}, M. and others},
  journal =       {A\&A},
  pages =         {A3},
  title =         {Euclid - III. The NISP Instrument},
  volume =        {697},
  year =          {2025},
  doi =           {10.1051/0004-6361/202450786},
  url =           {https://doi.org/10.1051/0004-6361/202450786},
}

@article{Banados25,
  author =        {{Ba{\~n}ados}, E. and {Le Brun}, V. and
                   {Belladitta}, S. and others},
  journal =       {\mnras},
  month =         sep,
  number =        {2},
  pages =         {1088-1102},
  title =         {{Euclid: the potential of slitless infrared
                   spectroscopy: a z = 5.4 quasar and new ultracool
                   dwarfs}},
  volume =        {542},
  year =          {2025},
  doi =           {10.1093/mnras/staf1274},
}

@article{EP-Lusso,
  author =        {{Euclid Collaboration: Lusso}, E. and
                   {Fotopoulou}, S. and {Selwood}, M. and others},
  journal =       {\aap},
  month =         may,
  pages =         {A108},
  title =         {{Euclid preparation. XXXVIII. Spectroscopy of active
                   galactic nuclei with NISP}},
  volume =        {685},
  year =          {2024},
  doi =           {10.1051/0004-6361/202348326},
  eid =           {A108},
}

@article{Q1-TP001,
  author =        {{Euclid Collaboration: Aussel}, H. and {Tereno}, I. and
                   {Schirmer}, M. and others},
  journal =       {A\&A, submitted (Euclid Q1 SI)},
  month =         mar,
  pages =         {arXiv:2503.15302},
  title =         {{Euclid Quick Data Release (Q1) - Data release
                   overview}},
  year =          {2025},
  eid =           {arXiv:2503.15302},
}

@article{Q1-SP027,
  author =        {{Euclid Collaboration: Matamoro Zatarain}, T. and
                   {Fotopoulou}, S. and {Ricci}, F. and others},
  journal =       {A\&A, in press (Euclid Q1 SI),
                   \url{https://doi.org/10.1051/0004-6361/202554619}},
  month =         mar,
  pages =         {arXiv:2503.15320},
  title =         {{Euclid Quick Data Release (Q1). The active galaxies
                   of Euclid}},
  year =          {2025},
  eid =           {arXiv:2503.15320},
}

@article{Q1-SP003,
  author =        {{Euclid Collaboration: Roster}, W. and {Salvato}, M. and
                   {Buchner}, J. and others},
  journal =       {A\&A, in press (Euclid Q1 SI),
                   \url{https://doi.org/10.1051/0004-6361/202554616}},
  month =         mar,
  pages =         {arXiv:2503.15316},
  title =         {{Euclid Quick Data Release (Q1). Optical and
                   near-infrared identification and classification of
                   point-like X-ray selected sources}},
  year =          {2025},
  eid =           {arXiv:2503.15316},
}

@article{2024A&A...682A..34M,
  author =        {{Merloni}, A. and {Lamer}, G. and {Liu}, T. and
                   {Ramos-Ceja}, M.~E. and {Brunner}, H. and
                   {Bulbul}, E. and {Dennerl}, K. and {Doroshenko}, V. and
                   {Freyberg}, M.~J. and {Friedrich}, S. and et al.},
  journal =       {\aap},
  month =         feb,
  pages =         {A34},
  title =         {{The SRG/eROSITA all-sky survey. First X-ray
                   catalogues and data release of the western Galactic
                   hemisphere}},
  volume =        {682},
  year =          {2024},
  doi =           {10.1051/0004-6361/202347165},
  eid =           {A34},
}

@article{2020A&A...641A.136W,
  author =        {{Webb}, N.~A. and {Coriat}, M. and {Traulsen}, I. and
                   {Ballet}, J. and {Motch}, C. and {Carrera}, F.~J. and
                   {Koliopanos}, F. and {Authier}, J. and
                   {de la Calle}, I. and {Ceballos}, M.~T. and et al.},
  journal =       {\aap},
  month =         sep,
  pages =         {A136},
  title =         {{The XMM-Newton serendipitous survey. IX. The fourth
                   XMM-Newton serendipitous source catalogue}},
  volume =        {641},
  year =          {2020},
  doi =           {10.1051/0004-6361/201937353},
  eid =           {A136},
}

@article{2024ApJS..274...22E,
  author =        {{Evans}, Ian N. and {Evans}, Janet D. and
                   {Mart{\'\i}nez-Galarza}, J. Rafael and
                   {Miller}, Joseph B. and {Primini}, Francis A. and
                   {Azadi}, Mojegan and {Burke}, Douglas J. and
                   {Civano}, Francesca M. and {D'Abrusco}, Raffaele and
                   {Fabbiano}, Giuseppina and et al.},
  journal =       {\apjs},
  month =         oct,
  number =        {2},
  pages =         {22},
  title =         {{The Chandra Source Catalog Release 2 Series}},
  volume =        {274},
  year =          {2024},
  doi =           {10.3847/1538-4365/ad6319},
  eid =           {22},
}

@article{2003MNRAS.342..467M,
  author =        {{Myers}, A.~D. and {Outram}, P.~J. and {Shanks}, T. and
                   {Boyle}, B.~J. and {Croom}, S.~M. and
                   {Loaring}, N.~S. and {Miller}, L. and {Smith}, R.~J.},
  journal =       {\mnras},
  month =         jun,
  number =        {2},
  pages =         {467-482},
  title =         {{The 2dF QSO Redshift Survey - X. Lensing of
                   background QSOs by galaxy groups}},
  volume =        {342},
  year =          {2003},
  doi =           {10.1046/j.1365-8711.2003.06584.x},
}

@article{2015MNRAS.449.4326P,
  author =        {{Pullen}, Anthony R. and {Alam}, Shadab and
                   {Ho}, Shirley},
  journal =       {\mnras},
  month =         jun,
  number =        {4},
  pages =         {4326-4335},
  title =         {{Probing gravity at large scales through CMB
                   lensing}},
  volume =        {449},
  year =          {2015},
  doi =           {10.1093/mnras/stv554},
}

@article{2023ApJ...946...27P,
  author =        {{Petter}, Grayson C. and {Hickox}, Ryan C. and
                   {Alexander}, David M. and {Myers}, Adam D. and
                   {Geach}, James E. and {Whalen}, Kelly E. and
                   {Andonie}, Carolina P.},
  journal =       {\apj},
  month =         mar,
  number =        {1},
  pages =         {27},
  title =         {{Host Dark Matter Halos of Wide-field Infrared Survey
                   Explorer-selected Obscured and Unobscured Quasars:
                   Evidence for Evolution}},
  volume =        {946},
  year =          {2023},
  doi =           {10.3847/1538-4357/acb7ef},
  eid =           {27},
}

@article{2023JCAP...11..043A,
  author =        {{Alonso}, David and {Fabbian}, Giulio and
                   {Storey-Fisher}, Kate and {Eilers}, Anna-Christina and
                   {Garc{\'\i}a-Garc{\'\i}a}, Carlos and
                   {Hogg}, David W. and {Rix}, Hans-Walter},
  journal =       {JCAP},
  month =         nov,
  number =        {11},
  pages =         {043},
  title =         {{Constraining cosmology with the Gaia-unWISE Quasar
                   Catalog and CMB lensing: structure growth}},
  volume =        {11},
  year =          {2023},
  doi =           {10.1088/1475-7516/2023/11/043},
  eid =           {043},
}

@article{1983ApJ...266..713O,
  author =        {{Oke}, J.~B. and {Gunn}, J.~E.},
  journal =       {\apj},
  month =         mar,
  pages =         {713-717},
  title =         {{Secondary standard stars for absolute
                   spectrophotometry.}},
  volume =        {266},
  year =          {1983},
  doi =           {10.1086/160817},
}

@misc{Q1cite,
  author =        {{Euclid Quick Release Q1}},
  howpublished =  {\url{https://doi.org/10.57780/esa-2853f3b}},
  year =          {2025},
}

@article{Q1-TP004,
  author =        {{Euclid Collaboration: Romelli}, E. and
                   {K\"ummel}, M. and {Dole}, H. and others},
  journal =       {A\&A, in press (Euclid Q1 SI),
                   \url{https://doi.org/10.1051/0004-6361/202554586}},
  month =         mar,
  pages =         {arXiv:2503.15305},
  title =         {{Euclid Quick Data Release (Q1). From images to
                   multiwavelength catalogues: the Euclid MERge
                   Processing Function}},
  year =          {2025},
  eid =           {arXiv:2503.15305},
}

@article{Q1-SP040,
  author =        {{Euclid Collaboration: Quilley}, L. and
                   {Damjanov}, I. and {de Lapparent}, V. and others},
  journal =       {A\&A, in press (Euclid Q1 SI),
                   \url{https://doi.org/10.1051/0004-6361/202554585}},
  month =         mar,
  pages =         {arXiv:2503.15309},
  title =         {{Euclid Quick Data Release (Q1). Exploring galaxy
                   morphology across cosmic time through Sersic fits}},
  year =          {2025},
  eid =           {arXiv:2503.15309},
}

@article{Q1-TP006,
  author =        {{Euclid Collaboration: Copin}, Y. and {Fumana}, M. and
                   {Mancini}, C. and others},
  journal =       {A\&A, in press (Euclid Q1 SI),
                   \url{https://doi.org/10.1051/0004-6361/202554627}},
  month =         mar,
  pages =         {arXiv:2503.15307},
  title =         {{Euclid Quick Data Release (Q1): From spectrograms to
                   spectra: the SIR spectroscopic Processing Function}},
  year =          {2025},
  eid =           {arXiv:2503.15307},
}

@incollection{Datalabscite,
  author =        {{Navarro}, Vicente and {del Rio}, Sara and
                   {Angel Diego}, Miguel and {Lopez-Caniego}, Marcos and
                   {Marinic}, Filip and {Kruk}, Sandor and
                   {Reerink}, Jan and {Arviset}, Christophe},
  booktitle =     {Space Data Management. Studies in Big Data},
  pages =         {1-13},
  title =         {{ESA Datalabs: Digital Innovation in Space Science}},
  volume =        {141},
  year =          {2024},
  doi =           {10.1007/978-981-97-0041-7_1},
}

@article{2004ApJS..154..501P,
  author =        {{Pirzkal}, N. and {Xu}, C. and {Malhotra}, S. and
                   {Rhoads}, J.~E. and {Koekemoer}, A.~M. and
                   {Moustakas}, L.~A. and {Walsh}, J.~R. and
                   {Windhorst}, R.~A. and {Daddi}, E. and {Cimatti}, A. and
                   {Ferguson}, H.~C. and {Gardner}, Jonathan P. and
                   {Gronwall}, C. and {Haiman}, Z. and {K{\"u}mmel}, M. and
                   {Panagia}, N. and {Pasquali}, A. and {Stiavelli}, M. and
                   {di Serego Alighieri}, S. and {Tsvetanov}, Z. and
                   {Vernet}, J. and {Yan}, H.},
  journal =       {\apjs},
  month =         oct,
  number =        {2},
  pages =         {501-508},
  title =         {{GRAPES, Grism Spectroscopy of the Hubble Ultra Deep
                   Field: Description and Data Reduction}},
  volume =        {154},
  year =          {2004},
  doi =           {10.1086/422582},
}

@article{GaiaCollaboration_2023_2023A&A...674A...1G,
  author =        {{Gaia Collaboration: Vallenari}, A. and
                   {Brown}, A.~G.~A. and {Prusti}, T. and
                   {de Bruijne}, J.~H.~J. and {Arenou}, F. and
                   {Babusiaux}, C. and {Biermann}, M. and
                   {Creevey}, O.~L. and {Ducourant}, C. and
                   {Evans}, D.~W. and {Eyer}, L. and {Guerra}, R. and
                   {Hutton}, A. and {Jordi}, C. and {Klioner}, S.~A. and
                   {Lammers}, U.~L. and {Lindegren}, L. and et al.},
  journal =       {\aap},
  month =         jun,
  pages =         {A1},
  title =         {{Gaia Data Release 3: Summary of the content and
                   survey properties}},
  volume =        {674},
  year =          {2023},
  doi =           {10.1051/0004-6361/202243940},
  eid =           {A1},
}

@article{GaiaCollaboration_2023_2023A&A...674A..41G,
  author =        {{Gaia Collaboration: Bailer-Jones}, C.~A.~L. and
                   {Teyssier}, D. and {Delchambre}, L. and
                   {Ducourant}, C. and {Garabato}, D. and
                   {Hatzidimitriou}, D. and {Klioner}, S.~A. and
                   {Rimoldini}, L. and {Bellas-Velidis}, I. and
                   {Carballo}, R. and {Carnerero}, M.~I. and
                   {Diener}, C. and {Fouesneau}, M. and {Galluccio}, L. and
                   {Gavras}, P. and {Krone-Martins}, A. and
                   {Raiteri}, C.~M. and {Teixeira}, R. and
                   {Brown}, A.~G.~A. and {Vallenari}, A. and
                   {Prusti}, T. and {de Bruijne}, J.~H.~J. and et al.},
  journal =       {\aap},
  month =         jun,
  pages =         {A41},
  title =         {{Gaia Data Release 3. The extragalactic content}},
  volume =        {674},
  year =          {2023},
  doi =           {10.1051/0004-6361/202243232},
  eid =           {A41},
}

@article{2023A&A...674A...2D,
  author =        {{De Angeli}, F. and {Weiler}, M. and
                   {Montegriffo}, P. and {Evans}, D.~W. and {Riello}, M. and
                   {Andrae}, R. and {Carrasco}, J.~M. and {Busso}, G. and
                   {Burgess}, P.~W. and {Cacciari}, C. and
                   {Davidson}, M. and {Harrison}, D.~L. and
                   {Hodgkin}, S.~T. and {Jordi}, C. and {Osborne}, P.~J. and
                   {Pancino}, E. and {Altavilla}, G. and
                   {Barstow}, M.~A. and {Bailer-Jones}, C.~A.~L. and
                   {Bellazzini}, M. and {Brown}, A.~G.~A. and
                   {Castellani}, M. and {Cowell}, S. and
                   {Delchambre}, L. and {De Luise}, F. and {Diener}, C. and
                   {Fabricius}, C. and {Fouesneau}, M. and
                   {Fr{\'e}mat}, Y. and {Gilmore}, G. and
                   {Giuffrida}, G. and {Hambly}, N.~C. and {Hidalgo}, S. and
                   {Holland}, G. and {Kostrzewa-Rutkowska}, Z. and
                   {van Leeuwen}, F. and {Lobel}, A. and {Marinoni}, S. and
                   {Miller}, N. and {Pagani}, C. and {Palaversa}, L. and
                   {Piersimoni}, A.~M. and {Pulone}, L. and
                   {Ragaini}, S. and {Rainer}, M. and {Richards}, P.~J. and
                   {Rixon}, G.~T. and {Ruz-Mieres}, D. and {Sanna}, N. and
                   {Sarro}, L.~M. and {Rowell}, N. and {Sordo}, R. and
                   {Walton}, N.~A. and {Yoldas}, A.},
  journal =       {\aap},
  month =         jun,
  pages =         {A2},
  title =         {{Gaia Data Release 3. Processing and validation of
                   BP/RP low-resolution spectral data}},
  volume =        {674},
  year =          {2023},
  doi =           {10.1051/0004-6361/202243680},
  eid =           {A2},
}

@article{2023A&A...674A..31D,
  author =        {{Delchambre}, L. and {Bailer-Jones}, C.~A.~L. and
                   {Bellas-Velidis}, I. and {Drimmel}, R. and
                   {Garabato}, D. and {Carballo}, R. and
                   {Hatzidimitriou}, D. and {Marshall}, D.~J. and
                   {Andrae}, R. and {Dafonte}, C. and {Livanou}, E. and
                   {Fouesneau}, M. and {Licata}, E.~L. and
                   {Lindstr{\o}m}, H.~E.~P. and {Manteiga}, M. and
                   {Robin}, C. and {Silvelo}, A. and
                   {Abreu Aramburu}, A. and {{\'A}lvarez}, M.~A. and
                   {Bakker}, J. and {Bijaoui}, A. and {Brouillet}, N. and
                   {Brugaletta}, E. and {Burlacu}, A. and
                   {Casamiquela}, L. and {Chaoul}, L. and
                   {Chiavassa}, A. and {Contursi}, G. and
                   {Cooper}, W.~J. and {Creevey}, O.~L. and
                   {Dapergolas}, A. and {de Laverny}, P. and
                   {Demouchy}, C. and {Dharmawardena}, T.~E. and
                   {Edvardsson}, B. and {Fr{\'e}mat}, Y. and
                   {Garc{\'\i}a-Lario}, P. and {Garc{\'\i}a-Torres}, M. and
                   {Gavel}, A. and {Gomez}, A. and
                   {Gonz{\'a}lez-Santamar{\'\i}a}, I. and {Heiter}, U. and
                   {Jean-Antoine Piccolo}, A. and {Kontizas}, M. and
                   {Kordopatis}, G. and {Korn}, A.~J. and
                   {Lanzafame}, A.~C. and {Lebreton}, Y. and {Lobel}, A. and
                   {Lorca}, A. and {Magdaleno Romeo}, A. and
                   {Marocco}, F. and {Mary}, N. and {Nicolas}, C. and
                   {Ordenovic}, C. and {Pailler}, F. and
                   {Palicio}, P.~A. and {Pallas-Quintela}, L. and
                   {Panem}, C. and {Pichon}, B. and {Poggio}, E. and
                   {Recio-Blanco}, A. and {Riclet}, F. and {Rybizki}, J. and
                   {Santove{\~n}a}, R. and {Sarro}, L.~M. and
                   {Schultheis}, M.~S. and {Segol}, M. and {Slezak}, I. and
                   {Smart}, R.~L. and {Sordo}, R. and {Soubiran}, C. and
                   {S{\"u}veges}, M. and {Th{\'e}venin}, F. and
                   {Torralba Elipe}, G. and {Ulla}, A. and {Utrilla}, E. and
                   {Vallenari}, A. and {van Dillen}, E. and {Zhao}, H. and
                   {Zorec}, J.},
  journal =       {\aap},
  month =         jun,
  pages =         {A31},
  title =         {{Gaia Data Release 3. Apsis. III. Non-stellar content
                   and source classification}},
  volume =        {674},
  year =          {2023},
  doi =           {10.1051/0004-6361/202243423},
  eid =           {A31},
}

@article{2023A&A...674A..14R,
  author =        {{Rimoldini}, Lorenzo and {Holl}, Berry and
                   {Gavras}, Panagiotis and {Audard}, Marc and
                   {De Ridder}, Joris and {Mowlavi}, Nami and
                   {Nienartowicz}, Krzysztof and
                   {Jevardat de Fombelle}, Gr{\'e}gory and
                   {Lecoeur-Ta{\"\i}bi}, Isabelle and {Karbevska}, Lea and
                   et al.},
  journal =       {\aap},
  month =         jun,
  pages =         {A14},
  title =         {{Gaia Data Release 3. All-sky classification of 12.4
                   million variable sources into 25 classes}},
  volume =        {674},
  year =          {2023},
  doi =           {10.1051/0004-6361/202245591},
  eid =           {A14},
}

@article{2023A&A...674A..24C,
  author =        {{Carnerero}, Maria I. and {Raiteri}, Claudia M. and
                   {Rimoldini}, Lorenzo and {Busonero}, Deborah and
                   {Licata}, Enrico and {Mowlavi}, Nami and
                   {Lecoeur-Ta{\"\i}bi}, Isabelle and {Audard}, Marc and
                   {Holl}, Berry and {Gavras}, Panagiotis and et al.},
  journal =       {\aap},
  month =         jun,
  pages =         {A24},
  title =         {{Gaia Data Release 3. The first Gaia catalogue of
                   variable AGN}},
  volume =        {674},
  year =          {2023},
  doi =           {10.1051/0004-6361/202244035},
  eid =           {A24},
}

@article{2023A&A...674A..11D,
  author =        {{Ducourant}, C. and {Krone-Martins}, A. and
                   {Galluccio}, L. and {Teixeira}, R. and
                   {Le Campion}, J. -F. and {Slezak}, E. and
                   {de Souza}, R. and {Gavras}, P. and {Mignard}, F. and
                   {Guiraud}, J. and et al.},
  journal =       {\aap},
  month =         jun,
  pages =         {A11},
  title =         {{Gaia Data Release 3. Surface brightness profiles of
                   galaxies and host galaxies of quasars}},
  volume =        {674},
  year =          {2023},
  doi =           {10.1051/0004-6361/202243798},
  eid =           {A11},
}

@article{2022A&A...667A.148G,
  author =        {{Gaia Collaboration: Klioner}, S.~A. and
                   {Lindegren}, L. and {Mignard}, F. and
                   {Hern{\'a}ndez}, J. and {Ramos-Lerate}, M. and
                   {Bastian}, U. and {Biermann}, M. and {Bombrun}, A. and
                   {de Torres}, A. and et al.},
  journal =       {\aap},
  month =         nov,
  pages =         {A148},
  title =         {{Gaia Early Data Release 3. The celestial reference
                   frame (Gaia-CRF3)}},
  volume =        {667},
  year =          {2022},
  doi =           {10.1051/0004-6361/202243483},
  eid =           {A148},
}

@article{Storey-Fisher2024,
  author =        {{Storey-Fisher}, Kate and {Hogg}, David W. and
                   {Rix}, Hans-Walter and {Eilers}, Anna-Christina and
                   {Fabbian}, Giulio and {Blanton}, Michael R. and
                   {Alonso}, David},
  journal =       {\apj},
  month =         mar,
  number =        {1},
  pages =         {69},
  title =         {{Quaia, the Gaia-unWISE Quasar Catalog: An All-sky
                   Spectroscopic Quasar Sample}},
  volume =        {964},
  year =          {2024},
  doi =           {10.3847/1538-4357/ad1328},
  eid =           {69},
}

@article{2019ApJS..240...30S,
  author =        {{Schlafly}, Edward F. and {Meisner}, Aaron M. and
                   {Green}, Gregory M.},
  journal =       {\apjs},
  month =         feb,
  number =        {2},
  pages =         {30},
  title =         {{The unWISE Catalog: Two Billion Infrared Sources
                   from Five Years of WISE Imaging}},
  volume =        {240},
  year =          {2019},
  doi =           {10.3847/1538-4365/aafbea},
  eid =           {30},
}

@article{Fu2024_CatNorth,
  author =        {{Fu}, Yuming and {Wu}, Xue-Bing and {Li}, Yifan and
                   {Pang}, Yuxuan and {Joshi}, Ravi and {Zhang}, Shuo and
                   {Wang}, Qiyue and {Yang}, Jing and {Ng}, FanLam and
                   {Liu}, Xingjian and {Qiu}, Yu and {Zhu}, Rui and
                   {Wang}, Huimei and {Wolf}, Christian and
                   {Zhang}, Yanxia and {Huo}, Zhi-Ying and {Ai}, Y.~L. and
                   {Ma}, Qinchun and {Feng}, Xiaotong and
                   {Bouwens}, R.~J.},
  journal =       {\apjs},
  month =         apr,
  number =        {2},
  pages =         {54},
  title =         {{CatNorth: An Improved Gaia DR3 Quasar Candidate
                   Catalog with Pan-STARRS1 and CatWISE}},
  volume =        {271},
  year =          {2024},
  doi =           {10.3847/1538-4365/ad2ae6},
  eid =           {54},
}

@article{Chambers2016,
  author =        {{Chambers}, K.~C. and {Magnier}, E.~A. and
                   {Metcalfe}, N. and {Flewelling}, H.~A. and
                   {Huber}, M.~E. and {Waters}, C.~Z. and {Denneau}, L. and
                   {Draper}, P.~W. and {Farrow}, D. and
                   {Finkbeiner}, D.~P. and {Holmberg}, C. and
                   {Koppenhoefer}, J. and {Price}, P.~A. and {Rest}, A. and
                   {Saglia}, R.~P. and {Schlafly}, E.~F. and
                   {Smartt}, S.~J. and {Sweeney}, W. and
                   {Wainscoat}, R.~J. and et al.},
  journal =       {arXiv e-prints},
  month =         dec,
  pages =         {arXiv:1612.05560},
  title =         {{The Pan-STARRS1 Surveys}},
  year =          {2016},
  eid =           {arXiv:1612.05560},
}

@article{2021ApJS..253....8M,
  author =        {{Marocco}, Federico and {Eisenhardt}, Peter R.~M. and
                   {Fowler}, John W. and {Kirkpatrick}, J. Davy and
                   {Meisner}, Aaron M. and {Schlafly}, Edward F. and
                   {Stanford}, S.~A. and {Garcia}, Nelson and
                   {Caselden}, Dan and {Cushing}, Michael C. and
                   {Cutri}, Roc M. and {Faherty}, Jacqueline K. and
                   {Gelino}, Christopher R. and {Gonzalez}, Anthony H. and
                   {Jarrett}, Thomas H. and {Koontz}, Renata and
                   {Mainzer}, Amanda and {Marchese}, Elijah J. and
                   {Mobasher}, Bahram and {Schlegel}, David J. and
                   {Stern}, Daniel and {Teplitz}, Harry I. and
                   {Wright}, Edward L.},
  journal =       {\apjs},
  month =         mar,
  number =        {1},
  pages =         {8},
  title =         {{The CatWISE2020 Catalog}},
  volume =        {253},
  year =          {2021},
  doi =           {10.3847/1538-4365/abd805},
  eid =           {8},
}

@article{2021ApJS..254....6F,
  author =        {{Fu}, Yuming and {Wu}, Xue-Bing and {Yang}, Qian and
                   {Brown}, Anthony G.~A. and {Feng}, Xiaotong and
                   {Ma}, Qinchun and {Li}, Shuyan},
  journal =       {\apjs},
  month =         may,
  number =        {1},
  pages =         {6},
  title =         {{Finding Quasars behind the Galactic Plane. I.
                   Candidate Selections with Transfer Learning}},
  volume =        {254},
  year =          {2021},
  doi =           {10.3847/1538-4365/abe85e},
  eid =           {6},
}

@article{Fu2025_catsouth,
  author =        {{Fu}, Yuming and {Wu}, Xue-Bing and {Bouwens}, R.~J. and
                   {Caputi}, Karina I. and {Pang}, Yuxuan and {Zhu}, Rui and
                   {Yang}, Da-Ming and {Qin}, Jin and {Wang}, Huimei and
                   {Wolf}, Christian and et al.},
  journal =       {\apjs},
  month =         aug,
  number =        {2},
  pages =         {54},
  title =         {{The CatSouth Quasar Candidate Catalog for the
                   Southern Sky and a Unified All-sky Catalog Based on
                   Gaia DR3}},
  volume =        {279},
  year =          {2025},
  doi =           {10.3847/1538-4365/ade999},
  eid =           {54},
}

@article{Onken_2024_2024PASA...41...61O,
  author =        {{Onken}, Christopher A. and {Wolf}, Christian and
                   {Bessell}, Michael S. and {Chang}, Seo-Won and
                   {Luvaul}, Lance C. and {Tonry}, John L. and
                   {White}, Marc C. and {Da Costa}, Gary S.},
  journal =       {\pasa},
  month =         oct,
  pages =         {e061},
  title =         {{SkyMapper Southern Survey: Data release 4}},
  volume =        {41},
  year =          {2024},
  doi =           {10.1017/pasa.2024.53},
  eid =           {e061},
}

@article{2006Msngr.126...41E,
  author =        {{Emerson}, J. and {McPherson}, A. and
                   {Sutherland}, W.},
  journal =       {The Messenger},
  month =         {December},
  pages =         {41-42},
  title =         {{Visible and Infrared Survey Telescope for Astronomy:
                   Progress Report}},
  volume =        {126},
  year =          {2006},
}

@article{2010NewA...15..433M,
  author =        {{Minniti}, D. and {Lucas}, P.~W. and {Emerson}, J.~P. and
                   {Saito}, R.~K. and {Hempel}, M. and
                   {Pietrukowicz}, P. and {Ahumada}, A.~V. and
                   {Alonso}, M.~V. and {Alonso-Garcia}, J. and
                   {Arias}, J.~I. and {Bandyopadhyay}, R.~M. and
                   {Barb{\'a}}, R.~H. and {Barbuy}, B. and
                   {Bedin}, L.~R. and {Bica}, E. and {Borissova}, J. and
                   {Bronfman}, L. and {Carraro}, G. and {Catelan}, M. and
                   {Clari{\'a}}, J.~J. and {Cross}, N. and
                   {de Grijs}, R. and {D{\'e}k{\'a}ny}, I. and
                   {Drew}, J.~E. and {Fari{\~n}a}, C. and
                   {Feinstein}, C. and {Fern{\'a}ndez Laj{\'u}s}, E. and
                   {Gamen}, R.~C. and {Geisler}, D. and {Gieren}, W. and
                   {Goldman}, B. and {Gonzalez}, O.~A. and
                   {Gunthardt}, G. and {Gurovich}, S. and
                   {Hambly}, N.~C. and {Irwin}, M.~J. and
                   {Ivanov}, V.~D. and {Jord{\'a}n}, A. and {Kerins}, E. and
                   {Kinemuchi}, K. and {Kurtev}, R. and
                   {L{\'o}pez-Corredoira}, M. and {Maccarone}, T. and
                   {Masetti}, N. and {Merlo}, D. and {Messineo}, M. and
                   {Mirabel}, I.~F. and {Monaco}, L. and {Morelli}, L. and
                   {Padilla}, N. and {Palma}, T. and {Parisi}, M.~C. and
                   {Pignata}, G. and {Rejkuba}, M. and {Roman-Lopes}, A. and
                   {Sale}, S.~E. and {Schreiber}, M.~R. and
                   {Schr{\"o}der}, A.~C. and {Smith}, M. and
                   {}, Jr., L. Sodr{\'e} and {Soto}, M. and {Tamura}, M. and
                   {Tappert}, C. and {Thompson}, M.~A. and {Toledo}, I. and
                   {Zoccali}, M. and {Pietrzynski}, G.},
  journal =       {\na},
  month =         {July},
  number =        {5},
  pages =         {433-443},
  title =         {{VISTA Variables in the Via Lactea (VVV): The public
                   ESO near-IR variability survey of the Milky Way}},
  volume =        {15},
  year =          {2010},
  doi =           {10.1016/j.newast.2009.12.002},
}

@article{2011A&A...527A.116C,
  author =        {{Cioni}, M. -R.~L. and {Clementini}, G. and
                   {Girardi}, L. and {Guandalini}, R. and
                   {Gullieuszik}, M. and {Miszalski}, B. and
                   {Moretti}, M. -I. and {Ripepi}, V. and {Rubele}, S. and
                   {Bagheri}, G. and {Bekki}, K. and {Cross}, N. and
                   {de Blok}, W.~J.~G. and {de Grijs}, R. and
                   {Emerson}, J.~P. and {Evans}, C.~J. and {Gibson}, B. and
                   {Gonzales-Solares}, E. and {Groenewegen}, M.~A.~T. and
                   {Irwin}, M. and {Ivanov}, V.~D. and {Lewis}, J. and
                   {Marconi}, M. and {Marquette}, J. -B. and
                   {Mastropietro}, C. and {Moore}, B. and
                   {Napiwotzki}, R. and {Naylor}, T. and
                   {Oliveira}, J.~M. and {Read}, M. and {Sutorius}, E. and
                   {van Loon}, J. Th. and {Wilkinson}, M.~I. and
                   {Wood}, P.~R.},
  journal =       {\aap},
  month =         {March},
  pages =         {A116},
  title =         {{The VMC survey. I. Strategy and first data}},
  volume =        {527},
  year =          {2011},
  doi =           {10.1051/0004-6361/201016137},
  eid =           {A116},
}

@article{2013Msngr.154...35M,
  author =        {{McMahon}, R.~G. and {Banerji}, M. and {Gonzalez}, E. and
                   {Koposov}, S.~E. and {Bejar}, V.~J. and {Lodieu}, N. and
                   {Rebolo}, R. and {VHS Collaboration}},
  journal =       {The Messenger},
  month =         {December},
  pages =         {35-37},
  title =         {{First Scientific Results from the VISTA Hemisphere
                   Survey (VHS)}},
  volume =        {154},
  year =          {2013},
}

@article{2013Msngr.154...32E,
  author =        {{Edge}, A. and {Sutherland}, W. and {Kuijken}, K. and
                   {Driver}, S. and {McMahon}, R. and {Eales}, S. and
                   {Emerson}, J.~P.},
  journal =       {The Messenger},
  month =         {December},
  pages =         {32-34},
  title =         {{The VISTA Kilo-degree Infrared Galaxy (VIKING)
                   Survey: Bridging the Gap between Low and High
                   Redshift}},
  volume =        {154},
  year =          {2013},
}

@article{2010AJ....140.1868W,
  author =        {{Wright}, Edward L. and {Eisenhardt}, Peter R.~M. and
                   {Mainzer}, Amy K. and {Ressler}, Michael E. and
                   {Cutri}, Roc M. and {Jarrett}, Thomas and
                   {Kirkpatrick}, J. Davy and {Padgett}, Deborah and
                   {McMillan}, Robert S. and {Skrutskie}, Michael and
                   {Stanford}, S.~A. and {Cohen}, Martin and
                   {Walker}, Russell G. and {Mather}, John C. and
                   {Leisawitz}, David and {Gautier}, Thomas N., III and
                   {McLean}, Ian and {Benford}, Dominic and
                   {Lonsdale}, Carol J. and {Blain}, Andrew and
                   {Mendez}, Bryan and {Irace}, William R. and
                   {Duval}, Valerie and {Liu}, Fengchuan and
                   {Royer}, Don and {Heinrichsen}, Ingolf and
                   {Howard}, Joan and {Shannon}, Mark and
                   {Kendall}, Martha and {Walsh}, Amy L. and
                   {Larsen}, Mark and {Cardon}, Joel G. and
                   {Schick}, Scott and {Schwalm}, Mark and
                   {Abid}, Mohamed and {Fabinsky}, Beth and
                   {Naes}, Larry and {Tsai}, Chao-Wei},
  journal =       {\aj},
  month =         dec,
  number =        {6},
  pages =         {1868-1881},
  title =         {{The Wide-field Infrared Survey Explorer (WISE):
                   Mission Description and Initial On-orbit
                   Performance}},
  volume =        {140},
  year =          {2010},
  doi =           {10.1088/0004-6256/140/6/1868},
}

@article{2011ApJ...731...53M,
  author =        {{Mainzer}, A. and {Bauer}, J. and {Grav}, T. and
                   {Masiero}, J. and {Cutri}, R.~M. and {Dailey}, J. and
                   {Eisenhardt}, P. and {McMillan}, R.~S. and
                   {Wright}, E. and {Walker}, R. and {Jedicke}, R. and
                   {Spahr}, T. and {Tholen}, D. and {Alles}, R. and
                   {Beck}, R. and {Brandenburg}, H. and {Conrow}, T. and
                   {Evans}, T. and {Fowler}, J. and {Jarrett}, T. and
                   {Marsh}, K. and {Masci}, F. and {McCallon}, H. and
                   {Wheelock}, S. and {Wittman}, M. and {Wyatt}, P. and
                   {DeBaun}, E. and {Elliott}, G. and {Elsbury}, D. and
                   {Gautier}, T., IV and {Gomillion}, S. and
                   {Leisawitz}, D. and {Maleszewski}, C. and
                   {Micheli}, M. and {Wilkins}, A.},
  journal =       {\apj},
  month =         apr,
  number =        {1},
  pages =         {53},
  title =         {{Preliminary Results from NEOWISE: An Enhancement to
                   the Wide-field Infrared Survey Explorer for Solar
                   System Science}},
  volume =        {731},
  year =          {2011},
  doi =           {10.1088/0004-637X/731/1/53},
  eid =           {53},
}

@article{2018ApJS..234...23A,
  author =        {{Assef}, R.~J. and {Stern}, D. and {Noirot}, G. and
                   {Jun}, H.~D. and {Cutri}, R.~M. and
                   {Eisenhardt}, P.~R.~M.},
  journal =       {\apjs},
  month =         feb,
  number =        {2},
  pages =         {23},
  title =         {{The WISE AGN Catalog}},
  volume =        {234},
  year =          {2018},
  doi =           {10.3847/1538-4365/aaa00a},
  eid =           {23},
}

@article{2018A&A...616A...6S,
  author =        {{Sartoretti}, P. and {Katz}, D. and {Cropper}, M. and
                   {Panuzzo}, P. and {Seabroke}, G.~M. and {Viala}, Y. and
                   {Benson}, K. and {Blomme}, R. and {Jasniewicz}, G. and
                   {Jean-Antoine}, A. and et al.},
  journal =       {\aap},
  month =         aug,
  pages =         {A6},
  title =         {{Gaia Data Release 2. Processing the spectroscopic
                   data}},
  volume =        {616},
  year =          {2018},
  doi =           {10.1051/0004-6361/201832836},
  eid =           {A6},
}

@article{Glikman_2006,
  author =        {{Glikman}, Eilat and {Helfand}, David J. and
                   {White}, Richard L.},
  journal =       {\apj},
  month =         apr,
  number =        {2},
  pages =         {579-591},
  title =         {{A Near-Infrared Spectral Template for Quasars}},
  volume =        {640},
  year =          {2006},
  doi =           {10.1086/500098},
}

@article{2001AJ....122..549V,
  author =        {{Vanden Berk}, Daniel E. and {Richards}, Gordon T. and
                   {Bauer}, Amanda and {Strauss}, Michael A. and
                   {Schneider}, Donald P. and {Heckman}, Timothy M. and
                   {York}, Donald G. and {Hall}, Patrick B. and
                   {Fan}, Xiaohui and {Knapp}, G.~R. and
                   {Anderson}, Scott F. and {Annis}, James and
                   {Bahcall}, Neta A. and {Bernardi}, Mariangela and
                   {Briggs}, John W. and {Brinkmann}, J. and
                   {Brunner}, Robert and {Burles}, Scott and
                   {Carey}, Larry and {Castander}, Francisco J. and
                   {Connolly}, A.~J. and {Crocker}, J.~H. and
                   {Csabai}, Istv{\'a}n and {Doi}, Mamoru and
                   {Finkbeiner}, Douglas and {Friedman}, Scott and
                   {Frieman}, Joshua A. and {Fukugita}, Masataka and
                   {Gunn}, James E. and {Hennessy}, G.~S. and
                   {Ivezi{\'c}}, {\v{Z}}eljko and {Kent}, Stephen and
                   {Kunszt}, Peter Z. and {Lamb}, D.~Q. and
                   {Leger}, R. French and {Long}, Daniel C. and
                   {Loveday}, Jon and {Lupton}, Robert H. and
                   {Meiksin}, Avery and {Merelli}, Aronne and
                   {Munn}, Jeffrey A. and {Newberg}, Heidi Jo and
                   {Newcomb}, Matt and {Nichol}, R.~C. and
                   {Owen}, Russell and {Pier}, Jeffrey R. and
                   {Pope}, Adrian and {Rockosi}, Constance M. and
                   {Schlegel}, David J. and {Siegmund}, Walter A. and
                   {Smee}, Stephen and {Snir}, Yehuda and
                   {Stoughton}, Chris and {Stubbs}, Christopher and
                   {SubbaRao}, Mark and {Szalay}, Alexander S. and
                   {Szokoly}, Gyula P. and {Tremonti}, Christy and
                   {Uomoto}, Alan and {Waddell}, Patrick and
                   {Yanny}, Brian and {Zheng}, Wei},
  journal =       {\aj},
  month =         aug,
  number =        {2},
  pages =         {549-564},
  title =         {{Composite Quasar Spectra from the Sloan Digital Sky
                   Survey}},
  volume =        {122},
  year =          {2001},
  doi =           {10.1086/321167},
}

@inproceedings{2020ASPC..527..461B,
  author =        {{Bertin}, E. and {Schefer}, M. and {Apostolakos}, N. and
                   {{\'A}lvarez-Ayll{\'o}n}, A. and {Dubath}, P. and
                   {K{\"u}mmel}, M.},
  booktitle =     {Astronomical Data Analysis Software and Systems XXIX},
  editor =        {{Pizzo}, R. and {Deul}, E.~R. and {Mol}, J.~D. and
                   {de Plaa}, J. and {Verkouter}, H.},
  month =         jan,
  pages =         {461},
  series =        {Astronomical Society of the Pacific Conference
                   Series},
  title =         {{The SourceXtractor++ Software}},
  volume =        {527},
  year =          {2020},
}

@article{2022arXiv221202428K,
  author =        {{K{\"u}mmel}, M. and {{\'A}lvarez-Ayll{\'o}n}, A. and
                   {Bertin}, E. and {Dubath}, P. and {Gavazzi}, R. and
                   {Hartley}, W. and {Schefer}, M.},
  journal =       {arXiv e-prints},
  month =         dec,
  pages =         {arXiv:2212.02428},
  title =         {{Using the SourceXtractor++ package for data
                   reduction}},
  year =          {2022},
  doi =           {10.48550/arXiv.2212.02428},
  eid =           {arXiv:2212.02428},
}

@article{2012MNRAS.426.3369J,
  author =        {{Jauzac}, Mathilde and {Jullo}, Eric and
                   {Kneib}, Jean-Paul and {Ebeling}, Harald and
                   {Leauthaud}, Alexie and {Ma}, Cheng-Jiun and
                   {Limousin}, Marceau and {Massey}, Richard and
                   {Richard}, Johan},
  journal =       {\mnras},
  month =         nov,
  number =        {4},
  pages =         {3369-3384},
  title =         {{A weak lensing mass reconstruction of the
                   large-scale filament feeding the massive galaxy
                   cluster MACS J0717.5+3745}},
  volume =        {426},
  year =          {2012},
  doi =           {10.1111/j.1365-2966.2012.21966.x},
}

@article{2022arXiv220705709S,
  author =        {{Sharon}, Keren and {Cerny}, Catherine and
                   {Rigby}, Jane R. and {Florian}, Michael K. and
                   {Bayliss}, Matthew B. and {Dahle}, Hakon and
                   {Gladders}, Michael D. and {Mahler}, Guillaume},
  journal =       {arXiv e-prints},
  month =         jul,
  pages =         {arXiv:2207.05709},
  title =         {{HST-Based Lens Model of SDSS J1226+2152, in
                   Preparation for JWST-ERS TEMPLATES}},
  year =          {2022},
  doi =           {10.48550/arXiv.2207.05709},
  eid =           {arXiv:2207.05709},
}

@article{2023A&A...671A.146E,
  author =        {{Estrada}, N. and {Mercurio}, A. and {Vulcani}, B. and
                   {Rodighiero}, G. and {Nonino}, M. and
                   {Annunziatella}, M. and {Rosati}, P. and {Grillo}, C. and
                   {Caminha}, G.~B. and {Angora}, G. and et al.},
  journal =       {\aap},
  month =         mar,
  pages =         {A146},
  title =         {{VST-GAME: Galaxy assembly as a function of mass and
                   environment with VST. Photometric assessment and
                   density field of MACSJ0416}},
  volume =        {671},
  year =          {2023},
  doi =           {10.1051/0004-6361/202245070},
  eid =           {A146},
}

@article{DESI_DR1,
  author =        {{DESI Collaboration: Abdul-Karim}, M. and
                   {Adame}, A.~G. and {Aguado}, D. and {Aguilar}, J. and
                   {Ahlen}, S. and {Alam}, S. and {Aldering}, G. and
                   {Alexander}, D.~M. and {Alfarsy}, R. and et al.},
  journal =       {arXiv e-prints},
  month =         mar,
  pages =         {arXiv:2503.14745},
  title =         {{Data Release 1 of the Dark Energy Spectroscopic
                   Instrument}},
  year =          {2025},
  doi =           {10.48550/arXiv.2503.14745},
  eid =           {arXiv:2503.14745},
}

@article{2023OJAp....6E..49F,
  author =        {{Flesch}, Eric Wim},
  journal =       {The Open Journal of Astrophysics},
  month =         dec,
  pages =         {49},
  title =         {{The Million Quasars (Milliquas) Catalogue, v8}},
  volume =        {6},
  year =          {2023},
  doi =           {10.21105/astro.2308.01505},
  eid =           {49},
}

@article{Q1-SP015,
  author =        {{Euclid Collaboration: Margalef-Bentabol}, B. and
                   {Wang}, L. and {La Marca}, A. and others},
  journal =       {A\&A, in press (Euclid Q1 SI),
                   \url{https://doi.org/10.1051/0004-6361/202554583}},
  month =         mar,
  pages =         {arXiv:2503.15318},
  title =         {{Euclid Quick Data Release (Q1). First Euclid
                   statistical study of the active galactic nuclei
                   contribution fraction}},
  year =          {2025},
  eid =           {arXiv:2503.15318},
}

@article{2012ApJ...753...30S,
  author =        {{Stern}, Daniel and {Assef}, Roberto J. and
                   {Benford}, Dominic J. and {Blain}, Andrew and
                   {Cutri}, Roc and {Dey}, Arjun and {Eisenhardt}, Peter and
                   {Griffith}, Roger L. and {Jarrett}, T.~H. and
                   {Lake}, Sean and et al.},
  journal =       {\apj},
  month =         jul,
  number =        {1},
  pages =         {30},
  title =         {{Mid-infrared Selection of Active Galactic Nuclei
                   with the Wide-Field Infrared Survey Explorer. I.
                   Characterizing WISE-selected Active Galactic Nuclei
                   in COSMOS}},
  volume =        {753},
  year =          {2012},
  doi =           {10.1088/0004-637X/753/1/30},
  eid =           {30},
}

@article{2012AJ....144...49W,
  author =        {{Wu}, Xue-Bing and {Hao}, Guoqiang and
                   {Jia}, Zhendong and {Zhang}, Yanxia and
                   {Peng}, Nanbo},
  journal =       {\aj},
  month =         aug,
  number =        {2},
  pages =         {49},
  title =         {{SDSS Quasars in the WISE Preliminary Data Release
                   and Quasar Candidate Selection with Optical/Infrared
                   Colors}},
  volume =        {144},
  year =          {2012},
  doi =           {10.1088/0004-6256/144/2/49},
  eid =           {49},
}

@article{2012MNRAS.426.3271M,
  author =        {{Mateos}, S. and {Alonso-Herrero}, A. and
                   {Carrera}, F.~J. and {Blain}, A. and {Watson}, M.~G. and
                   {Barcons}, X. and {Braito}, V. and {Severgnini}, P. and
                   {Donley}, J.~L. and {Stern}, D.},
  journal =       {\mnras},
  month =         nov,
  number =        {4},
  pages =         {3271-3281},
  title =         {{Using the Bright Ultrahard XMM-Newton survey to
                   define an IR selection of luminous AGN based on WISE
                   colours}},
  volume =        {426},
  year =          {2012},
  doi =           {10.1111/j.1365-2966.2012.21843.x},
}

@article{Q1-SP023,
  author =        {{Euclid Collaboration: Tarsitano}, F. and
                   {Fotopoulou}, S. and {Banerji}, M. and others},
  journal =       {A\&A, in press (Euclid Q1 SI),
                   \url{https://doi.org/10.1051/0004-6361/202554591}},
  month =         mar,
  pages =         {arXiv:2503.15319},
  title =         {{Euclid Quick Data Release (Q1) First study of red
                   quasars selection}},
  year =          {2025},
  eid =           {arXiv:2503.15319},
}

@article{2018PASJ...70S...1M,
  author =        {{Miyazaki}, Satoshi and {Komiyama}, Yutaka and
                   {Kawanomoto}, Satoshi and {Doi}, Yoshiyuki and
                   {Furusawa}, Hisanori and {Hamana}, Takashi and
                   {Hayashi}, Yusuke and {Ikeda}, Hiroyuki and
                   {Kamata}, Yukiko and {Karoji}, Hiroshi and et al.},
  journal =       {\pasj},
  month =         jan,
  pages =         {S1},
  title =         {{Hyper Suprime-Cam: System design and verification of
                   image quality}},
  volume =        {70},
  year =          {2018},
  doi =           {10.1093/pasj/psx063},
  eid =           {S1},
}

@article{2018PASJ...70S...4A,
  author =        {{Aihara}, Hiroaki and {Arimoto}, Nobuo and
                   {Armstrong}, Robert and {Arnouts}, St{\'e}phane and
                   {Bahcall}, Neta A. and {Bickerton}, Steven and
                   {Bosch}, James and {Bundy}, Kevin and
                   {Capak}, Peter L. and {Chan}, James H.~H. and et al.},
  journal =       {\pasj},
  month =         jan,
  pages =         {S4},
  title =         {{The Hyper Suprime-Cam SSP Survey: Overview and
                   survey design}},
  volume =        {70},
  year =          {2018},
  doi =           {10.1093/pasj/psx066},
  eid =           {S4},
}

@article{2016A&A...596A.109P,
  author =        {{Planck Collaboration} and {Aghanim}, N. and
                   {Ashdown}, M. and {Aumont}, J. and {Baccigalupi}, C. and
                   {Ballardini}, M. and {Banday}, A.~J. and
                   {Barreiro}, R.~B. and {Bartolo}, N. and {Basak}, S. and
                   {Benabed}, K. and {Bernard}, J. -P. and
                   {Bersanelli}, M. and {Bielewicz}, P. and
                   {Bonavera}, L. and {Bond}, J.~R. and {Borrill}, J. and
                   {Bouchet}, F.~R. and {Boulanger}, F. and
                   {Burigana}, C. and {Calabrese}, E. and
                   {Cardoso}, J. -F. and {Carron}, J. and
                   {Chiang}, H.~C. and {Colombo}, L.~P.~L. and
                   {Comis}, B. and {Couchot}, F. and et al.},
  journal =       {\aap},
  month =         dec,
  pages =         {A109},
  title =         {{Planck intermediate results. XLVIII. Disentangling
                   Galactic dust emission and cosmic infrared background
                   anisotropies}},
  volume =        {596},
  year =          {2016},
  doi =           {10.1051/0004-6361/201629022},
  eid =           {A109},
}

@article{2023ApJ...950...86G,
  author =        {{Gordon}, Karl D. and {Clayton}, Geoffrey C. and
                   {Decleir}, Marjorie and {Fitzpatrick}, E.~L. and
                   {Massa}, Derck and {Misselt}, Karl A. and
                   {Tollerud}, Erik J.},
  journal =       {\apj},
  month =         jun,
  number =        {2},
  pages =         {86},
  title =         {{One Relation for All Wavelengths: The
                   Far-ultraviolet to Mid-infrared Milky Way
                   Spectroscopic R(V)-dependent Dust Extinction
                   Relationship}},
  volume =        {950},
  year =          {2023},
  doi =           {10.3847/1538-4357/accb59},
  eid =           {86},
}

@article{2018JOSS....3..695G,
  author =        {{Green}, Gregory M.},
  journal =       {The Journal of Open Source Software},
  month =         jun,
  number =        {26},
  pages =         {695},
  title =         {{dustmaps: A Python interface for maps of
                   interstellar dust}},
  volume =        {3},
  year =          {2018},
  doi =           {10.21105/joss.00695},
}

@article{2024JOSS....9.7023G,
  author =        {{Gordon}, Karl},
  journal =       {The Journal of Open Source Software},
  month =         aug,
  number =        {100},
  pages =         {7023},
  title =         {{dust\_extinction: Interstellar Dust Extinction
                   Models}},
  volume =        {9},
  year =          {2024},
  doi =           {10.21105/joss.07023},
  eid =           {7023},
}

@article{2002PASP..114..144F,
  author =        {{Fruchter}, A.~S. and {Hook}, R.~N.},
  journal =       {\pasp},
  month =         feb,
  number =        {792},
  pages =         {144-152},
  title =         {{Drizzle: A Method for the Linear Reconstruction of
                   Undersampled Images}},
  volume =        {114},
  year =          {2002},
  doi =           {10.1086/338393},
}

@article{2011ApJS..197...36K,
  author =        {{Koekemoer}, Anton M. and {Faber}, S.~M. and
                   {Ferguson}, Henry C. and {Grogin}, Norman A. and
                   {Kocevski}, Dale D. and {Koo}, David C. and
                   {Lai}, Kamson and {Lotz}, Jennifer M. and
                   {Lucas}, Ray A. and {McGrath}, Elizabeth J. and
                   et al.},
  journal =       {\apjs},
  month =         dec,
  number =        {2},
  pages =         {36},
  title =         {{CANDELS: The Cosmic Assembly Near-infrared Deep
                   Extragalactic Legacy Survey{\textemdash}The Hubble
                   Space Telescope Observations, Imaging Data Products,
                   and Mosaics}},
  volume =        {197},
  year =          {2011},
  doi =           {10.1088/0067-0049/197/2/36},
  eid =           {36},
}

@article{2023ApJS..268...64W,
  author =        {{Williams}, Christina C. and {Tacchella}, Sandro and
                   {Maseda}, Michael V. and {Robertson}, Brant E. and
                   {Johnson}, Benjamin D. and {Willott}, Chris J. and
                   {Eisenstein}, Daniel J. and
                   {Willmer}, Christopher N.~A. and {Ji}, Zhiyuan and
                   {Hainline}, Kevin N. and et al.},
  journal =       {\apjs},
  month =         oct,
  number =        {2},
  pages =         {64},
  title =         {{JEMS: A Deep Medium-band Imaging Survey in the
                   Hubble Ultra Deep Field with JWST NIRCam and NIRISS}},
  volume =        {268},
  year =          {2023},
  doi =           {10.3847/1538-4365/acf130},
  eid =           {64},
}

@article{2023ApJ...946L..12B,
  author =        {{Bagley}, Micaela B. and {Finkelstein}, Steven L. and
                   {Koekemoer}, Anton M. and {Ferguson}, Henry C. and
                   {Arrabal Haro}, Pablo and {Dickinson}, Mark and
                   {Kartaltepe}, Jeyhan S. and {Papovich}, Casey and
                   {P{\'e}rez-Gonz{\'a}lez}, Pablo G. and {Pirzkal}, Nor and
                   et al.},
  journal =       {\apjl},
  month =         mar,
  number =        {1},
  pages =         {L12},
  title =         {{CEERS Epoch 1 NIRCam Imaging: Reduction Methods and
                   Simulations Enabling Early JWST Science Results}},
  volume =        {946},
  year =          {2023},
  doi =           {10.3847/2041-8213/acbb08},
  eid =           {L12},
}

@article{2008ApJS..174..282L,
  author =        {{Landt}, Hermine and {Bentz}, Misty C. and
                   {Ward}, Martin J. and {Elvis}, Martin and
                   {Peterson}, Bradley M. and {Korista}, Kirk T. and
                   {Karovska}, Margarita},
  journal =       {\apjs},
  month =         feb,
  number =        {2},
  pages =         {282-312},
  title =         {{The Near-Infrared Broad Emission Line Region of
                   Active Galactic Nuclei. I. The Observations}},
  volume =        {174},
  year =          {2008},
  doi =           {10.1086/522373},
}

@article{2011MNRAS.414..218L,
  author =        {{Landt}, Hermine and {Elvis}, Martin and
                   {Ward}, Martin J. and {Bentz}, Misty C. and
                   {Korista}, Kirk T. and {Karovska}, Margarita},
  journal =       {\mnras},
  month =         jun,
  number =        {1},
  pages =         {218-240},
  title =         {{The near-infrared broad emission line region of
                   active galactic nuclei - II. The
                   1-{\ensuremath{\mu}}m continuum}},
  volume =        {414},
  year =          {2011},
  doi =           {10.1111/j.1365-2966.2011.18383.x},
}

@article{2013MNRAS.432..113L,
  author =        {{Landt}, Hermine and {Ward}, Martin J. and
                   {Peterson}, Bradley M. and {Bentz}, Misty C. and
                   {Elvis}, Martin and {Korista}, Kirk T. and
                   {Karovska}, Margarita},
  journal =       {\mnras},
  month =         jun,
  number =        {1},
  pages =         {113-126},
  title =         {{A near-infrared relationship for estimating black
                   hole masses in active galactic nuclei}},
  volume =        {432},
  year =          {2013},
  doi =           {10.1093/mnras/stt421},
}

@inproceedings{2008SPIE.7014E..0US,
  author =        {{Simcoe}, Robert A. and {Burgasser}, Adam J. and
                   {Bernstein}, Rebecca A. and {Bigelow}, Bruce C. and
                   {Fishner}, Jason and {Forrest}, William J. and
                   {McMurtry}, Craig and {Pipher}, Judith L. and
                   {Schechter}, Paul L. and {Smith}, Matthew},
  booktitle =     {Ground-based and Airborne Instrumentation for
                   Astronomy II},
  editor =        {{McLean}, Ian S. and {Casali}, Mark M.},
  month =         jul,
  pages =         {70140U},
  series =        {Society of Photo-Optical Instrumentation Engineers
                   (SPIE) Conference Series},
  title =         {{FIRE: a near-infrared cross-dispersed echellette
                   spectrometer for the Magellan telescopes}},
  volume =        {7014},
  year =          {2008},
  doi =           {10.1117/12.790414},
  eid =           {70140U},
}

@article{2022ApJS..261....8R,
  author =        {{Ricci}, Federica and {Treister}, Ezequiel and
                   {Bauer}, Franz E. and
                   {Mej{\'\i}a-Restrepo}, Julian E. and
                   {Koss}, Michael J. and {den Brok}, Jakob S. and
                   {Balokovi{\'c}}, Mislav and {B{\"a}r}, Rudolf and
                   {Bessiere}, Patricia and {Caglar}, Turgay and et al.},
  journal =       {\apjs},
  month =         jul,
  number =        {1},
  pages =         {8},
  title =         {{BASS. XXIX. The Near-infrared View of the Broad-line
                   Region (BLR): The Effects of Obscuration in BLR
                   Characterization}},
  volume =        {261},
  year =          {2022},
  doi =           {10.3847/1538-4365/ac5b67},
  eid =           {8},
}

@misc{fu_2025_15571037,
  author =        {Fu, Yuming},
  howpublished =  {v1.2.2, Zenodo,
                   \url{https://doi.org/10.5281/zenodo.15571037}},
  month =         jun,
  publisher =     {Zenodo},
  title =         {QSOFITMORE: a python package for fitting UV-optical
                   spectra of quasars},
  year =          {2025},
  doi =           {10.5281/zenodo.15571037},
  url =           {https://doi.org/10.5281/zenodo.15571037},
}

@article{Q1-SP047,
  author =        {{Euclid Collaboration: Walmsley}, M. and
                   {Huertas-Company}, M. and {Quilley}, L. and others},
  journal =       {A\&A, accepted (Euclid Q1 SI)},
  month =         mar,
  pages =         {arXiv:2503.15310},
  title =         {{Euclid Quick Data Release (Q1): First visual
                   morphology catalogue}},
  year =          {2025},
  eid =           {arXiv:2503.15310},
}

@article{2003ApJS..147....1C,
  author =        {{Conselice}, Christopher J.},
  journal =       {\apjs},
  month =         jul,
  number =        {1},
  pages =         {1-28},
  title =         {{The Relationship between Stellar Light Distributions
                   of Galaxies and Their Formation Histories}},
  volume =        {147},
  year =          {2003},
  doi =           {10.1086/375001},
}

@article{2014ARA&A..52..291C,
  author =        {{Conselice}, Christopher J.},
  journal =       {\araa},
  month =         aug,
  pages =         {291-337},
  title =         {{The Evolution of Galaxy Structure Over Cosmic Time}},
  volume =        {52},
  year =          {2014},
  doi =           {10.1146/annurev-astro-081913-040037},
}

@inproceedings{2025ASPC..541...77J,
  author =        {{Juneau}, St{\'e}phanie and {Jacques}, Alice and
                   {Pothier}, Steve and {Bolton}, Adam S. and
                   {Weaver}, Benjamin A. and {Pucha}, Ragadeepika and
                   {McManus}, Sean and {Nikutta}, Robert and
                   {Olsen}, Knut},
  booktitle =     {Astronomical Data Analysis Software and Systems
                   XXXIII},
  editor =        {{Jacques}, Alice and {Seaman}, Robert and
                   {Gandilo}, Natalie and {Linder}, Tyler},
  month =         oct,
  pages =         {77},
  series =        {Astronomical Society of the Pacific Conference
                   Series},
  title =         {{SPARCL: SPectra Analysis and Retrievable Catalog
                   Lab}},
  volume =        {541},
  year =          {2025},
  doi =           {10.26624/FMJS9195},
}

@inproceedings{2014SPIE.9149E..1TF,
  author =        {{Fitzpatrick}, Michael J. and {Olsen}, Knut and
                   {Economou}, Frossie and {Stobie}, Elizabeth B. and
                   {Beers}, T.~C. and {Dickinson}, Mark and
                   {Norris}, Patrick and {Saha}, Abi and
                   {Seaman}, Robert and {Silva}, David R. and et al.},
  booktitle =     {Observatory Operations: Strategies, Processes, and
                   Systems V},
  editor =        {{Peck}, Alison B. and {Benn}, Chris R. and
                   {Seaman}, Robert L.},
  month =         aug,
  pages =         {91491T},
  series =        {Society of Photo-Optical Instrumentation Engineers
                   (SPIE) Conference Series},
  title =         {{The NOAO Data Laboratory: a conceptual overview}},
  volume =        {9149},
  year =          {2014},
  doi =           {10.1117/12.2057445},
  eid =           {91491T},
}

@article{2020A&C....3300411N,
  author =        {{Nikutta}, R. and {Fitzpatrick}, M. and {Scott}, A. and
                   {Weaver}, B.~A.},
  journal =       {Astronomy and Computing},
  month =         oct,
  pages =         {100411},
  title =         {{Data Lab-A community science platform}},
  volume =        {33},
  year =          {2020},
  doi =           {10.1016/j.ascom.2020.100411},
  eid =           {100411},
}

@article{1992ApJ...397..442B,
  author =        {{Boroson}, Todd A. and {Meyers}, Karie A.},
  journal =       {\apj},
  month =         oct,
  pages =         {442},
  title =         {{The Optical Properties of IR-selected and MG II
                   Broad Absorption Line Quasars}},
  volume =        {397},
  year =          {1992},
  doi =           {10.1086/171800},
}

@article{2002ApJS..141..267H,
  author =        {{Hall}, Patrick B. and {Anderson}, Scott F. and
                   {Strauss}, Michael A. and {York}, Donald G. and
                   {Richards}, Gordon T. and {Fan}, Xiaohui and
                   {Knapp}, G.~R. and {Schneider}, Donald P. and
                   {Vanden Berk}, Daniel E. and {Geballe}, T.~R. and
                   {Bauer}, Amanda E. and {Becker}, Robert H. and
                   {Davis}, Marc and {Rix}, Hans-Walter and
                   {Nichol}, R.~C. and {Bahcall}, Neta A. and
                   {Brinkmann}, J. and {Brunner}, Robert and
                   {Connolly}, A.~J. and {Csabai}, Istv{\'a}n and
                   {Doi}, Mamoru and {Fukugita}, Masataka and
                   {Gunn}, James E. and {Haiman}, Zoltan and
                   {Harvanek}, Michael and {Heckman}, Timothy M. and
                   {Hennessy}, G.~S. and {Inada}, Naohisa and
                   {Ivezi{\'c}}, {\v{Z}}eljko and {Johnston}, David and
                   {Kleinman}, S. and {Krolik}, Julian H. and
                   {Krzesinski}, Jurek and {Kunszt}, Peter Z. and
                   {Lamb}, D.~Q. and {Long}, Daniel C. and
                   {Lupton}, Robert H. and {Miknaitis}, Gajus and
                   {Munn}, Jeffrey A. and {Narayanan}, Vijay K. and
                   {Neilsen}, Eric and {Newman}, P.~R. and
                   {Nitta}, Atsuko and {Okamura}, Sadanori and
                   {Pentericci}, Laura and {Pier}, Jeffrey R. and
                   {Schlegel}, David J. and {Snedden}, S. and
                   {Szalay}, Alexander S. and {Thakar}, Anirudda R. and
                   {Tsvetanov}, Zlatan and {White}, Richard L. and
                   {Zheng}, Wei},
  journal =       {\apjs},
  month =         aug,
  number =        {2},
  pages =         {267-309},
  title =         {{Unusual Broad Absorption Line Quasars from the Sloan
                   Digital Sky Survey}},
  volume =        {141},
  year =          {2002},
  doi =           {10.1086/340546},
}

@article{2009ApJ...698.1095U,
  author =        {{Urrutia}, Tanya and {Becker}, Robert H. and
                   {White}, Richard L. and {Glikman}, Eilat and
                   {Lacy}, Mark and {Hodge}, Jacqueline and
                   {Gregg}, Michael D.},
  journal =       {\apj},
  month =         jun,
  number =        {2},
  pages =         {1095-1109},
  title =         {{The FIRST-2MASS Red Quasar Survey. II. An
                   Anomalously High Fraction of LoBALs in Searches for
                   Dust-Reddened Quasars}},
  volume =        {698},
  year =          {2009},
  doi =           {10.1088/0004-637X/698/2/1095},
}

@article{LEYS2013764,
  author =        {Christophe Leys and Christophe Ley and Olivier Klein and
                   Philippe Bernard and Laurent Licata},
  journal =       {Journal of Experimental Social Psychology},
  number =        {4},
  pages =         {764-766},
  title =         {Detecting outliers: Do not use standard deviation
                   around the mean, use absolute deviation around the
                   median},
  volume =        {49},
  year =          {2013},
  doi =           {https://doi.org/10.1016/j.jesp.2013.03.013},
  issn =          {0022-1031},
  url =           {https://www.sciencedirect.com/science/article/pii/
                  S0022103113000668},
}

@misc{iso11664-4,
  author =        {{International Organization for Standardization} and
                   {International Commission on Illumination}},
  howpublished =  {\url{https://www.iso.org/standard/74166.html}},
  title =         {{ISO/CIE 11664-4:2019} Colorimetry — Part 4: CIE
                   1976 L*a*b* Colour Space},
  year =          {2019},
}

\begin{appendix}
\nolinenumbers

\section{Sources with initially discrepant redshifts in the DESI comparison}
\label{app:desi_check}

We illustrate the 16 initially discrepant sources in the DESI consistency check, defined as
$\left|\zvi-\zdesi\right|/(1+\zdesi)>0.15$ among the 454 quasars in common between our Q1 sample and DESI DR1. For each of the 16 sources, we retrieve the DESI DR1 coadded \ac{1d} spectrum and the corresponding redshift (\zdesi) from the DESI redshift catalogue using the SPectra Analysis and Retrievable Catalog Lab \citep[SPARCL;][]{2025ASPC..541...77J} and the Astro Data Lab \citep{2014SPIE.9149E..1TF,2020A&C....3300411N}, and compare them to the \Euclid \RGE spectrum used in our visual inspection. The assessment focuses on whether the main emission features are mutually consistent between the DESI and \Euclid spectra at the proposed redshift solution.

\begin{figure*}
  \centering
  \includegraphics[width=\textwidth]{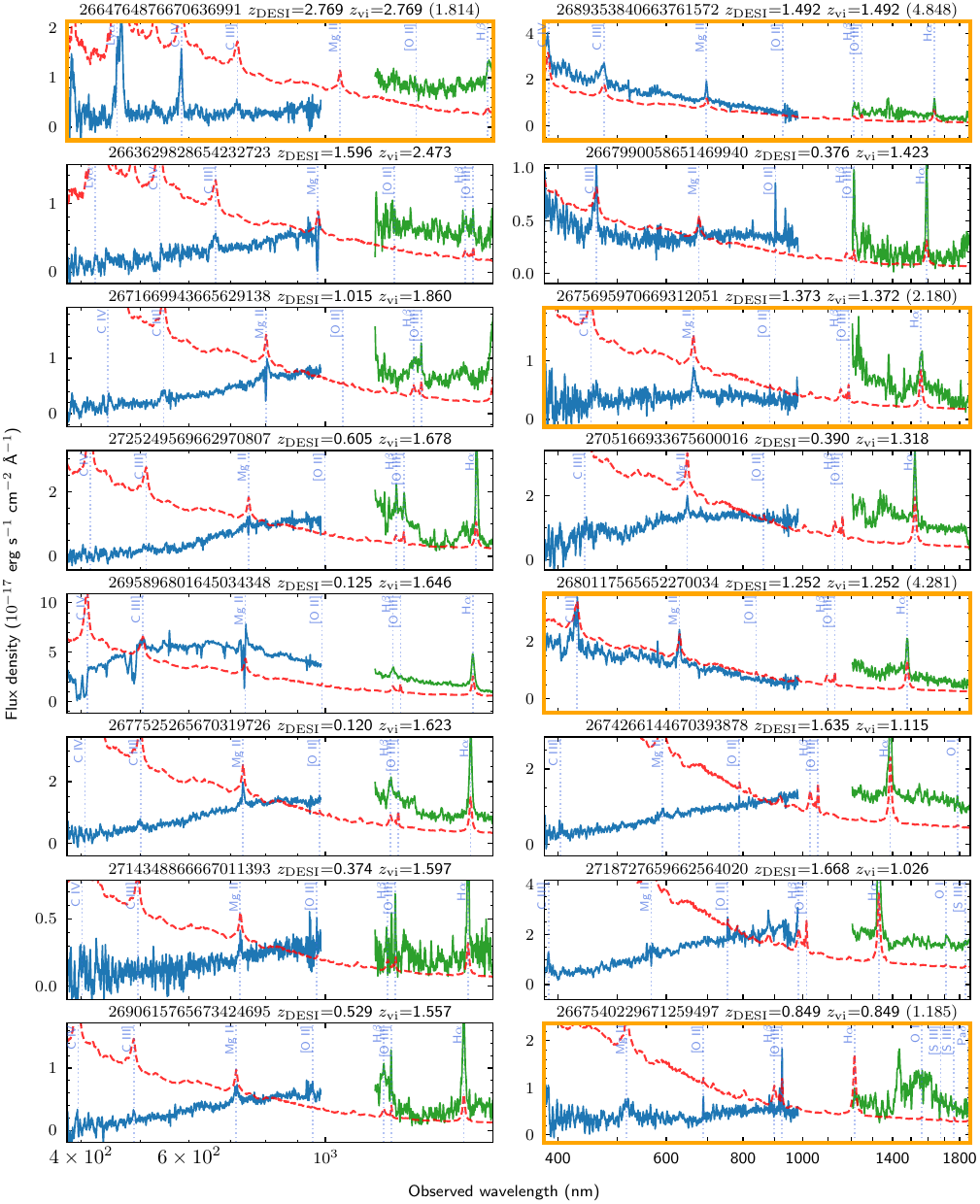}
  \caption{DESI (blue) and \Euclid (green) spectra for the 16 objects that initially satisfy $\left|\zvi-\zdesi\right|/(1+\zdesi)>0.15$ among the 454 quasars in common with DESI DR1. The dashed red curve shows the quasar template spectrum evaluated at the final $\zvi$ (scaled for display). Vertical dotted lines mark the expected observed wavelengths of common quasar emission lines at the final $\zvi$. Panels outlined in orange denote the five sources for which we revise $\zvi$ after reinspection; for these, the initial (incorrect) $\zvi$ is reported in parentheses after the updated $\zvi$ in the panel title.}
  \label{fig:desi_outliers_spectra}
\end{figure*}

\Cref{fig:desi_outliers_spectra} shows the DESI and \Euclid spectra together with the template spectrum evaluated at the final $\zvi$. For 11 sources, both DESI and \Euclid spectra support our visual redshift solution, and \zdesi is inconsistent with the spectra. For the remaining five sources whose visual redshifts are revised after reinspection, the panel titles list the updated $\zvi$ together with the initial (incorrect) $\zvi$ in parentheses; the updated values are adopted in the final catalogue.

As shown in \cref{fig:desi_outliers_spectra}, most of the 11 sources for which \zdesi is inconsistent with the spectra exhibit an unusually red rest-frame \ac{uv} continuum. In addition, several of these objects (\Euclid Q1 \texttt{object\_id}: 2663629828654232723, 2671669943665629138, 2725249569662970807, 2695896801645034348, and 2718727659662564020) show broad absorption troughs, including absorption associated with \ion{Mg}{ii} and other metal transitions, consistent with the unusual low-ionisation broad absorption line \citep[LoBAL; see e.g.][]{1992ApJ...397..442B,2002ApJS..141..267H,2009ApJ...698.1095U} quasars. Because such strongly reddened and LoBAL quasars are relatively rare in the training and template sets used by automated pipelines, and because their red, observed-frame optical continua often have lower \ac{snr} than those of typical blue quasars, automated redshift measurements can be less reliable for these objects.

For the remaining five sources, the main causes of the initial misidentifications of \zvi are (i) line-identification degeneracies when only a single prominent feature is present, and (ii) artefacts that complicate the interpretation of the spectrum. In particular, in two cases (\texttt{object\_id}: 2689353840663761572, and 2680117565652270034) an emission feature consistent with H\,$\alpha$ at the correct redshift was initially misidentified as \ion{Mg}{ii}, leading to spuriously high redshift estimates ($\zvi>4$). In another case (\texttt{object\_id}: 2664764876670636991), an emission feature consistent with H\,$\beta$ at the correct redshift was initially interpreted as H\,$\alpha$, leading to an overestimated redshift. Conversely, for object 2675695970669312051, H\,$\alpha$ was initially misidentified as H\,$\beta$, leading to an underestimated redshift. Finally, for object 2667540229671259497, the \Euclid spectrum is strongly affected by artefacts (anomalous spike and bump), likely related to contamination from a nearby source, and the initial redshift estimate was driven by the spike seen in the \Euclid spectrum.

\section{Description of the spectroscopically identified bright quasar catalogue}

The format of the catalogue of spectroscopically identified bright quasars from \Euclid Q1 is shown in \Cref{tab:main-metadata}. The full catalogue will be available at \url{https://cdsarc.cds.unistra.fr/}.

\begin{table*}[h!]
\centering
\small
\caption{Format of the catalogue of spectroscopically identified bright quasars from \Euclid Q1.}
\label{tab:main-metadata}
\begin{tabularx}{\textwidth}{@{}l l l l X@{}}
\toprule
Column & Name & Unit & Type & Description \\
\midrule
1 & object\_id & ... & long & Euclid Q1 unique source identifier \\
2 & name & ... & string & IAU-formatted source name (EUCL JHHMMSS.ss+DDMMSS.s) \\
3 & ra & deg & double & Source barycentre right ascension coordinate in decimal degrees \\
4 & dec & deg & double & Source barycentre declination coordinate in decimal degrees \\
5 & class\_vi & ... & string & Visual classification of the source \\
6 & z\_vi & ... & double & Visual redshift of the source \\
7 & sum\_mask & ... & int & Sum of the mask values of the spectrum within [12047, 18734] Å \\
8 & n\_invalid & ... & int & Number of invalid pixels in the spectrum within [12047, 18734] Å \\
9 & med\_snr & ... & double & Median signal-to-noise ratio of the spectrum within [12047, 18734] Å \\
10 & mumax\_minus\_mag & $\mathrm{mag~arcsec^{-2}}$ & double & The difference between mu\_max and mag\_stargal\_sep, valid even for NIR-only sources \\
11 & kron\_radius & pix & double & Major semi-axis (in pixels) of the elliptical aperture used for total (Kron) aperture photometry on the detection image \\
12 & gaia\_id & ... & long & The associated \gdr{3} source id \\
13 & f\_psf & ... & double & \ac{psf} fraction of the source from VIS image \citep{Q1-SP015}  \\
14 & mag\_vis\_psf & mag & double & VIS \IE band PSF-fitting AB magnitude (from flux\_vis\_psf) \\
15 & magerr\_vis\_psf & mag & double & Error on VIS \IE band PSF-fitting AB magnitude \\
16 & mag\_y\_templfit & mag & double & \YE band template-fit AB magnitude (from flux\_y\_templfit) \\
17 & magerr\_y\_templfit & mag & double & Error on \YE band template-fit AB magnitude \\
18 & mag\_j\_templfit & mag & double & \JE band template-fit AB magnitude (from flux\_j\_templfit) \\
19 & magerr\_j\_templfit & mag & double & Error on \JE band template-fit AB magnitude \\
20 & mag\_h\_templfit & mag & double & \HE band template-fit AB magnitude (from flux\_h\_templfit) \\
21 & magerr\_h\_templfit & mag & double & Error on \HE band template-fit AB magnitude \\
\bottomrule
\end{tabularx}%
\tablefoot{The catalogue described by \cref{tab:main-metadata} is available in its entirety at the CDS.}
\end{table*}

\section{Composite spectra generation with 1D drizzle}
\label{app:composite_method}

\subsection{Details of the 1D drizzle method}
\label{app:drizzle_details}

We construct \Euclid quasar composite spectra on a common rest-frame grid with a constant bin size of $\Delta\lambda = 4\,\AA$. The mapping preserves the integrated flux density in each bin through a linear, flux-conserving rebinning that partitions every input pixel across the overlapping output bins in proportion to the fractional overlap in wavelength. This procedure is mathematically equivalent to a one-dimensional version of the Drizzle algorithm \citep{2002PASP..114..144F} and matches the smallest native rest-frame pixel size at the blue end of our spectra, while mildly oversampling the data at longer wavelengths.

\paragraph{Preprocessing and normalisation.}
Each spectrum is shifted to the rest frame using the adopted redshift, then converted to rest-frame flux density in $F_\lambda$ via $F_{\lambda,\mathrm{rest}} = (1+z)\,F_{\lambda,\mathrm{obs}}$. The spectra are rebinned onto the common rest-frame grid with the flux-conserving drizzle scheme described below. Each rebinned spectrum is then normalised to the running composite within the spectral overlap so that relative shapes are combined consistently; the same normalisation factor is applied to the spectrum's per-pixel variances. We ignore pixels with non-finite or non-positive flux or variance values.

\paragraph{The 1D drizzle rebinning and variance propagation.}
Let $f_{p}$ and $\sigma_{p}^{2}$ be the flux and variance in input pixel $p$, whose wavelength interval is $[\lambda_{p-1/2},\lambda_{p+1/2}]$ with width $\Delta\lambda_{p} = \lambda_{p+1/2}-\lambda_{p-1/2}$. The $j$-th output bin covers $[\lambda_{j-1/2},\lambda_{j+1/2}]$ with width $\Delta\lambda_{j}$. We define the fractional overlap
\begin{equation}
A_{jp} = \frac{\max\bigl(0,\,\min[\lambda_{p+1/2},\lambda_{j+1/2}] - \max[\lambda_{p-1/2},\lambda_{j-1/2}]\bigr)}{\Delta\lambda_{p}}\,,
\end{equation}
so that $0 \le A_{jp} \le 1$ and $\sum_{j} A_{jp} = 1$ for fully covered pixels. The rebinned flux and variance for a single spectrum in bin $j$ are then
\begin{gather}
\hat f_{j} = \sum_{p} A_{jp}\, f_{p}\,, \\
\widehat{\sigma}_{j}^{2} = \sum_{p} A_{jp}^{2}\, \sigma_{p}^{2}\,,
\end{gather}
assuming uncorrelated input pixels. This operation introduces correlations between neighbouring output bins, but it preserves total flux and provides a well-defined per-bin variance. In the following, we treat $\widehat{\sigma}_{j}$ as the effective per-bin uncertainty and note the presence of correlations where relevant.

\paragraph{Equal-weight mean and its uncertainty.}
For the mean composite we combine, in each bin $j$, all rebinned spectra that have valid data in that bin with equal weights. If $n_{j}$ spectra contribute and their rebinned fluxes and variances are $\hat f_{ij}$ and $\widehat{\sigma}_{ij}^{2}$, the mean composite and its formal uncertainty are
\begin{gather}
\bar f_{j} = \frac{1}{n_{j}}\sum_{i=1}^{n_{j}} \hat f_{ij}\,, \\
\sigma_{\bar f_{j}}^{2} = \frac{1}{n_{j}^{2}}\sum_{i=1}^{n_{j}} \widehat{\sigma}_{ij}^{2}\,.
\end{gather}
In the implementation we apply sigma clipping in each wavelength bin to reduce the impact of outliers before computing the mean. The sigma-clipping is performed using \texttt{astropy.stats.sigma\_clip}, with a threshold of \texttt{sigma=6}, and a robust standard deviation estimator calculated as
\begin{equation}
    \sigma_{\mathrm{R}} \approx 1.4826 \ \textrm{MAD}\,,
\end{equation}
where $\textrm{MAD}$ is the median absolute deviation \citep{LEYS2013764}. For a univariate dataset $X_1$, $X_2$, ..., $X_n$, the MAD is defined as the median of the absolute deviations from the data's median
\begin{equation}
     \textrm{MAD} =\textrm{median} (|X_{i}-{\tilde {X}}|)\,.
\end{equation}

In addition, we track the observed object-to-object dispersion in each bin after sigma-clipping via the central second moment
\begin{equation}
s_{j}^{2} = \frac{1}{n_{j}}\sum_{i=1}^{n_{j}} \bigl(\hat f_{ij} - \bar f_{j}\bigr)^{2}\,,
\end{equation}
whose square root is stored as the `RMS' spectrum. This $s_{j}$ includes both intrinsic diversity and measurement noise and is quoted separately from the statistical uncertainty $\sigma_{\bar f_{j}}$ on the mean.

\paragraph{Median composite.}
We formed a median composite by taking, in each bin, the median of the contributing $\hat f_{ij}$ values. We applied the same sigma clipping as done when computing the mean composite. We did not attach formal uncertainties to the median composite.

\paragraph{Geometric-mean composite for continuum shape.}
To preserve the average continuum slope of quasars, we also computed a geometric-mean composite on the same rebinned spectra and wavelength grid, again using equal weights. For contributors with positive flux,
\begin{gather}
\mu_{j} = \frac{1}{n_{j}}\sum_{i=1}^{n_{j}} \ln \hat f_{ij}\,, \\
f^{\mathrm{geo}}_{j} = \exp(\mu_{j})\,.
\end{gather}
Using error propagation for $\ln \hat f_{ij}$ with $\mathrm{Var}[\ln \hat f_{ij}] \approx \widehat{\sigma}_{ij}^{2}/\hat f_{ij}^{2}$ when $\widehat{\sigma}_{ij} \ll \hat f_{ij}$, the uncertainty on $\mu_{j}$ and the geometric-mean flux are
\begin{gather}
\sigma_{\mu_{j}}^{2} = \frac{1}{n_{j}^{2}}\sum_{i=1}^{n_{j}} \frac{\widehat{\sigma}_{ij}^{2}}{\hat f_{ij}^{2}}\,, \\
\sigma^{\mathrm{geo}}_{j} = f^{\mathrm{geo}}_{j}\,\sigma_{\mu_{j}}\,.
\end{gather}
We masked bins where fewer than three spectra contribute or where negative flux values prevent a robust logarithmic mean from being determined.

\begin{figure}[htb]
\centering
\includegraphics[width=\hsize]{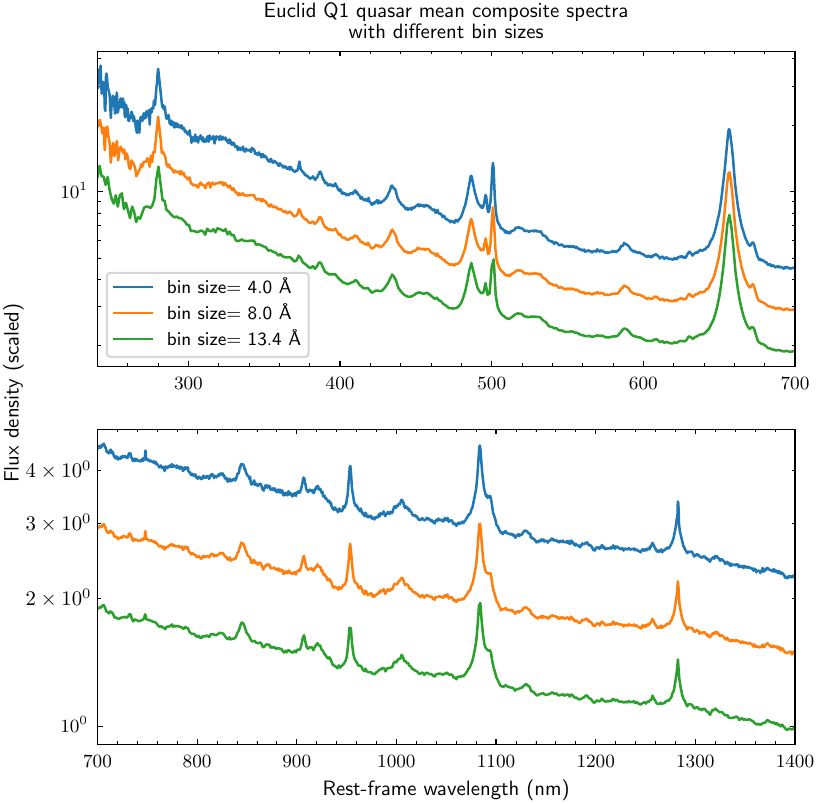}
\caption{Mean \Euclid Q1 quasar composite spectra constructed with different rest-frame wavelength bin sizes. The three curves show mean composites built on grids with $\Delta\lambda = 4.0\,\AA$ (blue), $8.0\,\AA$ (orange), and $13.4\,\AA$ (green). For clarity, the spectra are vertically offset by multiplicative factors, as indicated on the right-hand side. The top panel displays the optical range ($300$--$700\,\si{\nano\meter}$) and the bottom panel shows the near-infrared range ($700$--$1400\,\si{\nano\meter}$). The continua of the three spectra show close agreement, while the two spectra with larger bin sizes have flattened emission line peaks.}
\label{fig:composite_binsize_comp}
\end{figure}

\paragraph{Effect of wavelength bin size.}
In Sect.~\ref{subsec:spec_comps} we adopted a bin size of $\Delta\lambda = 4\,\AA$ for constructing the \Euclid mean composite spectrum. To assess the impact of this choice, we recomputed the mean composite using coarser grids with $\Delta\lambda = 8\,\AA$ and $13.4\,\AA$, keeping all other steps of the stacking procedure fixed. As shown in \cref{fig:composite_binsize_comp}, the continuum shapes of three mean composites are virtually identical in both the optical and near-infrared ranges, while the two larger bin sizes produce undersampled, flattened emission line peaks. This test demonstrates that the bin size of $\Delta\lambda = 4\,\AA$ is favourable for the preservation of details of the composite spectra.

\subsection{Construction of the piecewise quasar template}\label{app:piecewise_template}

The final rest-frame template combines
\begin{enumerate}
    \item \citet{2001AJ....122..549V} for $\lambda_{\rm rest}<320$ nm;
    \item The mean composite from this work for $320\le\lambda_{\rm rest}\le1550$ nm; and
    \item \citet{Glikman_2006} for $\lambda_{\rm rest}>1550$ nm.
\end{enumerate}
\noindent All three spectra are resampled onto a common rest-frame grid from 90 to 2100 nm with a spacing of 0.2 nm using a flux-conserving resampler (\texttt{specutils}.\texttt{FluxConservingResampler}). Before splicing, we scale the adjoining segments using the mean flux in 5 nm windows centred on each join. At 320 nm we scale our composite to match \citet{2001AJ....122..549V}; at 1550 nm we scale \citet{Glikman_2006} to match our composite. If $\langle F_{\mathrm{A}}\rangle$ and $\langle F_{\mathrm{B}}\rangle$ are the mean fluxes in the overlap window, the scale factor applied to $F_{\mathrm{B}}$ is $s=\langle F_{\mathrm{A}}\rangle/\langle F_{\mathrm{B}}\rangle$. After scaling, we combine the spectra piecewise at the join wavelengths (no additional cross-fade). The resulting template covers 90 to 2100 nm on a uniform grid.

\section{Example cutouts of quasars}
\label{app:euclid_cutouts}

\begin{figure*}[htbp]
\centering
\includegraphics[width=1\linewidth]{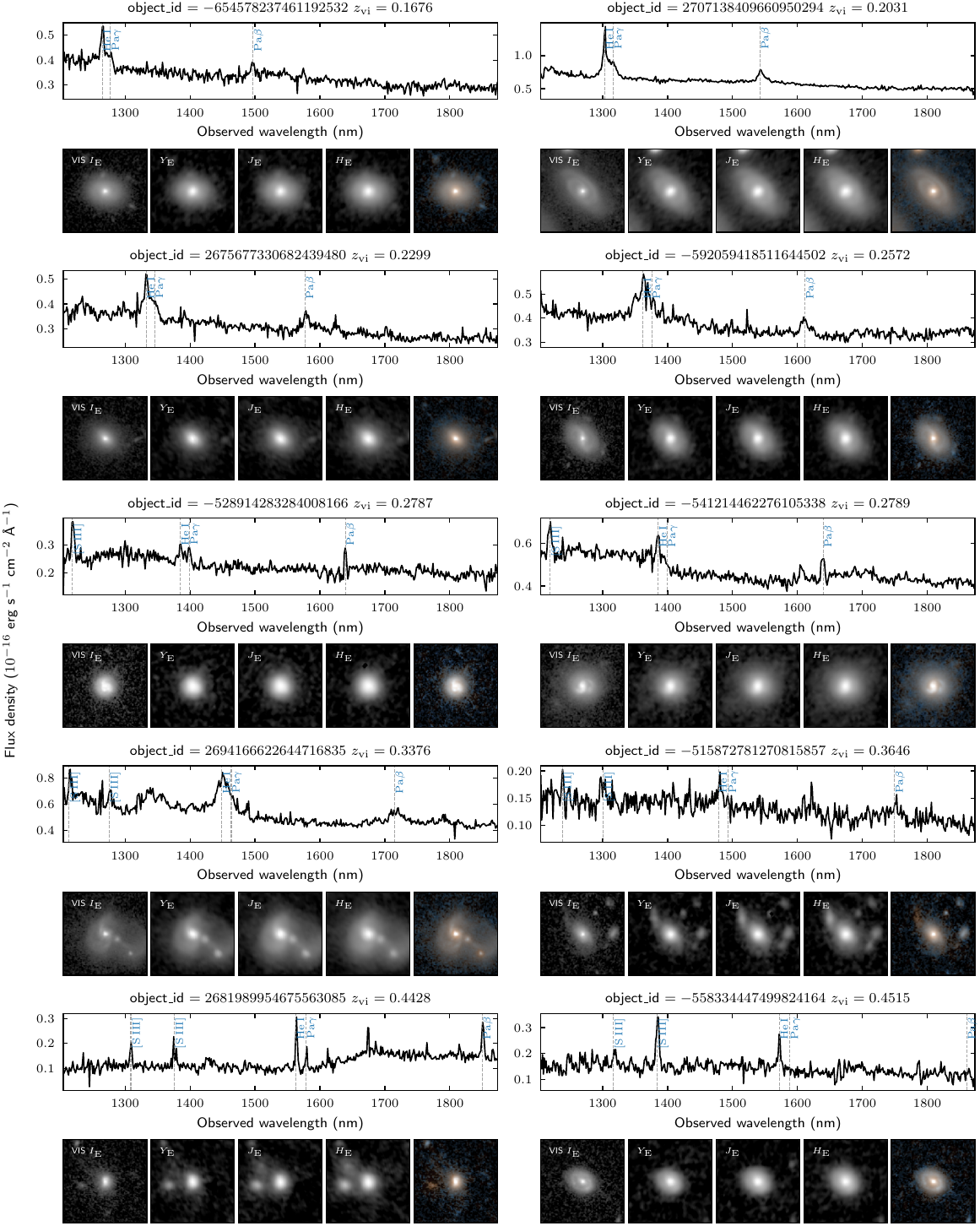}
\caption{\Euclid \ac{1d} spectra and imaging cutouts for a random sample of 10 low-redshift ($z<0.5$) sources. 
Each panel displays the \ac{1d} spectrum (top) and five imaging cutouts with $10''$ sizes (bottom) in the \IE, \YE, \JE, and \HE bands, as well as a VIS-\YE composite. The composite image is generated by mapping the VIS and \YE\ fluxes into the blue and red channels, respectively, with their mean used for green, and the VIS band used to define overall luminosity in the L*a*b* colour space \citep{iso11664-4} to enhance morphological detail. Major emission lines detected in the wavelength range [12\,047, 18\,734] \AA\ are marked. }
\label{fig:cutouts-lowz}
\end{figure*}

\begin{figure*}[htbp]
\centering
\includegraphics[width=1\linewidth]{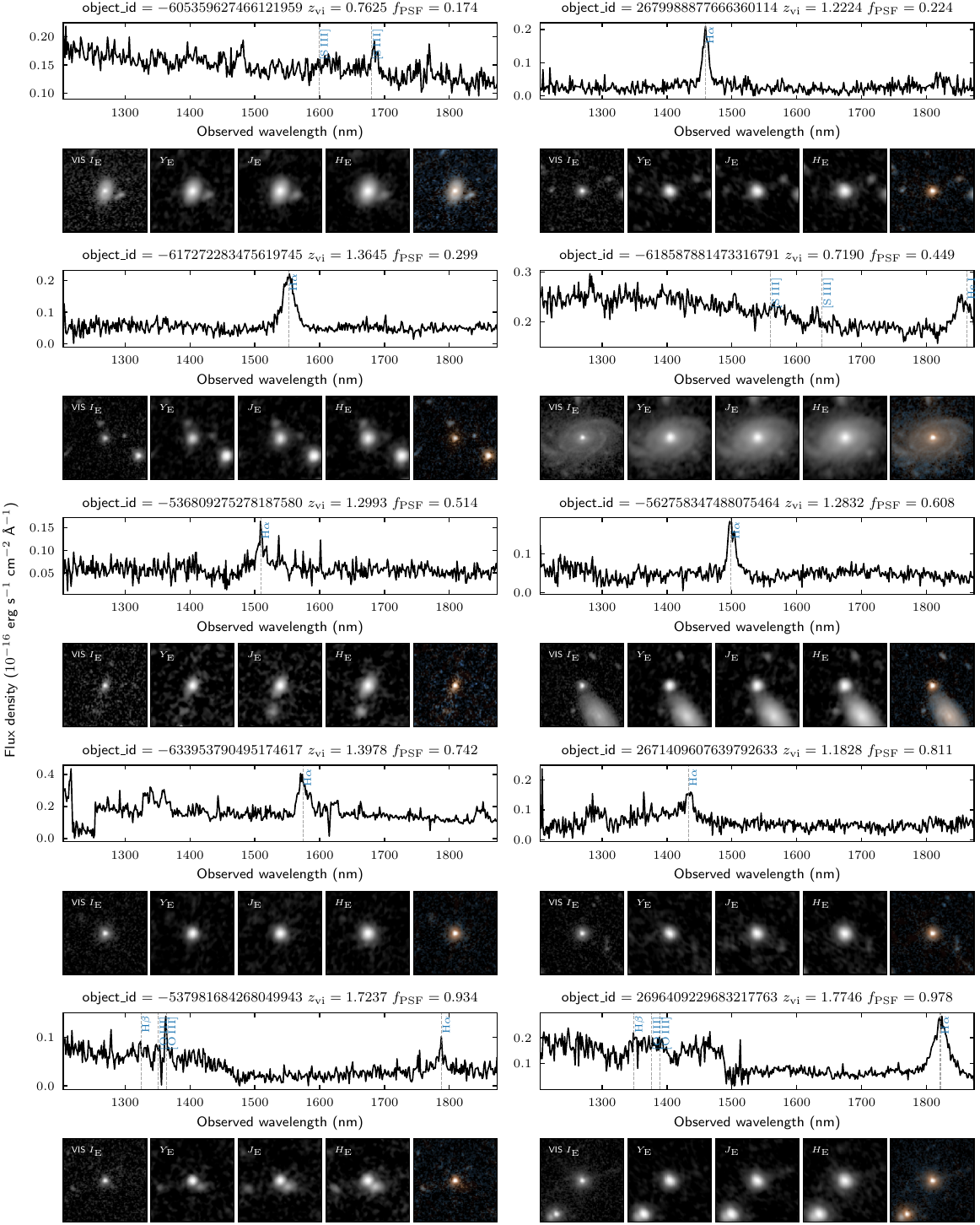}
\caption{Same as \cref{fig:cutouts-lowz}, but for 10 intermediate-redshift ($0.5<z<2$) quasars with different $f_{\sfont{PSF}}$ levels.}
\label{fig:cutouts-fpsf}
\label{LastPage}
\end{figure*}

We present the \Euclid spectra and image cutouts of a sample of low-redshift ($z<0.5$) quasars in \cref{fig:cutouts-lowz}, and a sample of intermediate-redshift ($0.5<z<2$) quasars in \cref{fig:cutouts-fpsf}.

\end{appendix}

\acrodef{smbh}[SMBH]{super massive black hole}
\acrodef{agn}[AGN]{active galactic nucleus}
\acrodefplural{agn}[AGNs]{active galactic nuclei}
\acrodef{uv}[UV]{ultraviolet}
\acrodef{nuv}[NUV]{near-ultraviolet}
\acrodef{bl}[BL]{broad-line}
\acrodef{lf}[LF]{luminosity function}
\acrodef{qlf}[QLF]{quasar luminosity function}
\acrodef{nir}[NIR]{near-infrared}
\acrodef{esa}[ESA]{European Space Agency}
\acrodef{snr}[S/N]{signal-to-noise ratio}
\acrodef{ew}[EW]{equivalent width}
\acrodef{rms}[RMS]{root mean square}
\acrodef{irtf}[IRTF]{Infrared Telescope Facility}
\acrodef{fwhm}[FWHM]{full width at half maximum}
\acrodef{nisp}[NISP]{Near-Infrared Spectrometer and Photometer}
\acrodef{psf}[PSF]{point spread function}
\acrodef{1d}[1D]{one-dimensional}
\acrodef{sed}[SED]{spectral energy distribution}
\end{document}